\documentclass{emulateapj}
\lefthead{Laha et al.}

\usepackage{times}
\usepackage{url}
\usepackage{graphicx}
\usepackage[usenames]{color}
\DeclareGraphicsExtensions{.pdf,.png,.jpg,.mps,.eps,.ps}
\usepackage{amsmath}
\usepackage{natbib}
\usepackage{pdflscape}
\usepackage{blindtext}
\usepackage{longtable}
\usepackage{tabu}

\def\unit #1{\,{\rm #1}}

\newcommand\kms{\rm \,\unit{km\,s^{-1}}}

\newcommand\cmsqi{\rm \,\unit{cm^{-2}}}

\newcommand\kev{\rm \,\unit{keV}}

\newcommand\funit{\rm \,erg\,cm^{-2}\,s^{-1}}

\newcommand\lunit{\rm \,erg \,s^{-1}}

\newcommand\lbol{L_{\rm \, bol}}

\newcommand\msolyi{M_{\odot}\,\rm yr^{-1}}
\newcommand\lsol{L_{\odot}}

\newcommand\nh{\rm N_{H}}

\newcommand\ks{\, \rm ks}

\newcommand\pc{\unit{pc}}

\newcommand\kpc{\unit{kpc}}

\newcommand\lsoft{L_{\rm 0.6-2 \kev}}
\newcommand\lhard{L_{\rm 2-10 \kev}}
\newcommand\lx{L_{\rm 2-10 \kev,\, X-ray \, spectra}}
\newcommand\lmu{L_{\rm 2-10 \kev,\, 12\mu m}}
\newcommand\lagn{L_{\rm AGN}}
\newcommand\lsb{L_{\rm Starburst}}

\newcommand\chandra{{\it Chandra}}

\newcommand\xmm{{\it XMM-Newton}}
\newcommand\nustar{{\it NuSTAR}}

\newcommand\msol{M_{\odot}}

\newcommand\mout{\dot{M}_{\rm out}}

\def\aap{A\&A} \def\apjl{ApJ}  
 \def\apjs{ApJS}

\usepackage{graphicx}
\usepackage{longtable}

\begin{document}

\title{A study of X-ray emission of galaxies hosting molecular outflows (MOX sample).}

\author{Sibasish Laha\altaffilmark{1}, Matteo Guainazzi\altaffilmark{2}, Enrico Piconcelli\altaffilmark{3}, Poshak Gandhi\altaffilmark{4}, Claudio Ricci\altaffilmark{5,}\altaffilmark{6,}\altaffilmark{7}, Ritesh Ghosh\altaffilmark{8}, \\ Alex G. Markowitz\altaffilmark{9,1}, and Joydeep Bagchi\altaffilmark{10}.} 

\altaffiltext{1}{{University of California, San Diego, Center for Astrophysics and Space Sciences, 9500 Gilman Dr, La Jolla, CA 92093-0424, USA.}; {\tt email: slaha@ucsd.edu, sib.laha@gmail.com}} 
\altaffiltext{2} {European Space Research and Technology Centre, Keplerlaan 1, 2201 AZ Noordwijk, Netherlands.}
\altaffiltext{3} {Observatorio Astronomico di Roma (INAF), via Frascati 33, 00040, Monte Porzio Catone, Roma, Italy.}
\altaffiltext{4}{Department of Physics and Astronomy, University of Southampton, Highfield, Southampton SO17 1BJ, UK.} 
\altaffiltext{5}{N\'ucleo de Astronom\'ia de la Facultad de Ingenier\'ia, Universidad Diego Portales, Av. Ej\'ercito Libertador 441, Santiago, Chile}
\altaffiltext{6}{Kavli Institute for Astronomy and Astrophysics, Peking University, Beijing 100871, China}
\altaffiltext{7}{Chinese Academy of Sciences South America Center for Astronomy, Camino El Observatorio 1515, Las Condes, Santiago, Chile}
\altaffiltext{8}{Visva-Bharati University, Santiniketan, Bolpur 731235, West Bengal, India.}
\altaffiltext{9}{Nicolaus Copernicus Astronomical Center, Polish Academy of Sciences, Bartycka 18, PL-00-716 Warszawa, Poland.}  
\altaffiltext{10}{Inter University Centre for Astronomy and Astrophysics, Post Bag 4, Ganeshkhind, Pune, India.}

\begin{abstract}
	
	We have carried out an extensive X-ray spectral analysis of a sample of galaxies exhibiting molecular outflows (MOX sample), to characterize the X-ray properties and investigate the effect of AGN on the dynamical properties of the molecular outflows. We find that the X-ray bolometric correction ($\lhard/\lagn$) of these sources ranges from $\sim10^{-4.5}$ to $10^{-0.5}$, with $\sim 70\%$ of the sources below $10^{-2}$, implying a weak X-ray emission relative to the AGN bolometric luminosity ($\lagn$). However, the upper limit on the $2-10\kev$ luminosity ($\lmu$) obtained from $12\mu$m flux, following the correlation derived by Asmus et al., are $\sim 0.5-3$ orders of magnitude larger than the $\lhard$ values estimated using X-ray spectroscopy, implying a possibility that the MOX sources host normal AGN (not X-ray weak), and their X-ray spectra are extremely obscured. We find that both $\lhard$, and $\lagn$ correlates strongly with the molecular outflow velocity as well as the mass outflow rates ($\mout$), implying that the central AGN plays an important role in driving these massive outflows. However, we also find statistically significant positive correlations between the starburst emission and MO mass outflow rate, $\lsb$ vs $\mout$, and $\lsoft$ vs $\mout$, which implies that starbursts can generate and drive the molecular outflows. The correlations of MO velocity and $\mout$ with AGN luminosities are found to be stronger compared to those with the starburst luminosities. We conclude that both starbursts and AGN play crucial role in driving the large scale MO.

 \end{abstract}

\keywords{galaxies: Seyfert, X-rays: galaxies, AGN, Molecular outflows, X-ray, Galaxies, Feedback.} 

\vspace{0.5cm}

\section{INTRODUCTION}

 The tight correlation between the mass of the central super massive black hole (SMBH) and the stellar bulge velocity distribution points to a coevolution of black hole and its host galaxy over cosmological times \citep{2000ApJ...539L...9F,2000ApJ...539L..13G}. However, the exact nature of the interaction is still not clearly understood. Energetic outflows detected in absorption and emission in different wavelength bands have been postulated to be important mechansims responsible for galaxy SMBH co-evolution \citep[see for e.g.,][ and references therein]{2012ARA&A..50..455F}. 

With the advent of high spatial resolution IR and radio telescopes in the last couple of decades, we have made rapid progress in understanding the nature of the molecular outflows (MO), which are outflows detected using the broad CO(I-J) emission lines, OH absorption lines, HCN and SiO emission lines, and several other molecular tracers \citep[][]{2011ApJ...733L..16S,2012A&A...546A..68A,2013ApJ...776...27V,2015A&A...578A..11B,2015A&A...583A..99F,2016ApJ...826..111S,2018A&A...612A..29B}. In several cases where the host galaxies can be spatially resolved, the outflows have been found to extend to a few $\kpc$, and these are believed to be one of the most powerful mechanisms by which the SMBH deposits matter onto its host galaxy. The relation between the SMBH and MO, as well as the mechanism through which MO would interact with and deposit energy into the interstellar medium (ISM) are still poorly understood.

The effect of the central AGN on the $\kpc$ scale MO is still debated. Several investigations \citep[see for e.g.,][]{2011ApJ...733L..16S,2013ApJ...776...27V,2014A&A...562A..21C}  have revealed that the presence of an AGN in the host galaxy boosts the power of the MO. However, we still do not have a consensus on how the AGN interacts with the host galaxy molecular clouds and drives the outflows at $\kpc$ scale. \citet{2015Natur.519..436T} and \citet{2015A&A...583A..99F} have suggested that the ultra fast outflows (UFOs) detected in X-rays may interact with the ISM of the host galaxy and generate MO in an energy conserving way \citep{2012MNRAS.425..605F,2012ApJ...745L..34Z}. A more recent study by \citet{2017A&A...601A.143F} have found tight correlations between the bolometric luminosity of the AGN with the mass outflow rates of molecular outflows.

X-ray emission from active galactic nuclei (AGN) probes the innermost energetic regions where matter is accreted onto a central SMBH. The emission from the accretion process for an SMBH of mass $\sim 10^7 - 10^8 \msol$ peaks in the UV, and these photons get inverse Comptonised by a corona in AGN to yield a powerlaw spectrum which extends into the hard X-rays. X-ray photons being less obscured by dust, is a good probe of the SMBH activity. In this work we carry out a systematic study of the $0.5-10\kev$ X-ray spectral properties of the sources exhibiting molecular outflows (MOX sample hereafter). The main aim in this work is to characterise the X-ray properties of the MOX sample and investigate the effect of AGN on the dynamical properties of the MO. Several sources in the MOX sample are luminous in infra-red (See Section \ref{sec:sample} for details). Previous studies on ultra luminous infra-red galaxies (ULIRGs) revealed that these galaxies are under-luminous in X-rays \citep{2004AJ....127..758I}. A recent hard X-ray survey of six nearby ULIRGS using \nustar{} data \citep{2015ApJ...814...56T} revealed similar findings. The unabsorbed $2-10\kev$ luminosity for these sources when compared with the bolometric luminosity $\lbol$ of the AGN or the mid-IR [O {\sc{IV}}] line luminosity, are found lower than that for Seyfert 1 galaxies. However, there are a few studies which have pointed out that the ULIRGs are not actually X-ray faint, but are extremely obscured. For example, a recent work on an ULIRG UGC~5101 \citep{2017ApJ...835..179O} with \nustar{} and Swift BAT telescopes have revealed that the intrinsic $2-10\kev$ luminosity of the source is $\lhard=1.4\times 10^{43}\lunit$, which is $\sim 2.5$ times larger than those obtained by previous estimates using X-ray spectra only up to $10\kev$. The new value of $\lhard$ luminosity of UGC~5101 when compared with the luminosity of the $26\mu$m forbidden emission lines of [O IV], were found to be similar to Seyfert galaxies indicating that the source is not X-ray weak. \citet{2017MNRAS.468.1273R} in a sample study of LIRGs and ULIRGs found that these sources are heavily obscured by dust and almost $65\%$ of the sources in their sample were Compton thick. We should clearly note that estimating the intrinsic X-ray luminosity in these massive dusty galaxies is not straight forward due to the largely unknown obscuration column density, and the unknown geometry and composition of the obscurer.

This paper aims at addressing the following important questions:

 \begin{enumerate}
	 \item{Are the AGN in the galaxies hosting molecular outflows intrinsically X-ray weak?}

	 \item{Is AGN the main driver of the large scale molecular outflows?}

 \end{enumerate}

 The paper is organized as follows: Section \ref{sec:sample} describes the sample selection. It also includes the bolometric luminosity, the $12\, \mu m$ luminosity, and molecular outflow properties of the MOX sample. Section \ref{sec:obs} describes the X-ray observations for the MOX sample. Section \ref{sec:data-analysis} describes the methods employed for data analysis. Section \ref{sec:correlation} describes the correlation analysis. Section \ref{sec:results} discusses the results from the extensive X-ray analysis, followed by conclusions in Section \ref{sec:conclusions}.


\section{ Sample description}\label{sec:sample}

We have selected a sample of 47 galaxies from published literature, which have exhibited molecular outflows, as on 20th October 2016. Table \ref{Table:sources} lists the sources, their redshift and spectral classification based on previous optical and X-ray studies. Table \ref{Table:MO} lists the MO velocity and the mass outflow rates ($\mout$). These 47 sources define the MOX sample.

The molecular outflows detected in the MOX sample are either in the form of OH absorption lines at $119, \,79, \, 65 \, \rm \mu m$ or CO rotational-vibrational emission line at $115$ GHz, CO(1-0), using several state-of-the-art IR and radio telescopes such as VLT-SINFONI, Herschel-PACS, ALMA, Noema, and IRAM-PDBI. For the sources IRAS~17208$-$0014 and NGC~1433, the MO properties were derived using the transitions CO(2-1) and CO(3-2) respectively (See Table \ref{Table:MO} and Section \ref{subsec:MOsample} for details). 


From Table \ref{Table:MO} we find that four sources have CO as well as OH detections of MO. In all cases the velocity measured by the two different outflows are consistent within errors, except for the source IRAS 17208-0014, where the CO measured a velocity of $600\kms$ while OH measured a velocity of $100\kms$. We consider only the highest velocity outflow in this case, which measures the maximum impact of the central engine on the host galaxy ISM. We should note that the OH absorption and the CO emission lines may be probing entirely different clumps of molecular gas at different locations in the host galaxy. However, we find that the distributions of MO velocity and mass outflow rates ($\mout$) estimated using OH absorption features or the CO emission lines are similar for the sources in the MOX sample. Hence, we treat the velocity and $\mout$ obtained using OH and CO methods on equal footings.  

\begin{figure}
  \centering 
\includegraphics[width=8cm,angle=0]{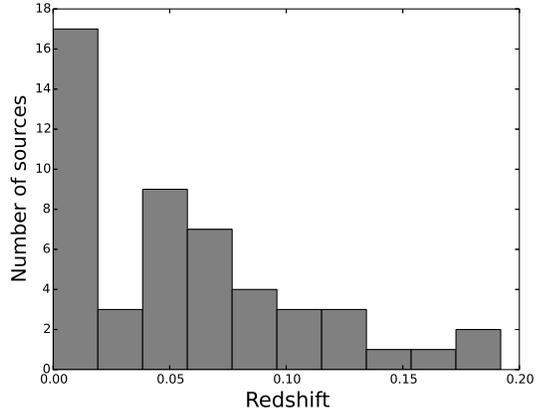} 

	\caption{ The redshift distribution of the MOX sources. } \label{fig:MOredshift}
\end{figure}

The MOX sample is not complete and can be biased towards infra-red bright objects as most of these are ULIRGS or LIRGs. Figure \ref{fig:MOredshift} left panel shows the distribution of the redshift of the galaxies in the sample, and we find that they are all in the local Universe ($z<0.2$).


\subsection{ Molecular outflow properties of the sample.}\label{subsec:MOsample}

As noted earlier, Table \ref{Table:MO} lists the molecular outflow properties of the sources along with the references from which they have been derived. We briefly describe in this section the methods used by different authors to estimate the MO properties and the threshold they have set for detecting a MO.

\citet{2011ApJ...733L..16S} detected the MO using the OH absorption lines at $\rm 79 \, \mu m$ and $\rm 119 \, \mu m$ observed using {\it Herschel}-PACS telescope. The average error on the velocity estimated by the authors is $150 \kms$. \citet{2013ApJ...776...27V} detected the MO using the OH absorption line $\rm 119 \, \mu m$ observed using {\it Herschel}-PACS telescope. The average error on the velocity estimated by the authors is $50 \kms$. The authors define a wind as an OH absorption profile whose median velocity ($v_{50}$) is more negative than $-50 \kms$ with respect to the systemic velocity. In our work, we use the quantity $v_{84}$ as the outflow velocity. $v_{84}$ is the velocity above which $84\%$ of the absorption of the OH profile takes place. \citet{2014A&A...562A..21C} studied the MO in a sample of galaxies using CO(1-0) emission lines at 115.271 GHz, observed by the IRAM-PDBI telescope. The authors have relied on the simultaneous detection of OH absorption as well as CO emission lines for a given galaxy to detect MO. In a few cases the authors could only put upper limits on velocity and hence could not effectively constrain the mass outflow rates. \citet{2016ApJ...826..111S} carried out a blind search for MO in a sample of 52 local Seyfert galaxies using the OH $119 \, \mu$m absorption line with {\it Herschel}-PACS data, and detected MO in three sources. 

\begin{figure}
  \centering 
\hbox{
\includegraphics[width=8cm,angle=0]{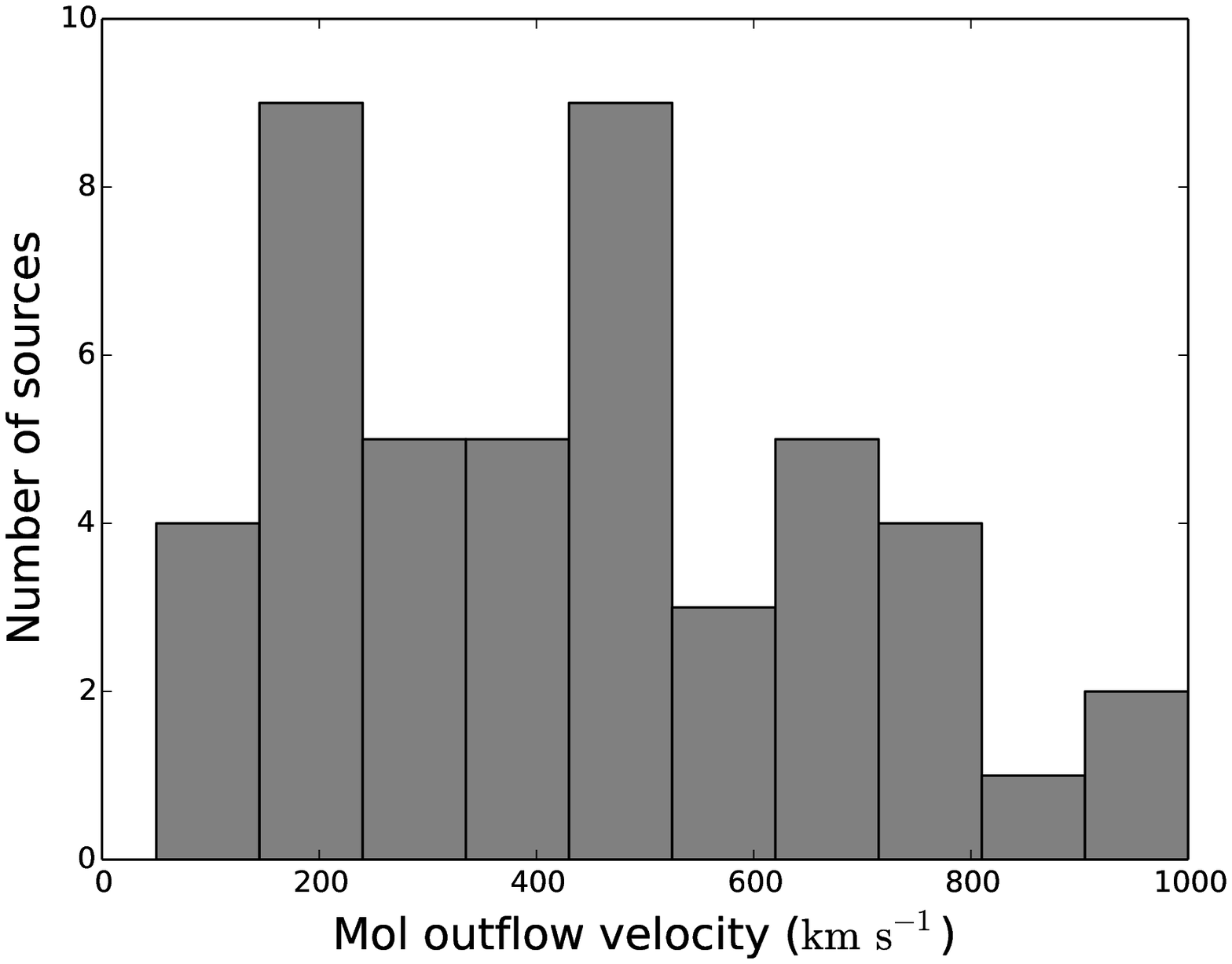} 
\includegraphics[width=8cm,angle=0]{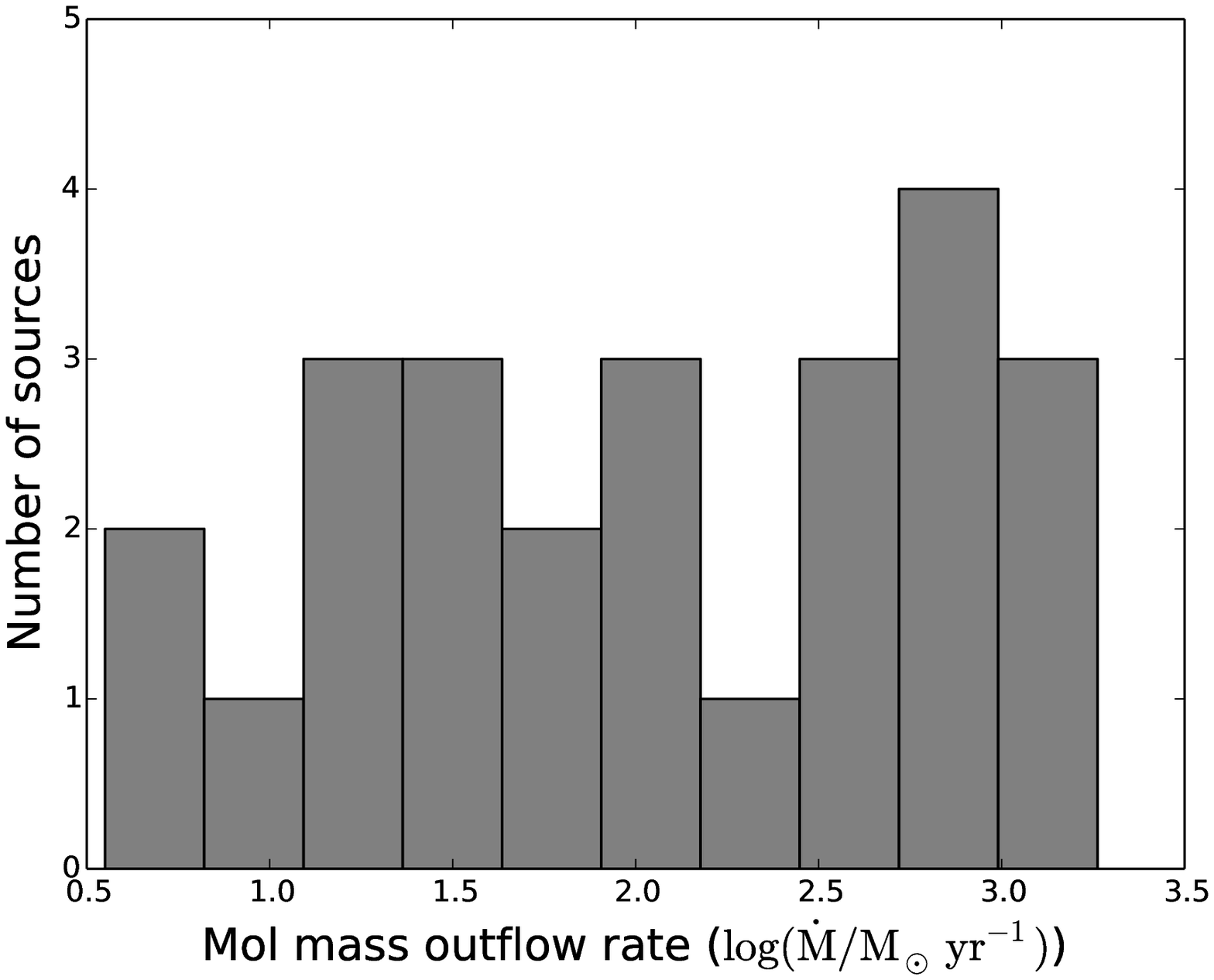} 

}\caption{{\it Left:} The distribution of MO velocity. {\it Right:} The distribution of the MO mass outflow rate. } \label{fig:MOvel}
\end{figure}

The MO velocities range from $\sim 50-1000\kms$ and the mass outflow rates $\sim 10^{0.20}-10^{3.26} \, \msolyi$. Figure \ref{fig:MOvel} left panel shows the distribution of MO velocity and right panel shows the distribution of mass outflow rates. From Table \ref{Table:MO} we note that the mass outflow rates of the MO are reported only for the first 27 sources, and for other sources they could not be calculated by the authors due to the lack of distance estimates, because of insufficient spatial resolution.

\subsection{ The bolometric luminosity of the sample.}\label{subsec:lbol}

The total bolometric luminosity $\lbol$ of the host galaxies along with the references are listed in Table \ref{Table:lbol}. The values of $\lbol$ include the AGN as well as the stellar contribution from the host galaxies calculated using the integrated infra-red luminosity ($8-1000 \mu m$) and following the scaling relation $\lbol= 1.12 L_{\rm IR}$ \citep{2014A&A...562A..21C,2013ApJ...776...27V}. We also list the AGN fraction, $\alpha_{\rm AGN}$, for each source calculated using the IR flux ratios, $f_{\rm 15 \mu m}/f_{\rm 30 \mu m}$ \citep[][]{2013ApJ...776...27V}. The bolometric luminosity of the central AGN is calculated as $\lagn=\alpha_{\rm AGN}\times \lbol$. The starburst luminosity from the galaxies are calculated using $\lsb=(1-\alpha_{\rm AGN})\times \lbol$.


\subsection{The 12 $\mu$m luminosity of the sample}\label{subsec:IRflux}

\citet{2009A&A...502..457G} have found a strong correlation between the $12\,\mu$m luminosity from the inner core of active galaxies and the $2-10\kev$ AGN X-ray luminosity ($\lhard$), indicating the effects of dust being heated by the central AGN which then re-emits in the IR. Thus the $12\,\mu$m luminosity can be used as a probe for the $\lhard$ AGN emission for the MOX sources. However, obtaining the $12\,\mu$m flux  of the spatially resolved inner core of the MOX galaxies is beyond the scope of the paper. Hence we use the $12\,\mu$m values quoted in NASA Extragalactic database (NED) obtained using the Infrared Astronomical Satellite (IRAS) observatory. Since this is the emission from the whole galaxy, we must remember that the $12\,\mu$m luminosity can have contributions from both the AGN and stellar emission. In order to obtain an estimate of the AGN contribution to the $12\,\mu$m emission we multiply the values obtained from NED by $\alpha_{\rm AGN}$ as described in the last section, and then use this quantity in the correlation in \citet{2015MNRAS.454..766A}, given by, $\log (\frac{L_{2-10\kev}}{10^{43}\lunit})=-0.32 + 0.95\times \log (\frac{L_{12\, \mu m} \times \alpha_{\rm AGN}}{10^{43}\lunit})$. However, we note that the $\lhard$ obtained using this method is possibly an upper limit to the intrinsic AGN emission as there can be other mechanisms in the host galaxy contributing to the $12\mu$m flux (See Section \ref{subsection:xrayweak} for a discussion). Table \ref{Table:l2to10} lists the values of the $\lhard$ obtained using this method. See Figure \ref{fig:L2to10} left and right panels for a comparison of the $\lhard$ estimated using the $12\mu$m luminosity and that directly measured via X-ray spectroscopy.


\section{X-ray observations and data reduction}\label{sec:obs}
We have used broadband X-ray spectra from \xmm{} EPIC-pn and \chandra{} ACIS CCD telescopes which give an energy coverage of $0.3-10\kev$ and $0.6-10\kev$ respectively. For sources where there are multiple observations, we have considered only the longest observation to maximize the signal to noise ratio irrespective of the flux or spectral state of the source. Table \ref{Table:xray} shows the list of X-ray observations used for the MOX sample. All the data used in this work are publicly available in the \xmm{} and \chandra{} archives.

The \xmm{} EPIC-pn data were downloaded and reduced using the standard Scientific-Analysis-System (SAS) software, version 15. Calibrated and concatenated event lists for the EPIC-pn camera were generated using the SAS task {\it epchain}. Good time intervals for the accumulation of scientific products were defined as those with  particle background count rate $R\le 1 \rm \, ct\, s^{-1}$ above $10 \kev$. The source region was selected using a circle of radius 40 arcsec with the centre of the circle fixed to the RA and Dec of the source obtained from NED (NASA extragalactic database). The background regions were selected from regions away from the source but from the same CCD. There was no photon pile up for any of the sources, which we checked using the command {\it epatplot}.

The \chandra{} data were reprocessed using the software {\it ciao}, version 4.7.1. The source regions were extracted from circles of radius 2.5 arc-secs with the centre of the circle fixed to the RA and Dec of the source. The background regions were selected from regions away from the source but from the same CCD. The  command {\it specextract} was used to extract the source+background spectra, the background spectra, the effective area (ARF) and the redistribution matrix (RMF). In the MOX sample there are 16 sources for which we have used \chandra{} observations. Appendix A lists the X-ray spectra, the best fit models, and the residuals of the MOX sources. In Appendix B we describe the previous studies of the sources in the MOX sample, as well as we list the details of the X-ray spectral modelling carried out in this work.


\section{ X-ray spectral analysis}\label{sec:data-analysis}

For the 39 out of 47 sources where the total photon count is $>200$ (see Table \ref{Table:xray}) we have used a combination of mostly phenomenological models, step by step, to fit the spectra. The signal to noise ratio (SNR) of several sources in the MOX sample is not sufficiently high to obtain statistically meaningful results with complex models. The simple baseline model consists of a power law absorbed by Galactic extinction \citep{2005A&A...440..775K}. A further intrinsic absorber ({\it ztbabs}) was added if the source exhibited obscuration. The model {\it APEC}  \citep{2001ApJ...556L..91S} was used to describe emission in the soft X-rays. In a few cases two {\it APEC} models were necessary to describe the soft X-ray emission. A simple black body model was used in two sources (I~ZW~1 and NGC~7172) where the {\it APEC} did not give a good fit. For sources with higher SNR, Gaussian profiles were used to describe the emission lines in soft as well as hard X-ray, specially the Fe K features in the $6-8\kev$ band. These narrow soft emission lines arise mostly due to photo-ionisation of the plasma by the central source, or reprocessing of the nuclear high-energy primary continuum by optically thick matter. The {\it diskline} profile was used to model the broad Fe K$\alpha$ profile, wherever present.

The \xmm{} EPIC-pn data were grouped using {\it specgroup} command in Scientific Analysis System (SAS), by which we ensured that each data bin has at least 20 counts and there are at most 5 data bins per resolution element. The \chandra{} observation were grouped by a minimum signal to noise ratio of $2$. We used $\chi^2$ statistics to fit the data. All errors quoted on the fitted parameters reflect the $90\%$ confidence interval for one interesting parameter corresponding to $\Delta \chi^2=2.7$ \citep{1976ApJ...208..177L}. The Interactive Spectral Interpretation System (ISIS) software \citep{2000ASPC..216..591H} was used in fitting the spectra.

Table \ref{Table:flux} lists the best fit parameters along with the $0.6-2\kev$ and $2-10\kev$ absorption-corrected flux and luminosity. The $0.6-2\kev$ luminosity quoted in the table is from the model {\it APEC} only and not the integrated continuum flux. We assume hereafter that the model {\it APEC} describes the thermal emission in the soft X-rays mostly arising from supernova remnants and star-bursts \citep[see for e.g.,][]{2013PASJ...65...44M,2013A&A...553A...7D}. The best fit temperatures of {\it APEC} lie in the range $\sim kT=0.08-1 \kev$. In most cases the AGN continuum is absorbed below $2\kev$, and from Table \ref{Table:flux} we find that almost all the sources with sufficient SNR have a neutral intrinsic column density of $\sim 10^{21}-10^{22}\cmsqi$. However, as a caveat we should note that in some cases the soft X-rays may also contain contributions from the reprocessed emission from the central AGN, such as ionized disk reflection, and can mimic thermal emission. Figures \ref{fig:7}-\ref{fig:11} show the data and the best fit model in the upper panels, and the residuals in the lower panels. Note that for the two sources M~82 and NGC~1068, the soft X-ray spectra were extremely complex and could not be fit using the simple baseline model. 

For the sources where the SNR does not permit us to constrain simultaneously the powerlaw slope and the neutral absorption column, we fix the slope to a value $\Gamma=1.8$, typical of Seyfert galaxies, and calculate the corresponding fluxes and absorption column. See Table \ref{Table:flux} for details. For the six Compton thick sources (NGC~6240, NGC~1068, NGC~1377, IRAS F08572+3915, IRAS F20551-4250, IRASF14348-1447, IRAS~13120-5453) in our sample identified from previous studies, we have multiplied the observed $\lhard$ with a factor of 100 to obtain the intrinsic $\lhard$ while carrying out the correlation analysis \citep[see for e.g.,][]{2009A&A...504...73L,2016A&A...585A.157P}. See Table \ref{Table:Finallhard} for the final $\lhard$ values of these sources and Section \ref{subsec:bestlhard} for details.  

For eight sources in the MOX sample, having total counts $<200$, we have used the $2-10\kev$ luminosity from previous studies who have employed hardness ratio method \citep[e.g.,][]{2010ApJ...725.1848T}. The hardness ratio is defined as HR=(H-S)/(H+S), where H and S are the number of counts in the hard ($2-8\kev$) and soft ($0.5-2.0 \kev$) bands respectively. The hardness ratios calculated from the data were then compared with hardness ratios generated with an absorbed power law model to estimate the model parameters \citep{2005ApJ...633..664T}.


\subsection{Assembling the best values of $\lhard$ estimated using X-ray spectroscopy}\label{subsec:bestlhard}

As inferred from Sec. \ref{sec:data-analysis}, most of the sources in the MOX sample are probably obscured by the intervening host galaxy dust and gas and therefore the $2-10\kev$ luminosity estimated from the X-ray spectral analysis of \xmm{} and \chandra{} may not give us the real picture. A better glimpse of the unobscured $\lhard$ luminosities can be obtained by analysing spectra at energies $>10\kev$ where the hard X-ray photons have lesser probability to get absorbed. \nustar{} operates in the energy range $\sim 3-40\kev$ and gives us the unique opportunity of such an intrinsic view of the $\lhard$ luminosity. In this section we discuss how we selected the best estimate of $\lhard$ available to us by different X-ray spectral analysis methods.

We assigned the highest preference to the unabsorbed $\lhard$ values estimated using \nustar{}, whenever available. Only 23 sources in the MOX sample have been observed by \nustar{} either as targets or serendipitously. We carried out a literature search on the analysis of \nustar{} data of these sources and found that out of these 23, only 10 sources have enough SNR to carry out a spectral study in the broad band $3-40\kev$. For e.g., the \nustar{} observations of the sources IRASF08572+3915 and IRASF10565+2448 found no detectable X-ray signatures in the $3-40\kev$ energy band \citep{2015ApJ...814...56T}. The intrinsic $\lhard$ values for the 10 sources obtained with \nustar{} have been quoted in Table \ref{Table:Finallhard}, along with the references from where they have been derived. For a list of the MOX sources not observed by \nustar{} see Appendix C. For the rest of the MOX sources we use the $\lhard$ estimated from the \xmm{} and \chandra{} spectroscopy and the HR method enumerated in Section \ref{sec:data-analysis}. For sources which have been previously identified as C-thick and have not been studied by \nustar{}, we have multiplied the $\lhard$ values obtained using the \xmm{} and \chandra{} spectroscopy by a factor of 100 \citep[see for e.g.,][]{2009A&A...504...73L,2016A&A...585A.157P}, to obtain an estimate of the intrinsic unabsorbed $2-10\kev$ luminosity.

\begin{figure*}
  \centering 

	\hbox{

\includegraphics[width=9.5cm,angle=0]{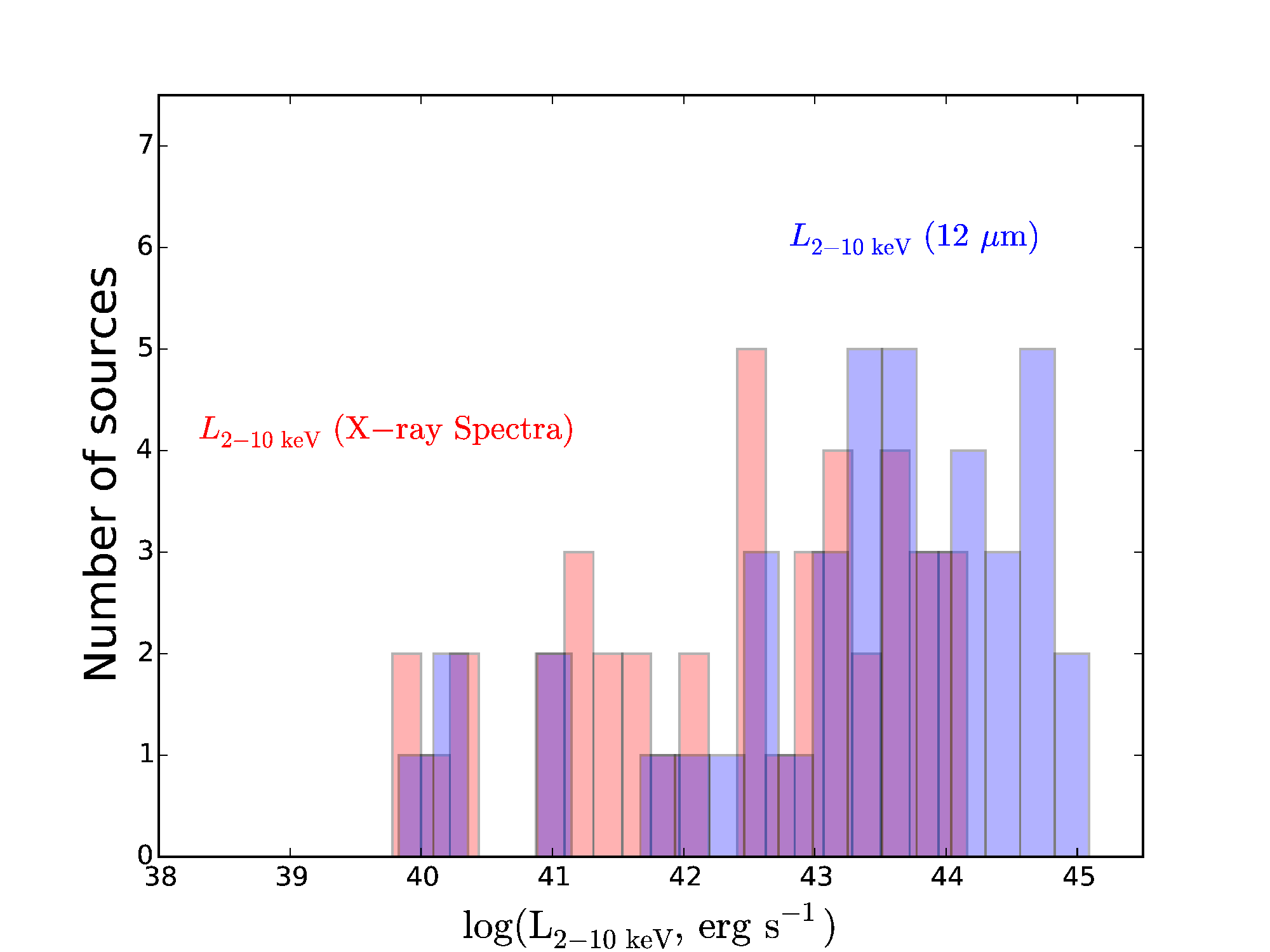}
\includegraphics[width=9.5cm,angle=0]{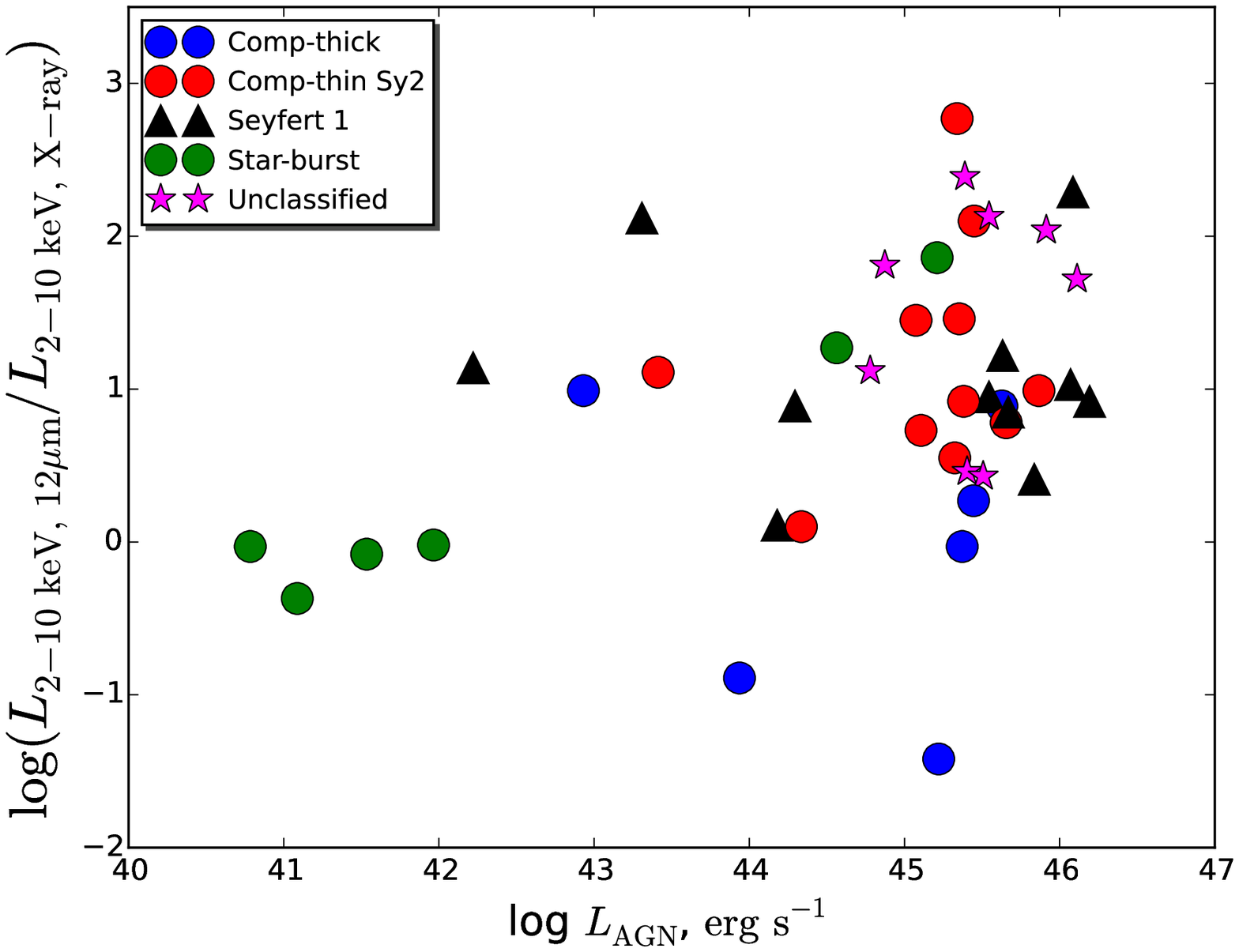} 
}

	\caption{{{\it Left:} The $2-10\kev$ luminosity, $\lhard$, distribution of the MOX sources calculated by the two methods, X-ray spectroscopy (in pink color) and $12\,\mu$m luminosity (in blue color), as described in Sections \ref{subsec:IRflux} and \ref{subsec:bestlhard}. {\it Right:} The ratio between the $\lhard$ calculated using the two methods plotted against the bolometric luminosity of the AGN $\lagn$ of the MOX sources. } The classification of source types are obtained from NED and from previous optical and X-ray studies. See Section \ref{sec:sample} and Table \ref{Table:sources} for details. The blue circles, red circles, black triangles, green circles and magenta stars denote Compton thick, Compton thin, Seyfert 1, starburst, and unclassified sources respectively. We use this classification consistently throughout the paper.} \label{fig:L2to10}
\end{figure*}

Table \ref{Table:Finallhard} column 3 lists the $\lhard$ values obtained using \xmm{} and \chandra{} spectroscopy, while column 4 lists the $\lhard$ values obtained using \nustar{}. The last column of Table \ref{Table:Finallhard} lists the best values of $\lhard$ we use in the rest of this work for analysis, which we refer to as $\lx$. In Table \ref{Table:l2to10} we compare the finally selected $\lhard$ values with those estimated using the $12\mu$m flux ($\lmu$). Fig \ref{fig:L2to10} left panel shows the distribution of the best $\lhard$ estimated above and the $\lhard$ estimated using $12\, \mu$m flux. The right panel of Figure \ref{fig:L2to10} shows the ratio $\lx/ \lmu$ plotted against the bolometric luminosity of the AGN ($\lagn$).
 
The bolometric corrections ($\lhard/\lbol$) corresponding to the two sets of $\lhard$ values are listed in Table \ref{Table:lbol}. Figure \ref{fig:Xraycorr} left and right panels show the bolometric corrections of the MOX sources with $\lx$ and $\lmu$ values respectively, plotted against the bolometric luminosity of the AGN. In the left panel of Fig. \ref{fig:Xraycorr} we have plotted in yellow triangles the bolometric corrections of the sources for which the $\lhard$ were obtained using \nustar{} broad band spectroscopy. \nustar{} provides an accurate estimate of the intrinsic $\lhard$ luminosity and hence the bolometric corrections obtained using those estimates are more reliable.


\section{Correlations}  \label{sec:correlation}

\begin{figure*}
  \centering 
\hbox{
\includegraphics[width=9cm,angle=0]{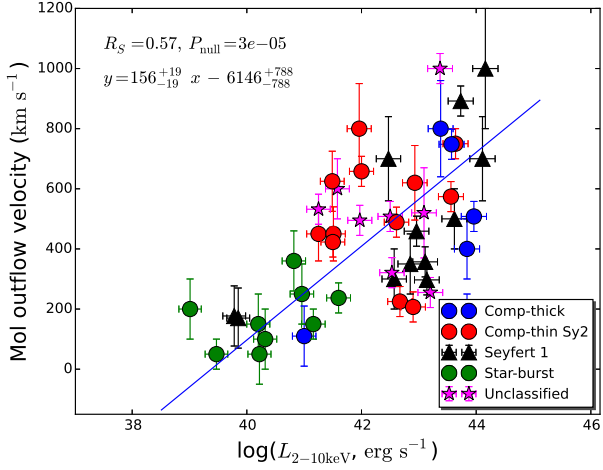} 
\includegraphics[width=9cm,angle=0]{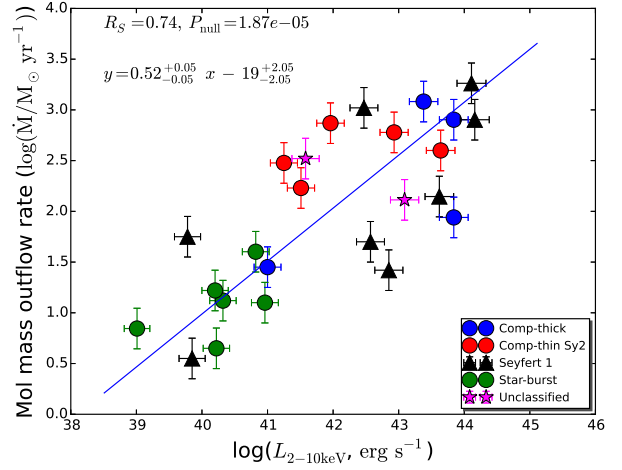}
}

	\caption{{\it Left:}The correlation between the molecular outflow velocity and the $2-10\kev$ unabsorbed luminosity of the MOX sources obtained using X-ray spectroscopy as described in Section \ref{subsec:bestlhard}.   {\it Right:} The correlation between the molecular mass outflow rate and the $2-10\kev$ unabsorbed luminosity of the sources. We have assumed an error of $0.2$ dex on the molecular mass outflow rates uniformly.  } \label{fig:corrMOLxray}
\end{figure*}

\begin{figure*}
  \centering 
\hbox{
\includegraphics[width=9cm,angle=0]{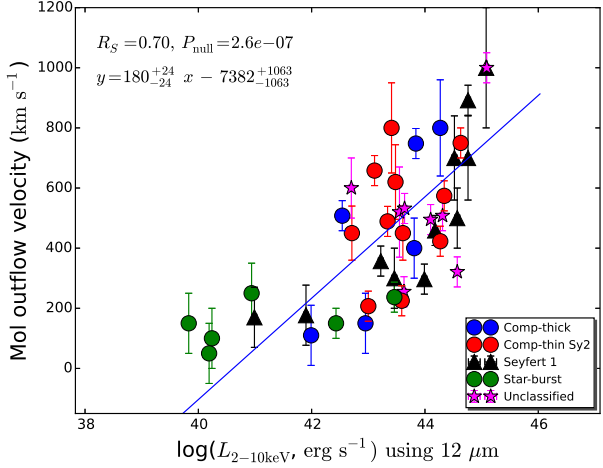} 
\includegraphics[width=9cm,angle=0]{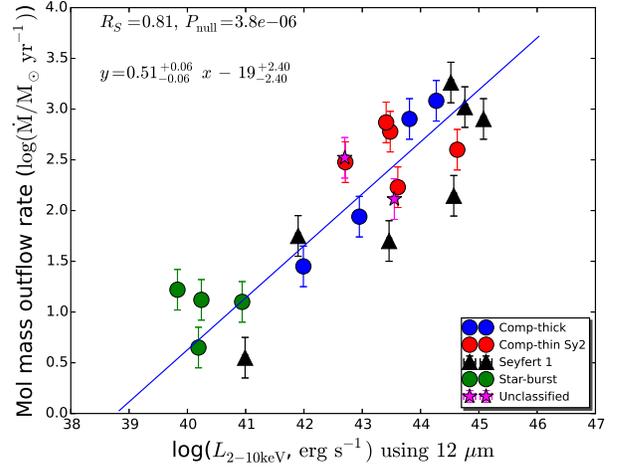}
}

	\caption{{\it Left:} Same as Figure \ref{fig:corrMOLxray} except that the $2-10\kev$ luminosity has been obtained using the $12\mu$m flux, as described in Section \ref{subsec:IRflux}. {\it Right:} Same as Figure \ref{fig:corrMOLxray} except that the $2-10\kev$ luminosity has been obtained using the $12\mu$m flux.  } \label{fig:corrMOLxray12micron}
\end{figure*}

\begin{figure*}
  \centering 
\hbox{
\includegraphics[width=9cm,angle=0]{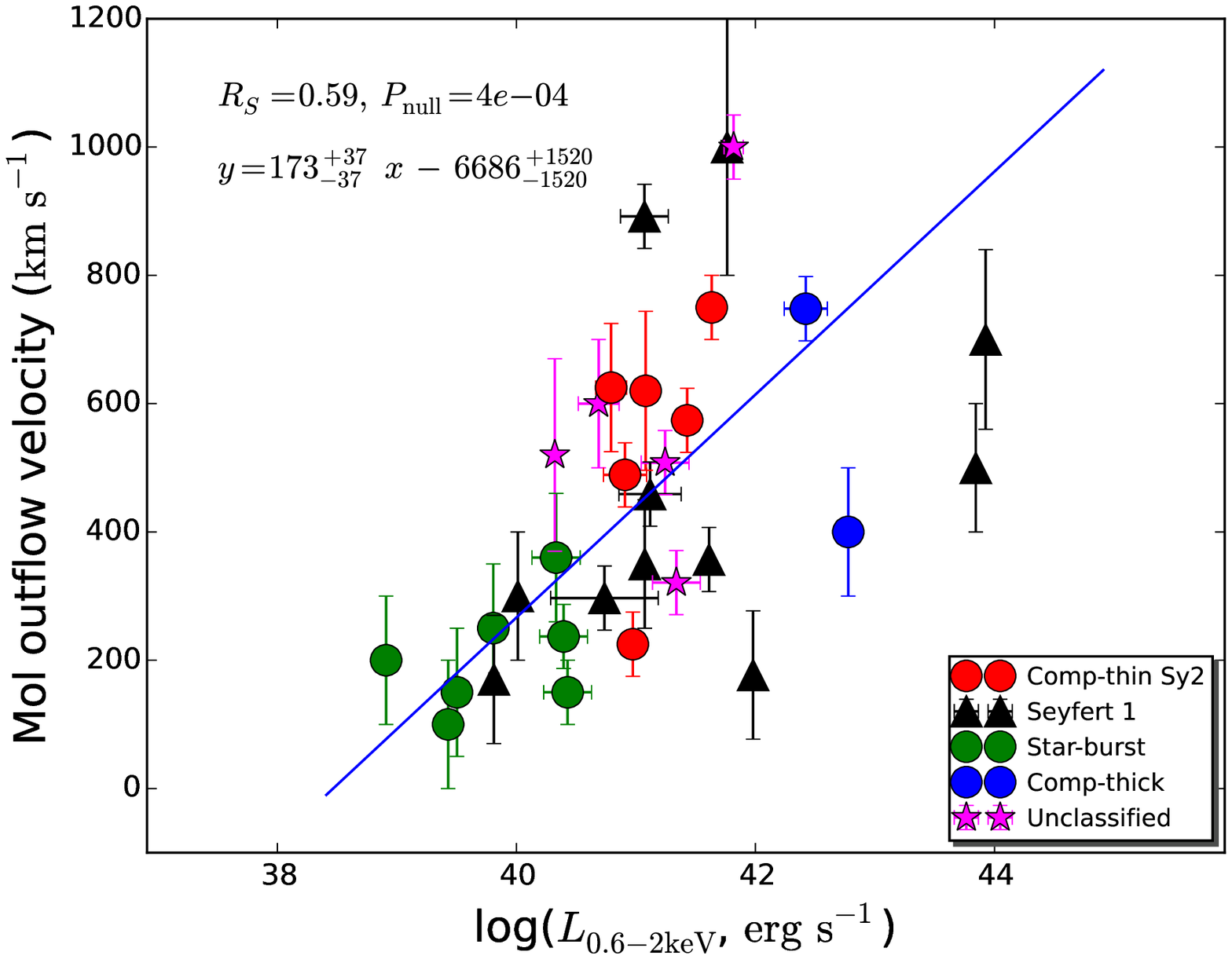} 
\includegraphics[width=9cm,angle=0]{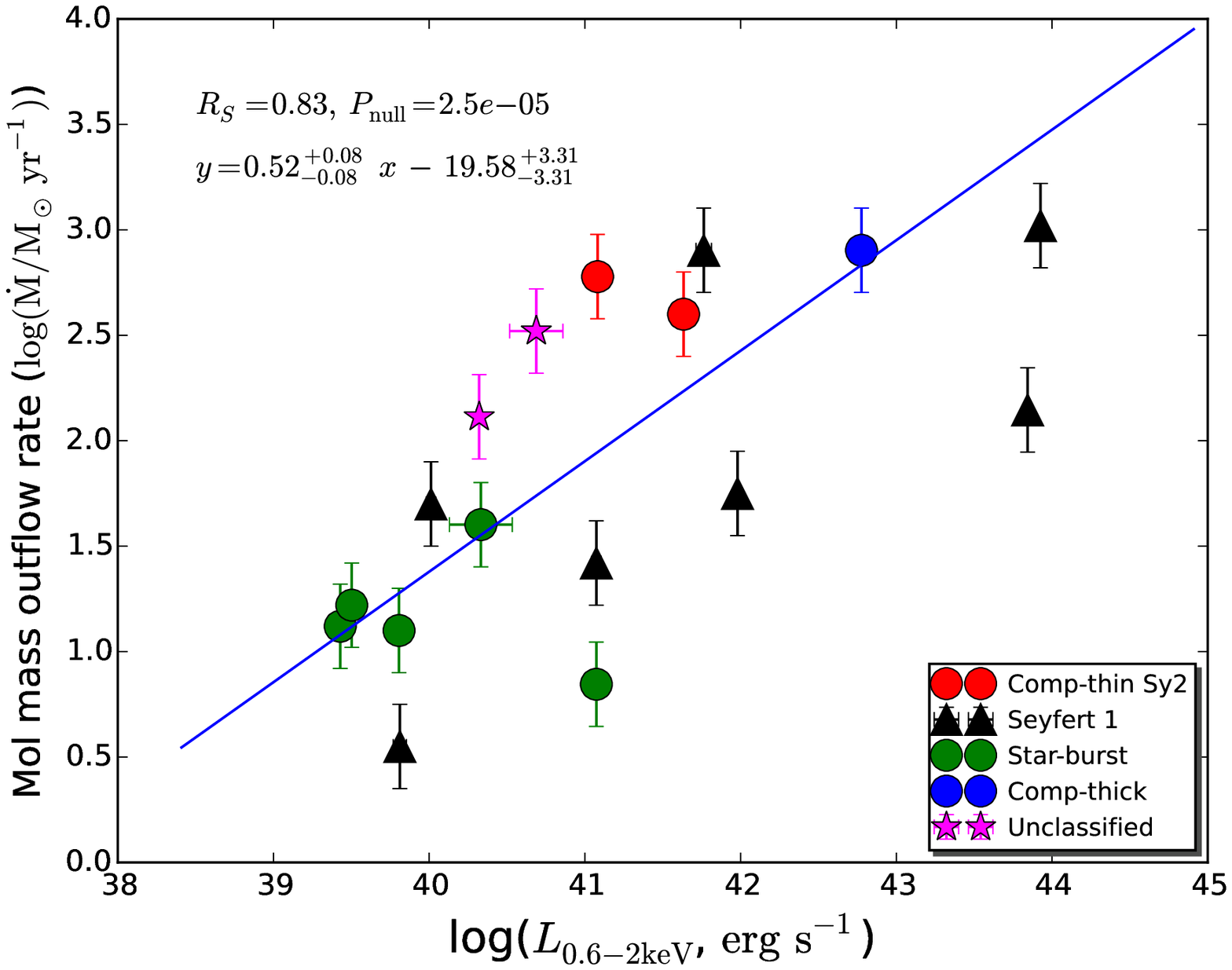}

}
	\caption{{\it Left:} The dependence of Mol outflow velocity on the $0.6-2\kev$ {\it APEC} luminosity. {\it Right:} The dependence of Mol mass outflow rate on the $0.6-2\kev$ {\it APEC} luminosity. Symbol description as in Figure \ref{fig:L2to10}} \label{fig:5}
\end{figure*}

\begin{figure*}
  \centering 
\hbox{
\includegraphics[width=9cm,angle=0]{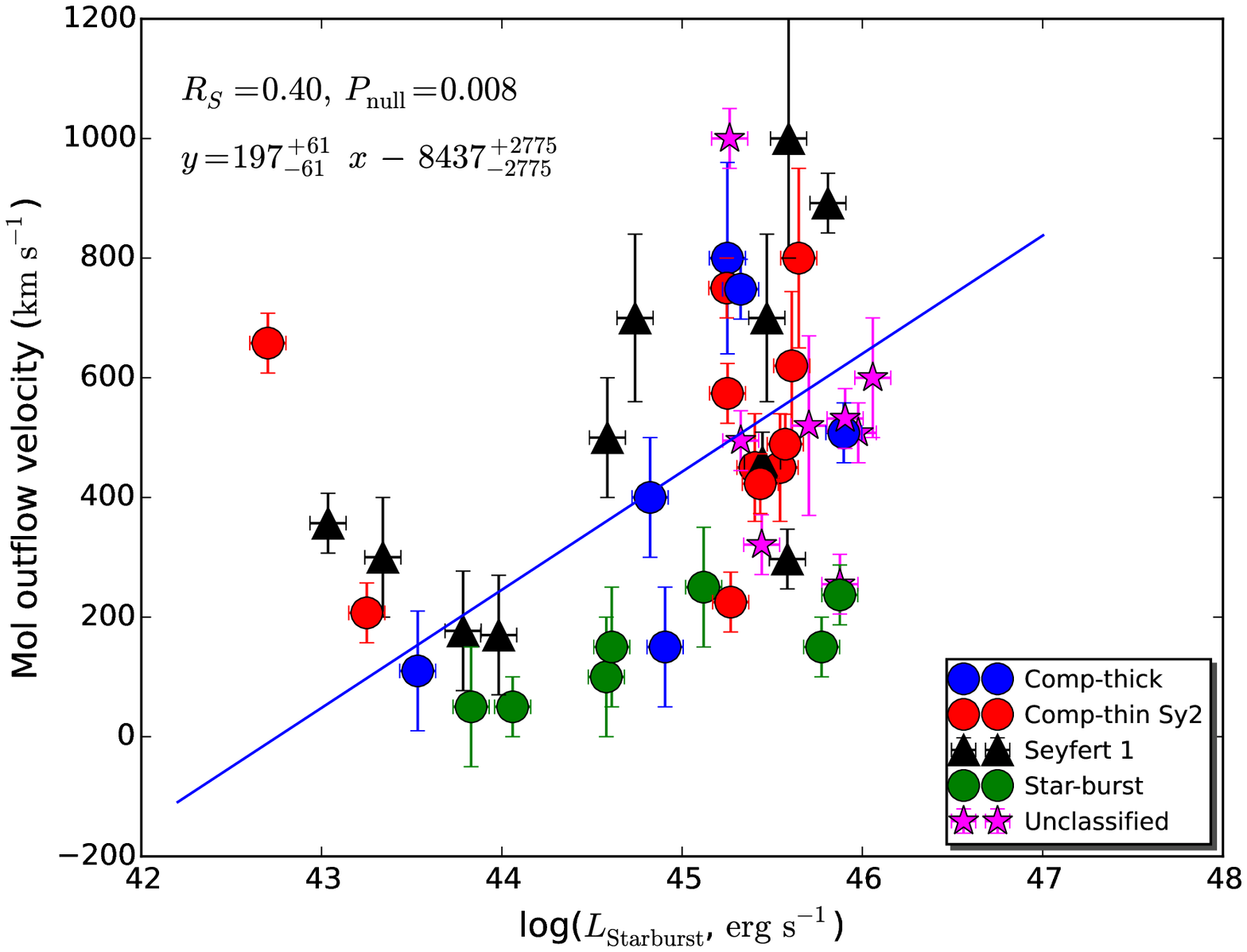} 
\includegraphics[width=9cm,angle=0]{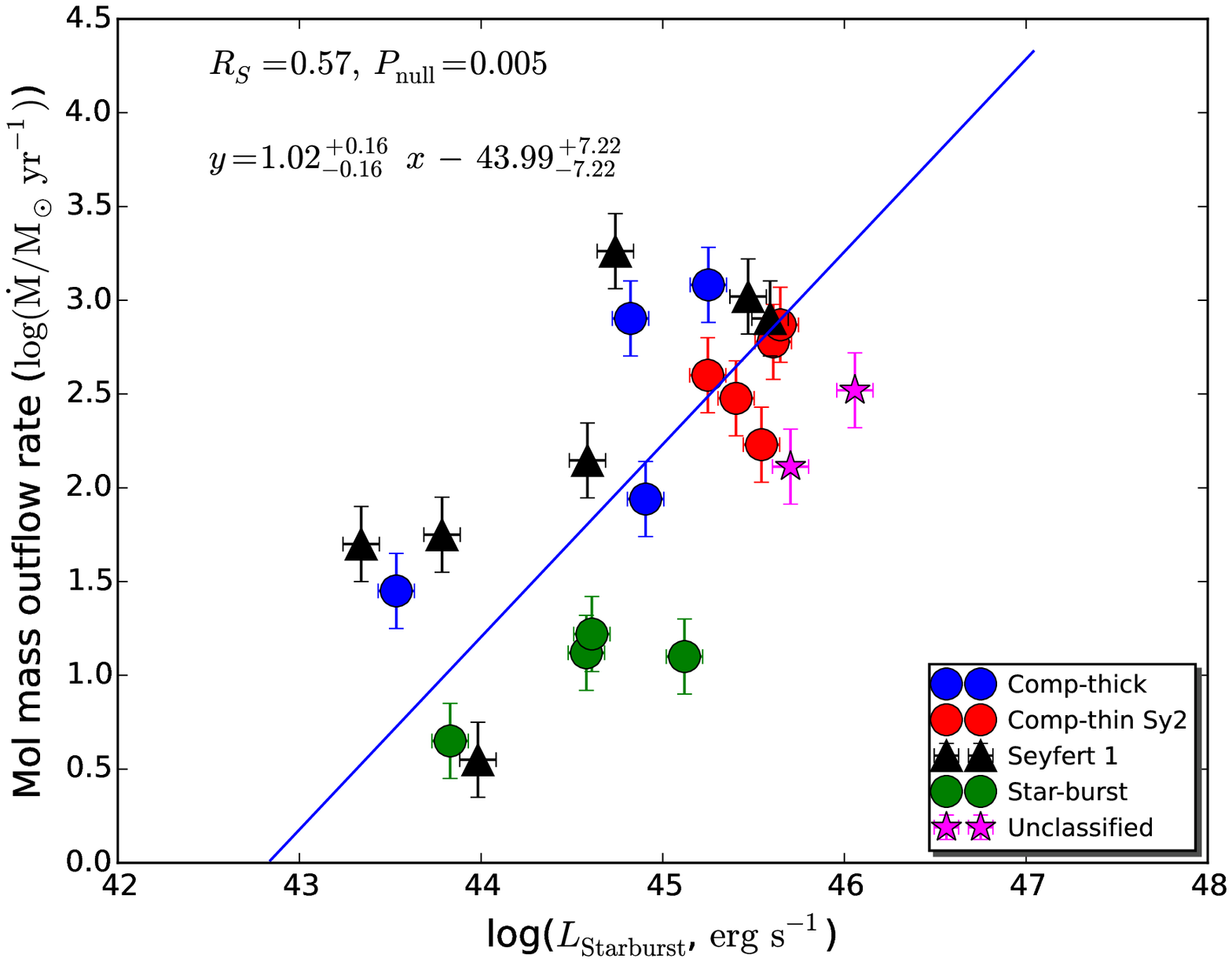}

}
	\caption{{\it Left:} The dependence of Mol outflow velocity on the starburst luminosity, $\lsb$, as calculated in Section \ref{subsec:lbol}. {\it Right:} The dependence of Mol mass outflow rate on the starburst luminosity. Symbol description as in Figure \ref{fig:L2to10}} \label{fig:starburst}
\end{figure*}

To test the dependence of molecular outflow kinematics on AGN activity, we have correlated the X-ray luminosity in the two energy bands, $\lsoft$ ({\it APEC}) and $\lhard$, as well as the AGN bolometric luminosity $\lagn$ with the MO velocity and mass outflow rates ($\mout$). We have also correlated the starburst luminosity, $\lsb$, with MO velocity and mass outflow rates. Table \ref{Table:corr} lists the non-parametric Spearman rank coefficient, the null hypothesis probability, as well as the linear regression slope and intercept for these correlations. The number of data points involved in each correlation are also quoted in Table \ref{Table:corr}. The difference in the number of data points arises due to the fact that some of the sources in the MOX sample do not have mass outflow rate estimates, and also for a few sources we do not have an estimate of the $\lsoft$ ({\it APEC}) and $\lagn$. Figures \ref{fig:corrMOLxray}-\ref{fig:starburst} show the correlation between the $\lsoft$ ({\it APEC}), $\lhard$, the $\lagn$ and the $\lsb$ luminosities with the MO dynamical parameters ($v$ and $\mout$). The starburst galaxies are plotted in green circles, and they occupy the phase space of lowest X-ray and AGN luminosity and lowest MO velocity as well as the mass outflow rates. The black triangles, red circles, and the blue circles denote the Seyfert 1 galaxies, Compton-thin Seyfert 2 galaxies, and the Compton-thick galaxies respectively. The magenta stars denote the unclassified sources. 

From Figures \ref{fig:corrMOLxray} and \ref{fig:corrMOLxray12micron} we find that the $2-10\kev$ luminosities of the MOX sources, $\lx$ and $\lmu$ respectively, show strong correlation with MO velocity and $\mout$, with a confidence $>99.99\%$. In both the figures we find that the mass outflow rate $\mout$ correlates better than that of the MO velocity. Similarly, Figure \ref{fig:corrMOLbol} shows that both the MO velocity and $\mout$ strongly correlate with the AGN bolometric luminosity, with a confidence $>99.99\%$. From Figures \ref{fig:5} and \ref{fig:starburst} we find that the $\lsoft$ and $\lsb$ correlates with the MO velocity and $\mout$ with a confidence $>99\%$, but the correlations are not as strong as those with the AGN X-ray and bolometric luminosity. We discuss the implications of these results in Section \ref{sec:results}

We have used the freely available Python code by \citet{2012Sci...338.1445N} using the BCES technique \citep{1996ApJ...470..706A} to carry out the linear regression analysis between the quantities mentioned above. In this method the errors in both variables defining a data point are taken into account, as is any intrinsic scatter that may be present in the data, in addition to the scatter produced by the random variables. The strength of the correlation analysis was tested using the non-parametric Spearman rank correlation method.


\section{ Results and discussion}\label{sec:results}

We have carried out a uniform X-ray spectral analysis of a sample of 47 sources exhibiting molecular outflows and obtained the best estimates of $\lhard$ values using X-ray spectroscopy. We have also estimated the $\lhard$ luminosity using $12\mu$m flux. As a caveat we note that estimating the intrinsic X-ray luminosity in these massive dusty galaxies is not straight forward due to large uncertainties in the obscuration along the line of sight, and most estimates are based on several assumptions. In this section we discuss the main results.


\subsection{Are the AGN in the MOX sources X-ray weak?}\label{subsection:xrayweak}

The MOX sources are bright in IR and hence it is possible that large columns of neutral gas and dust obscures our line of sight and we do not observe the intrinsic $\lhard$ for most of the galaxies. In this section we therefore investigate whether the MOX sources are extremely Compton thick or the AGN at the centre of the galaxies are indeed X-ray weak.

The NED classification of the MOX sources as listed in Table \ref{Table:sources} shows that 33 out of 47 sources are ULIRGs or LIRGs, implying that they have large columns of gas  and dust emitting in the infra-red. A systematic study of the ULIRGs in the X-rays using the broad band \chandra{} and \xmm{} data were carried out by \citet{2010ApJ...725.1848T}, and the authors noted that possibly in the ULIRGs we are capturing the nascent stages of AGN activity \citep{1988ApJ...325...74S}, in which case the central AGN emission could be weak and starburst emission dominates the total power. In a more recent study by \citet{2015ApJ...814...56T} using \nustar{} observations of six ULIRGs, the authors conclude that these sources are indeed X-ray weak and not obscured. The typical example is that of MRK~231, which is a merger remnant containing both intense starburst as well as a luminous AGN at its centre. MRK~231 which was earlier thought to be a Compton thick AGN, was found by the authors to be intrinsically X-ray weak using the $4-80\kev$ \nustar{} spectra \citep{2014ApJ...785...19T}. The X-ray bolometric correction ($\lhard/\lagn$) estimated by the authors for the six sources in their sample were found in the range from $8\times 10^{-4}$ to $10^{-2}$ indicating that the AGN at the centre of these sources are X-ray weak. Particularly for the two sources MRK~231 and IRAS~08572+3915 they are remarkably low, at $\sim 5\times 10^{-4}$ and $<10^{-4}$ respectively. Normally for Seyfert galaxies these values lie in the range $0.02-0.15$ \citep[see for e.g.,][and references therein]{1994ApJS...95....1E,2009MNRAS.399.1553V,2010MNRAS.402.1081V}. The authors rule out obscuration as the cause for the X-ray weakness. They conclude that possibly the AGN is accreting at super Eddington rates, in which case the UV bump dominates, or else the presence of large scale outflows may have quenched the X-ray emission in the AGN. 

\begin{figure*}
  \centering 

	\hbox{
\includegraphics[width=9.5cm,angle=0]{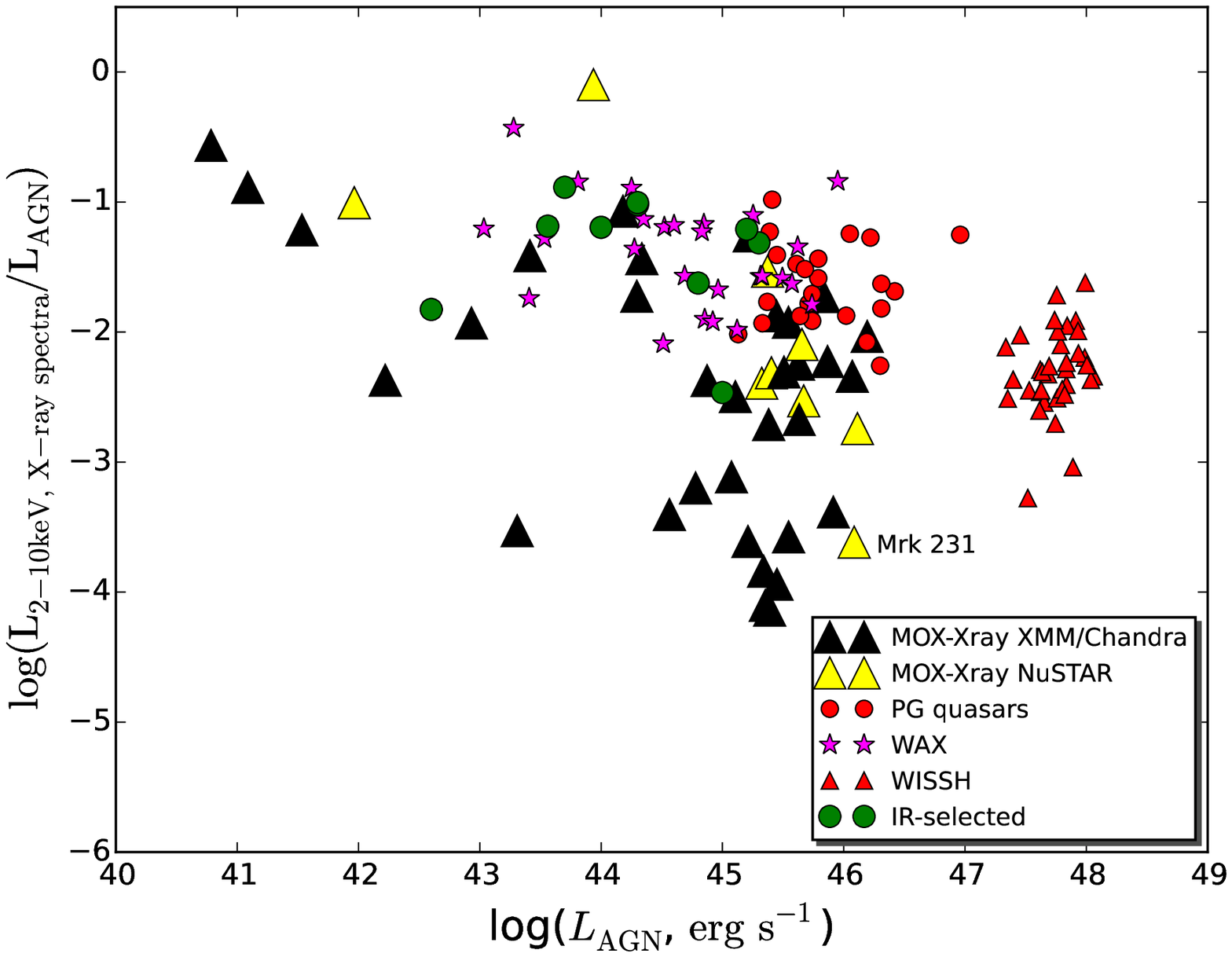} 
\includegraphics[width=9.5cm,angle=0]{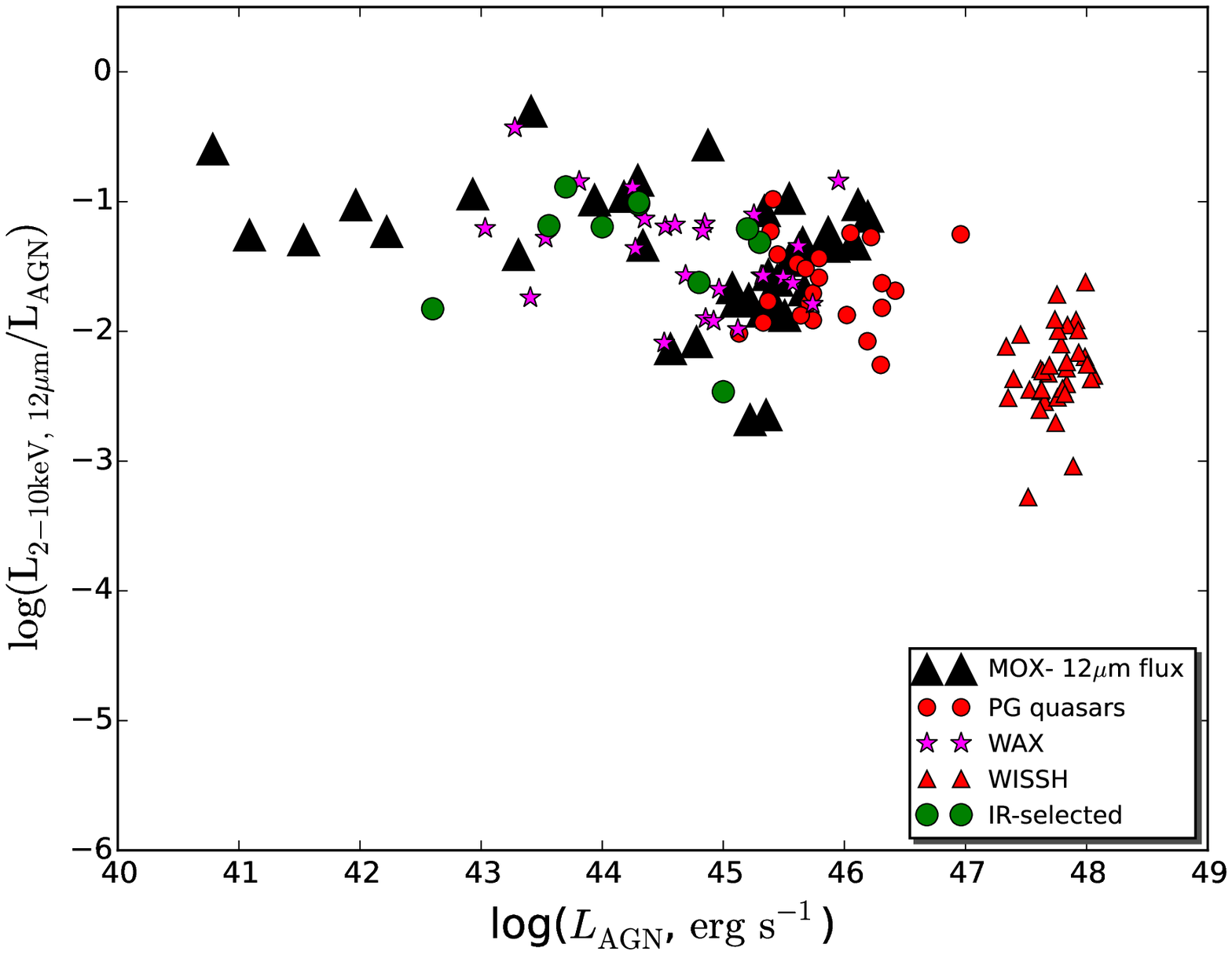} 
}

	\caption{{\it Left:}The X-ray bolometric correction $\log(L_{2-10\kev}/\lagn)$ vs $\lagn$ of the MOX sources along with the WAX \citep{2014MNRAS.441.2613L}, PG quasars \citep{1994ApJ...435..611L}, WISSH quasars \citep{2017A&A...608A..51M} and the $12\mu$m selected AGN \citep{2011MNRAS.413.1206B}, as described in section \ref{subsection:xrayweak}. Here the $\lhard$ values of the MOX sources have been estimated using the method described in Section \ref{subsec:bestlhard} and listed in Table \ref{Table:Finallhard}. As \nustar{} gives us the best estimate of the unabsorbed $\lhard$, we have plotted those sources in yellow triangles to separate them from those estimated using \xmm{} and \chandra{} spectroscopy. {\it Right:} Same as left, but here the $\lhard$ of the MOX sources are estimated using $12\,\mu$m luminosity.} \label{fig:Xraycorr}
\end{figure*}

From Figure \ref{fig:Xraycorr} left panel and Table \ref{Table:lbol}, we find that the distribution of the bolometric correction ($\lhard/\lagn$) of the MOX sample ranges from $10^{-4.5}$ to $10^{-0.5}$, with $70\%$ of the sources having X-ray bolometric correction below $10^{-2}$. To compare the bolometric correction of the MOX sources with Seyfert and quasars we selected four AGN samples at different redshift ranges and well studied in X-rays: 1. The warm absorbers in X-rays (WAX) sample, \citet{2014MNRAS.441.2613L}. This sample consists of 26 nearby ($z<0.06$) Seyfert 1 galaxies, with an X-ray luminosity of $10^{42}<L_{2-10\kev}<10^{45} \lunit$ .  2) `Palomar Green (PG) quasars', \citet{1994ApJ...435..611L}, consists of quasars in a redshift range $z=0.06-1.72$, with an X-ray luminosity of $10^{43}<L_{2-10\kev}<10^{46} \lunit$, 3) The WISSH quasar sample \citep{2017A&A...608A..51M}, consisting of WISE-SDSS selected high redshift quasars ($z=3-4$) with an X-ray luminosity of $10^{44}<L_{2-10\kev}<10^{46} \lunit$ . 4) The $12\mu$m selected AGN sample by \citet{2011MNRAS.413.1206B}, for which we have used a subsample of 10 sources which are type-1 AGN having well estimated values of $\lhard$ and $\lbol$ \citep{2009MNRAS.392.1124V}. The type-1 constraint on these IR bright sources ensures that we obtain an unobscured view of the central engine. In Figure \ref{fig:Xraycorr} left panel we have over-plotted the bolometric correction ($\lhard/\lbol$) vs  $\lbol$ of these comparison samples along with the MOX sample. We find that the X-ray bolometric corrections of most of the MOX sources are orders of magnitude lower than that of the Seyfert galaxies and the quasars. We also find that the \nustar{} estimates of the bolometric correction for the MOX sources (plotted as yellow triangles) are nearly similar to the Seyfert galaxies and quasars except for the source Mrk~231, which has a correction of $\log(\lx/\lagn)=-3.62$. Therefore, it may be possible that most of the MOX sources are not X-ray weak, instead they are heavily obscured.

The broad absorption line quasars (BAL) have also been found to be extremely X-ray weak. A study of two BAL quasars, PG~1004+130 (radio loud) and PG~1700+518 (radio quiet), by \citet{2013ApJ...772..153L}, using \nustar{} data has revealed that although they are among the optically brightest BAL quasars, their $2-10\kev$ luminosity is 16-120 times weaker as compared to typical quasars. Another study by \citet{2014ApJ...794...70L} of six optically bright BAL quasars using \nustar{} and \chandra{} revealed that the $2\kev$ luminosity of the sources are almost $>330$ times fainter than normal Seyfert galaxies, while the overall hard X-ray $8-24 \kev$ luminosity is consistently weak for all the six sources. Extreme Compton thick absorption ($\nh >10^{25} \cmsqi$) is ruled out from the analysis of the stacked \chandra{} spectra, confirming the sources to be bonafide X-ray weak. One possibility for the X-ray weakness of the BAL quasars is the failed winds, which are ionised clouds which do not get enough radiative push to get out of the gravitational field of the SMBH and falls back on the central engine. These failed winds obscure substantial fraction of the AGN luminosity in the X-rays making them X-ray weak. Another possibility is that the BAL outflows remove the feeding gas near the SMBH thereby quenching the central AGN. The latter possiblity can hold true for molecular outflows. Observations by \citet{2014A&A...562A..21C} have shown that MO kinetic energy can be as large as $\sim 5\%$ of the AGN bolometric luminosity, which according to feedback models \citep{2010MNRAS.401....7H} is enough to blow away the gases in the host galaxies. \citet{2017A&A...608A..51M} studied the X-ray properties of a sample of hyper luminous quasars ($\lbol\ge 2\times 10^{47}\lunit$) at redshift of $z\sim 2-4$. They found that the X-ray bolometric correction for these sources lies in the range $\sim 10^{-3}-10^{-2}$, which are orders of magnitude lower than low luminosity AGN. They conjecture that possibly the X-ray weakness could be due to the powerful high ionization emission line driven winds which perturb the X-ray corona and weaken their emission. On a similar vein we find that although the MOX sources have a wide range of X-ray bolometric corrections, on an average they are mostly lower compared to the other quasar samples, and possibly the molecular outflows are responsible for their lower X-ray bolometric corrections.

   It is not very straight forward to understand why the AGN at the centre of the MOX galaxies can be X-ray weak, given the fact that the AGN bolometric luminosities of these galaxies are comparable to local Seyfert galaxies and quasars ($\lagn \sim 10^{41}-10^{46}\lunit$). The X-ray coronal emission is very unlikely to be affected by molecular outflows because the AGN corona is confined to a location $<<\pc$ while the MO are detected at distances of $\kpc$ scales. Therefore, a direct link between the molecular outflows and quenching of X-ray emission does not seem feasible. In the light of this argument we probe in detail the possibility of extreme Compton thick obscuration of the MOX galaxies. In the scenario where the X-ray photons find it hard to escape out of the dust, the $12\,\mu$m flux gives us an approximate upper limit on the $\lhard$ luminosity (See Section \ref{subsec:IRflux}). Figure \ref{fig:Xraycorr} right panel shows the bolometric correction ($\lhard/\lagn$) vs the bolometric luminosity $\lagn$, where the $\lhard$ values have been calculated using the $12\, \mu$m luminosity. We find that the range of the bolometric corrections calculated using $12\, \mu$m flux are similar with those of local Seyfert galaxies, and also the MOX galaxies follow the trend of having a lower bolometric correction for sources with higher bolometric luminosity, as also detected in Seyfert galaxies and quasars \citep[See for e.g.,][ and references therein]{2017A&A...608A..51M}. The $\lhard$ values estimated using $12\mu$m flux may therefore be good indicators of the intrinsic $\lhard$ luminosity as it shows that the AGN central engine at the centre of the MOX sources functions similarly as that of the Seyfert galaxies and quasars. From Figure \ref{fig:L2to10} right panel we find that the ratio between the $\lhard$ of the MOX sources obtained using the $12\mu$m flux and using X-ray spectroscopy ranges from $10^{-1}-10^{3}$, indicating that these estimates differ by orders of magnitude. The possible reasons behind this discrepancy could be any or all of the following: 1. The X-ray spectra does not give us the correct estimate of intrinsic X-ray luminosity due to uncertainties in the obscuring column, 2. The $12\mu$m flux can contain emission from polycyclic aromatic hydrocarbon (PAH) from the host galaxy \citep[see for e.g.,][, and the references therein]{2015ApJ...803..109H}, which mostly affects the mid-IR energy band, 3. The AGN emission factor $\alpha_{\rm AGN}$ may have an intrinsic uncertainty leading to uncertainties in the estimates of $\lhard$. In a future work we intend to address these uncertainties with a more comprehensive multi-wavelength approach.

  In summary, we find that on an average the AGN at the centres of the MOX sources may not actually be X-ray weak. The apparent X-ray weakness could be due to the large obscuration column of the intervening dust and gas. The \nustar{} estimates as well as the $12\mu$m estimates of the bolometric corrections of the MOX sources mostly lie in the range spanned by Seyfert galaxies and quasars. Therefore, the AGN at the heart of these galaxies may be functioning similar to that of the local Seyfert galaxies and quasars, and the X-ray emission is weak due to obscuration. As a caveat we must remember that the $\lhard$ estimated using $12\mu$m  flux is an indirect measurement and there can be other contributors to the $12\mu$m flux apart from the AGN and the starburst processes, such as the PAH emission from galaxies.


\subsection{Is the AGN the main driver of molecular outflows?}

\begin{figure*}
  \centering 

\hbox{
\includegraphics[width=9cm,angle=0]{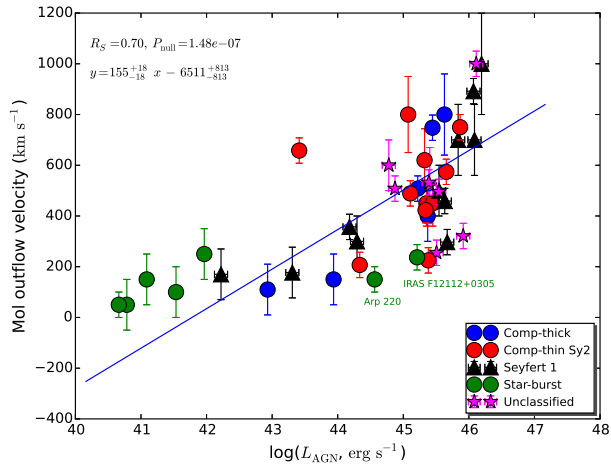} 
\includegraphics[width=9cm,angle=0]{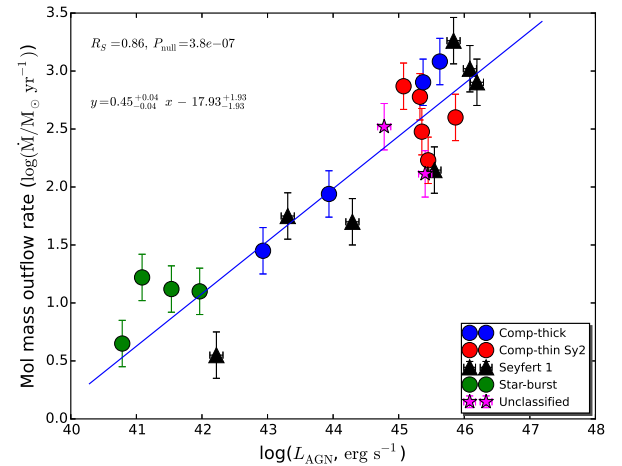} 

}
	\caption{{\it Left:} The correlation between MO velocity and the AGN bolometric luminosity, $\lagn$. {\it Right:} The correlation between the mass outflow rate and the AGN bolometric luminosity. Symbol description as in Figure \ref{fig:L2to10}} \label{fig:corrMOLbol}
\end{figure*}

Figure \ref{fig:corrMOLxray} left panel shows the correlation between the $\lhard$ and MO velocity in the MOX sample while the right panel shows the correlation between $\lhard$ and MO mass outflow rate. The correlations are statistically significant (See Table \ref{Table:corr}) and the positive slope indicates that a stronger AGN emission drives faster and more powerful MO. We find that the correlations between $\lhard$ and MO dynamical quantities become stronger when we use the $\lhard$ estimated using $12\mu$m flux (See Figure \ref{fig:corrMOLxray12micron}). Figure \ref{fig:corrMOLbol} shows that the MO outflow velocity and the mass outflow rates correlate very strongly with the AGN bolometric luminosity. In the left panel of Figure \ref{fig:corrMOLbol} we find that two SB dominated sources Arp~220 and IRASF12112+0305 have larger AGN luminosity, $\lagn$, compared to other SB galaxies, although their AGN fraction is small, $5.8\%$ and $17.8\%$ respectively (See Table \ref{Table:lbol}). However, the MO velocity in those sources are $\le 400\kms$, comparable with the other SB galaxies. From the right panel of Figure \ref{fig:corrMOLbol} we find a tight correlation between the $\lagn$ and the MO mass outflow rate with a probability $>99.99\%$. A recent study by \citet{2017A&A...601A.143F} found similar strong correlations not only between the molecular outflows and $\lagn$, but also with ionised outflows and $\lagn$. The linear regression slope derived by them for $\log(\lagn)$ vs $\log(\mout)$ is $0.76\pm 0.06$ for molecular winds. We find a more flat slope of $0.45\pm 0.04$ probably due to the fact that the SB dominated galaxies skewes the correlation. We note from Figures \ref{fig:corrMOLxray}-\ref{fig:corrMOLbol} that the starburst dominated sources (in green circles) have the lowest MO velocity and $\mout$. These results indicate that the central AGN plays a dominant role in driving these large scale molecular outflows.

\citet{2011ApJ...733L..16S} in a sample of six galaxies detected molecular outflows and found that the MO velocity scales positively with the strength of the AGN. They concluded that the central AGN plays a definitive role in driving these large scale outflows. Moreover, the authors predicted that we can distinguish between an AGN driven MO with a SB driven by noting the velocity of the outflow. Typically AGN driven flows are faster $\sim 1000\kms$ while the SB driven outflows are slower $200-400\kms$. More recent studies by \citet{2014A&A...562A..21C} on a sample of 19 sources with molecular outflows show that the molecular mass outflow rates increases with the strength of the central AGN. The starburst dominated sources on the other hand harbour outflows with lower mass outflow rates. These point to the fact that the central AGN plays a dominant role in driving these outflows.

Although we find that the presence of an AGN boosts the MO velocity and $\mout$, yet, the physical nature of the interaction between the central AGN and the MO is still not clear. One possibility investigated by previous studies is the effect of highly ionised high velocity outlfows (UFOs) striking the inter-stellar medium (ISM). \citet{2015A&A...583A..99F} detected the presence of ultra-fast outflows (UFO) as well as molecular outflow in the galaxy MRK~231. The MO extends to $1\kpc$, which the authors conjectured could be driven by the UFOs by transferring the kinetic energy to the inter-stellar medium. \citet{2015Natur.519..436T} found similar trends of energy conserving interactions of the faster UFOs and the slower MO for the source IRAS~F11119+3257, suggesting that the UFOs could be the mechanism generating large molecular outflows at $\kpc$ scale. This theory is however, still debated \citep{2017ApJ...843...18V}. Moreover, except for two MOX sources, MRK~231 and IRAS~F11119+3257, no other sources exhibit simultaneous detections of MO and UFO, which can also be due to low SNR in the spectral range of $7-9\kev$ where the UFOs are found. Another mechanism that may produce large scale molecular outflows is the radiative thrust from the central AGN, much similar to UV line driven disk winds \citep{2004ApJ...616..688P}. The presence of dust enhances the possibility of coupling the AGN radiation with the inter-stellar matter and thereby transfering the radiative thrust onto the gas leading to the MO. However, it is not clear how the AGN emission from $<\pc$ radial distance influences molecular gas clouds at $\kpc$ scales and what physical mechanism tranfers momentum and energy efficiently in the region $\pc-\kpc$ of the host galaxy.

The question therefore remains, whether the presence of an AGN is necessary to generate and drive a MO? \citet{2014Natur.516...68G} have detected molecular outflows in a compact massive starburst galaxy at a redshift of $\sim 0.7$ which are mainly driven by stellar radiation pressure. The authors demonstrated that nuclear bursts of star formation can eject large amounts of cold gas from the centre of the galaxies which truncates the star formation and affects their evolution. Similarly, \citet{2014MNRAS.441.3417S} in a sample of 12 massive galaxies, at $z\sim 0.6$, exhibiting signs of rapid quenching of star formation rate, have shown that the quenching is happening likely due to feedback from the fast outflows generated by star formation rather than AGN. For 9/12 galaxies the authors rule out the presence of any AGN at the centre of the galaxies. \citet{2012ApJ...755L..26D} in a sample of starburst galaxies at $z\sim 0.6$ also find that radiation pressure from massive stars and ram pressure from supernova and stellar winds is sufficient to produce high velocity outflows and the presence of an AGN is not needed in such cases. Theoretical studies by \citet{2013ApJ...763...17S} have also suggested that starbursts can play an active role in driving massive galactic winds.

From Table \ref{Table:lbol} we find that more than $50\%$  (27 out of 47) of the sources in the MOX sample have an AGN fraction of $<50\%$, implying that the total galactic emission is dominated by star bursts in more than half of the sources. Very interestingly we also find statistically significant positive correlations between the soft X-ray APEC luminosity $\lsoft \, ({\rm APEC})$ and MO velocity and  $\mout$. In this work we assume {\it APEC} luminosity in the energy range $0.6-2\kev$ probes the strength of starburst (SB) activity. As a caveat we note that this may not be true for a few sources where the primary or the reflected emission from AGN may also contribute to the $0.6-2\kev$ luminosity. We also find statistically strong positive correlations between $\lsb$ vs MO velocity and $\lsb$ vs $\mout$ (See Figure \ref{fig:starburst}). These correlations indicate that SB also can play a significant role in generating and driving the molecular outflows. The SB emission arises from extended regions of the galaxies (compared to the size of the central AGN) and are sometimes cospatial with the MO ($\sim \kpc$), and hence has a good probability to generate the MO. However, we should note that the correlations of the molecular outflow velocity and $\mout$ with the $\lsb$ and $\lsoft$ are weaker compared to those of the AGN X-ray and bolometric luminosities. It is possible that both star bursts and AGN generate and drive these massive MO.

In summary we confirm that the AGN power is well correlated with the power of the MO. However, the fact that the powerful MO are also found in sources whose contribution to the AGN bolometric luminosity is small, and the strong correlations between $\lsoft$ vs $\mout$, and $\lsb$ vs $\mout$ indicate that powerful star bursts are equally probable to generate and drive the large scale MO.


\section{Conclusions}\label{sec:conclusions}

We have carried out an extensive X-ray spectral analysis of a sample of 47 galaxies exhibiting molecular outflows (the MOX sample), using observations from \chandra{} and \xmm{}. Below we list the main conclusions:

\begin{itemize}

	\item From the X-ray spectra of the MOX sources we find that they are generally X-ray weak, with an X-ray bolometric correction ranging from $\lhard/\lagn \sim$ $10^{-4.5}$ to $10^{-0.5}$, with $70\%$ of the sources below $10^{-2}$. Possibly the MOX sources have AGN with weaker X-ray emission compared to local Seyfert galaxies and quasars. However it is not physically clear why and how should the X-ray emission be selectively quenched relative to the overall AGN bolometric luminosity.

	\item	We obtain an upper limit on the $\lhard$ emission from the AGN ($\lmu$) in the MOX sources using the $12\mu$m flux emitted from the galaxies, following the correlation by Asmus et al., $\log (\frac{L_{2-10\kev}}{10^{43}\lunit})=-0.32 + 0.95\times \log (\frac{L_{12\, \mu m} \times \alpha_{\rm AGN}}{10^{43}\lunit})$. The factor $\alpha_{\rm AGN}$ ensures that we consider the $12\mu$m flux from the central AGN only. The $\lmu$ values obtained using this method are $0.5-3$ orders of magnitude larger than the $\lhard$ values obtained using X-ray spectroscopy. Moreover, the $\lmu$ values are consistent with local Seyfert galaxies and quasars. Speculatively we can say that the AGN at the heart of the MOX sources may have similar $\lhard$ as local Seyfert galaxies and quasars, but their weak X-ray emission is due to the high column of obscuration along the line of sight. As a caveat we must note that the galactic PAH emission also contributes to the $12\mu$m flux which are unaccounted for, and hence we refer to the $\lmu$ obtained using the $12\mu$m flux as an upper limit on the $2-10\kev$ emission from the AGN.

	\item The relation ($\lmu/\lagn$) vs $\lagn$ of the MOX sources also shows a similar trend as that of the local Seyfert galaxies and quasars, that is, with increasing bolometric luminosity of AGN ($\lagn$) the X-ray bolometric correction decreases. This  may imply that at the heart of these galaxies the AGN functions similarly as that of the quasars, and their apparent X-ray weakness is due to extreme obscuration.

	\item{We find statistically significant positive correlations between $\lhard$ and $\lagn$ with the molecular outflow velocity and $\mout$ in the MOX sample, indicating that the presence of an AGN boosts the molecular outflow velocity and power.}

	\item{We find that the starburst emission in the host galaxies of the MOX sample, $\lsb$ correlates strongly with the molecular outflow velocity and $\mout$. The starburst emission, measured in the soft X-rays ($0.6-2\kev$) with the model {\it APEC}, also shows significant correlation with the MO velocity and $\mout$. These correlations points to the fact that starburst has the potential to generate and drive the molecular outflows. The starburst emission arises from regions that are more extended (compared to the size of AGN central engine) and hence may sometimes be co-spatial with the molecular outflows, and therefore can play more important role in driving the outflows. Supporting our claim above, we also find that 27 of the 47 sources in the MOX sample have an AGN fraction $<50\%$, implying that the starburst are dominant in these galaxies, and they drive can these large scale molecular outflows. However, we should note that the correlations of the molecular outflow velocity and $\mout$ with the $\lsb$ and $\lsoft$ are weaker compared to those of the AGN X-ray and bolometric luminosities. It is possible that although starburst can drive massive molecular outflows, the presence of an AGN always boosts the power of the outflows.}

\end{itemize}



\clearpage

%



\begin{table*}

{\footnotesize
\centering
  \caption{The list of sources, their general properties, previous X-ray studies and the 12 $\mu$m flux. \label{Table:sources}}
  \begin{tabular}{llllllllllll} \hline\hline \\

	  Index   &  Source 		&Other names& z	& R.A.		& Dec.		&Classification$^{\rm A}$	&Galaxy $^{\rm B}$	&Ref$^{\rm C}$	& $F_{12\, \rm \mu m}^{\rm D}$		&	 \\ 
	&			&		&	&		&		&			&activity	&		&mJy \\ 
1	&2			&3		&4	&5		&6		&7		&8		&9		&10	\\  \hline 

1	&IRAS~F08572+3915	&-		&0.0583	&09h00m25.3s	&+39d03m54.4s	&ULIRG		&CT		&1,2 		&$325\pm30$		\\			

2	&IRAS~F10565+2448	&-		&0.0431	&10h59m18.1s	&+24d32m34s	&ULIRG		&OA		& 2		&$200\pm30$			\\

3	&IRAS~23365+3604	&-		&0.0645	&23h39m01s	&+36d21m08s	&ULIRG		&OA/LINER	&1		&$<0.09$		\\

4	&Mrk~273		&-		&0.0377	&13h44m42.1s	&+55d53m13s	&ULIRG		&Sy2/OA		&2		&$240\pm 17$		\\

5	&Mrk~876		&-		&0.129	&16h13m57.2s	&+65d43m10s	&-		&Sy1		&		&$87\pm12$		\\

6	&I~Zw~1			&UGC~00545	&0.0589	&00h53m34.9s	&+12d41m36s	&Sy1		&NLSy1		&		&$549\pm 11$		\\

7	&MrK~231		&-		&0.0421	&12h56m14.2s	&+56d52m25s	&ULIRG/RL	&Sy1/SB		&2		&$1830 \pm 17$		\\		

8	& NGC~1266		&-		&0.0072	&03h16m00.7s	&-02d25m38s	&Sy		&AGN		&3		&$250\pm 30$		\\

9	&M~82			&-		&0.0006	&09h55m52.7s	&+69d40m46s	&-		&SB		&4		&$63000\pm 3150$		\\

10	&NGC~1377		&-		&0.0059	&03h36m39.1s	&-20d54m08s	&-		&NC		&5,6		&$560\pm 20$		\\

11	&NGC~6240		&-		&0.0244	&16h52m58.9s	&+02d24m03s	&LIRG		&CT/GM/SB	&1		&$590\pm 25$		\\

12	&NGC~3256		&-		&0.0093	&10h27m51.3s	&-43d54m13s	&LIRG		&SB		&7		&$3570\pm 31$		\\

13	&NGC~3628		&-		&0.0028	&11h20m17.0s	&+13d35m23s	&RL		&SB		&8		&$3130\pm 48$		\\

14	&NGC~253		&-		&0.0008	&00h47m33.1s	&-25d17m18s	&-		&Variable SB	&9		&$41000\pm 35$		\\

15	&NGC~6764		&-		&0.0081	&19h08m16.4s	&+50d56m00s	&-		&AGN+SB		&10		&$310\pm 47$		\\

16	&NGC~1068		&-		&0.0038	&02h42m40.7s	&-00d00m48s	&LIRG		&CT/Sy2		&11		&$39800\pm 76$		\\

17	&IC~5063		&-		&0.0113	&20h52m02.3s	&-57d04m08s	&Sy1/RL		&NLSy2		&12		&$1110\pm 23$		\\

18	&NGC~2146		&-		&0.0029	&06h18m37.7s	&+78d21m25s	&LIRG		&SB		&13		&$7360\pm 800$		\\	

19	&IRAS~17208-0014 	&-		&0.0428	&17h23m21.9s	&-00d17m01s	&ULIRG/LINER	&ULIRG		&2		&$200\pm 25$		\\		

20	&NGC~1614	   	&-		&0.0159	&04h33m59.8s	&-08d34m44s	&LIRG/SB	&SB		&14		&$1210\pm 111$		\\	

21	&IRAS~05083+7936	&VII~Zw~031	&0.0536	&05h16m46.1s	&+79d40m13s	&LIRG		&OA		&-		&$200\pm 26$		\\	

22	&Iras~13451+1232	&4C~+12.50	&0.1217	&13h47m33.3s	&+12d17m24s	&ULIRG/RL	&Sy2		&2		&$<143$		\\

23	&3C~293			&UGC~08782	&0.0450	&13h52m17.8s	&+31d26m46s	&Sy/RL		&NC		&15		&$19\pm 2$	\\

24	&NGC~1433		&		&0.0035	&03h42m01.5s	&-47d13m19s	&SB		&NC		&-		&$237\pm 17$		\\ 

25	&IRAS~13120-5453	&WKK~2031	&0.0308	&13h15m06.3s	&-55d09m23s	&ULIRG		&NC		&16		&$440\pm 27$		\\

26	&IRASF~14378-3651	&-		&0.0676	&14h40m59s	&-37d04m32s	&ULIRG		&Sy2		&1		&$<100$		\\

27	&IRAS~F11119+3257	&B2~1111+32	&0.1890	&11h14m38.9s	&+32d41m33s	&ULIRG		&NC		&17		&$167\pm 27$		\\


28	&IRAS~F01572+0009	&MRK~1014	&0.1631	&01h59m50.2s	&+00d23m41s	&ULIRG/Sy 1.5	&NC	 	&18		&$134\pm 40$		\\
29	&IRAS~F05024-1941	&-		&0.1920	&05h04m36.5s	&-19d37m03s	&ULIRG		&NC		&2		&$<276$		\\

30	&IRAS~F05189-2524	&-		&0.0425	&05h21m45s	&-25d21m45s	&ULIRG		&Sy2		&2		&$740\pm 16$		\\
31	&IRAS 07251-0248	&-		&0.0875	&07h27m37.5s	&-02d54m55s	&ULIRG		&Faint src	&1		&$<7$	 \\

32	&IRAS~F07599+6508	&-		&0.1483	&08h04m33.1s	&+64d59m49s	&ULIRG		&NC		&2		&$264\pm 23$	\\

33	&IRAS 09022-3615	&-		&0.0596	&09h04m12.7s	&-36d27m01s	&ULIRG		&AGN		&1		&$200\pm 32$		\\
34	&IRAS~F09320+6134	&UGC~05101	&0.0393	&09h35m51.6s	&+61d21m11s	&ULIRG		&OA		&1		&$179\pm 16$		\\

35	&IRAS~F12072-0444	&-		&0.1284	&12h09m45.1s	&-05d01m14s	&ULIRG/Sy2	&NC		&2		&$<119$		\\

36	&IRAS~F12112+0305	&-		&0.0733	&12h13m46.0s	&+02d48m38s	&ULIRG		&SB		&1		&$<110$		\\


37	&IRAS~F14348-1447	& -		&0.0830	&14h37m38.4s	&-15d00m20s	&ULIRG		&CT/SB		&1		&$108\pm 32$		\\

38	&IRAS~F14394+5332	&-		&0.1045	&14h41m04.4s	&+53d20m09s	&ULIRG		&NC		&-		&$<72$		\\


39	&IRAS~F15327+2340	&ARP~220	&0.0181	&15h34m57.2s	&+23d30m11s	&ULIRG/Sy	&OA		&2		&$496\pm 45$		\\

40	&IRAS~F15462-0450	&-		&0.0997	&15h48m56.8s	&-04d59m34s	&ULIRG/NLSy1	&NC		&2		&$100\pm 30$		\\

41	&IRAS~F19297-0406	&-		&0.0857	&19h32m21.2s	&-03d59m56s	&ULIRG		&NC		&1		&$<100$	\\

42	&IRAS 19542+1110	&-		&0.0649	&19h56m35.4s	&+11d19m03s	&ULIRG		&OA		&1		&$80$		 \\

43	&IRAS~F20551-4250	&ESO~286IG019	&0.0429	&20h58m26.8s	&-42d39m00s	&ULIRG		&CT		&1		&$280\pm 21$		\\


44	&IRAS~F23233+2817  	&-		&0.1140	&23h25m49.4s	&+28d34m21s	&ULIRG/Sy2	&NC		&-		&$<129$		\\

45	&NGC~5506		&-		&0.0062	&14h13m14.9s	&-03d12m27s	&Sy		&NC		&19		&$1480\pm 90$		 \\

46	&NGC~7479		& -		&0.0079	&23h04m56.6s	&+12d19m22s	&SB/Sy1.9	&NC		&20		&$1390\pm 30$		\\

47	&NGC~7172		& 		&0.0087	&22h02m01.9s	&-31d52m11s	&Sy2		&NC		&21		&$720 \pm 60$	\\

\hline \hline

\end{tabular}  

{$^{\rm A}$ The classification as obtained from NED.}\\

{$^{\rm B}$ The galaxy activity as identified by previous X-ray and optical studies: CT= Compton thick, CL=Changing Look, OA=Obscured AGN,  LINER= Low ionisation nuclear emission line region, Sy2= Seyfert 2, NLSy1= Narrow line Seyfert 1, SB= starburst, GM=Galaxy mergers , NC=Not classified. }

{$^{\rm C}$ The references to the previous X-ray studies: 1=\citet{2011A&A...529A.106I}, 2=\citet{2010ApJ...725.1848T}, 3= \citet{2015ApJ...798...31A}, 4=\citet{2014MNRAS.437L..76L}, 5= \citet{2016A&A...594A.114C},6=\citet{2016A&A...590A..73A},7=\citet{2015ApJ...806..126L}, 8=\citet{2012ApJ...752...38T}, 9=\citet{2016A&A...592L...3K}, 10=\citet{2008ApJ...688..190C}, 11=\citet{2016MNRAS.456L..94M}, 12= \citet{2014A&A...562A..21C}, 13=\citet{2005PASJ...57..135I}, 14= \citet{2014ApJ...786..156H}, 15=\citet{2015ApJ...801...17L}, 16=\citet{2015ApJ...814...56T}, 17=\citet{2015Natur.519..436T}, 18=\citet{2014A&A...567A.142R}, 19=\citet{2010MNRAS.406.2013G}, 2-= \citet{2009A&A...500..999A} , 21=\citet{1998MNRAS.298..824G}\\

{$^{\rm D}$ The $12\mu$m monochromatic flux of the galaxies obtained from NASA Extragalactic Database.}

}
}
\end{table*}


\begin{table*}

{\footnotesize
\centering
  \caption{The list of sources and their molecular outflow properties. \label{Table:MO}}
  \begin{tabular}{llllllllllll} \hline\hline 

	  Index   &  Source 		&Method		&  Reference$^{\rm B}$ & Outflow velocity$^{\rm C}$	&   $\mout$		&SFR	&\\ 
	&			&used		&		&$\kms$			& log($\msol/yr$)	&[$\msol/yr$]	&			\\  \hline \\ 

	  1	&IRAS~F08572+3915$^{\rm A}$	&CO(1-0)		&1 		&$800\pm 160$		      &3.082			&20		&	\\

	&"			&OH 		&2 		&$700 \pm 140$		      &2.98			&-		&		\\

	  2	&IRAS~F10565+2448	&CO(1-0)		&1 		&$450\pm 90$		      &2.477			&95		&		\\

	  3	&IRAS~23365+3604	&CO(1-0)		&1 		&$450 \pm 90$		      &2.230			&137		&		\\

4	&Mrk~273		&CO(1-0)		&1 		&$620 \pm 124$		      &2.778			&139		&		\\

5	&Mrk~876		&CO(1-0)		&1 		&$700\pm 140$		      &$\le 3.262$		&6.5		&		\\

6	&I~Zw~1			&CO(1-0)		&1 		&$500\pm 100$		      &$\le 2.146$		&36		&		\\

7	&MrK~231$^1$		&CO(1-0)		&1 		&$700\pm 140$		      &$3.02$			&234		&		\\

	&"			&OH		&2 			&$600 \pm 120$		      &3.07			&-		&		\\

8	& NGC~1266		&CO(1-0)		&1,3 		&$177\pm 100$		      &$ 1.518-2.255$		&1.6		&		\\

9	&M~82			&CO(1-0)		&1,4 		&$100\pm 100$		      &$ 1.079 - 1.255$		&10		&		\\

10	&NGC~1377		&CO(1-0)		&1,5 		&$110\pm 100$		      &$1.146- 1.881$		&0.9		&		\\

11	&NGC~6240		&CO(1-0)		&1,6 		&$400\pm 100$		      &2.903			&16		&	\\

	  12	&NGC~3256	&CO(1-0)		&1,7 		&$250 \pm 100$		      &$1.041- 1.204$		&36		&	\\

13	&NGC~3628		&CO(1-0)		&1,8 		&$50 \pm 100$		      &$ 0.653-0.826$		&1.8		&	\\

14	&NGC~253$^1$		&CO(1-0)		&1,9 		&$50 \pm 100$		      &$0.623 - 0.799$		&3		&	\\

	&"			&OH		&2 		&$75\pm 100$		      &$0.20$			&-		&-		\\

15	&NGC~6764		&CO(1-0)		&1,10 	 	&$170\pm 100$		      &$0.491-0.672$		&2.6		&	\\

16	&NGC~1068		&CO(1-0)		&1,11		&$150 \pm 100$		      &1.924			&18		&	\\

17	&IC~5063		&CO(1-0)		&1,12 		&$300 \pm 100$		      &$1.361-2.103$		&0.6		&	\\

18	&NGC~2146		&CO(1-0)		&1,13 		&$150 \pm 100$		      &$1.146-1.342$		&12		&	\\	

19	&IRAS~17208-0014$^1$ 	&CO(2-1)		&14 		&$600 \pm 100$		      &$2.518$			&- 		&	\\

	&"			& OH		&2 		&$100 \pm 100$		      &$1.954$			&-		&		\\		

20	&NGC~1614	   	&CO(1-0)		& 14 		&$360\pm 100$		      &$1.602$			&-		&	\\	

21	&IRAS~05083+7936	&CO(1-0)		&15		&$750 \pm 100$	      		& -			&-		&	\\	

22	&IRAS~13451+1232	&CO(1-0)		&16  		&$750 \pm 50$	      		&$2.361-2.903$		&-		&\\

23	&3C~293			&CO(1-0)		&17 		&$350\pm 100$ 		      &$1.397-1.477$		&-		&	\\

24	&NGC~1433		&CO(3-2)		&18  		&$200 \pm 100$		      &0.845			&-		&	\\

25	&IRAS~13120-5453	&OH		&2 		&$520 \pm 150$		      &2.113			&-		&	\\

26	&IRAS~14378-3651	&OH		&2 		&$800 \pm 150$		      &2.869			&-		&	\\

27	&IRAS~F11119+3257	&OH 		&19 		&$1000\pm 200$	      &$2.903_{-0.501}^{+0.400}$	&-		&	\\


28	&IRAS~F01572+0009	&OH		&20	 	&$892\pm 50$			&-			&-		&		\\
29	&IRAS~F05024-1941	&OH		&20		&$508 \pm 50$			&-			&-		&	\\

30	&IRAS~F05189-2524	&OH		&20		&$574 \pm 50$			&-			&-		&	\\
31	&IRAS 07251-0248	&OH		&20		&$255 \pm 50$			&-			&-		&	\\

32	&IRAS~F07599+6508	&OH		&20		&$1000 \pm 50$			&-			&-		&	\\

33	&IRAS 09022-3615	&OH		&20		&$297\pm 50$			&-			&-		&	\\
34	&IRAS~F09320+6134	&OH		&20		&$225\pm 50$			&-			&-		&	\\

35	&IRAS~F12072-0444	&OH		&20		&$321 \pm 50$			&-			&-		&\\

36	&IRAS~F12112+0305	&OH		&20		&$237\pm 50$			&-			&-		&\\


37	&IRAS~F14348-1447	&OH		&20		&$508\pm 50$			&-			&-		&	\\

38	&IRAS~F14394+5332	&OH		&20		&$495 \pm 50$			&-			&-		&	\\


39	&IRAS~F15327+2340	&OH		&20		&$153 \pm 50$			&-			&-		&	\\

40	&IRAS~F15462-0450	&OH		&20		&$459 \pm 50$			&-			&-		&	\\

41	&IRAS~F19297-0406	&OH		&20		&$532 \pm 50$			&-			&-		&\\

42	&IRAS 19542+1110	&OH		&20		&$489 \pm 50$			&-	 		&-		&	\\

43	&IRAS~F20551-4250	&OH		&20		&$748 \pm 50$			&-			&-		&	\\


44	&IRAS~F23233+2817  	&OH		&20		&$423 \pm 50$			&-			&-		&\\

45	&NGC~5506		&OH		&21 		&$357 \pm 50$	 		&-     			&-		&		\\

46	&NGC~7479		&OH		&21		&$658 \pm 50$			&-			&-		&	\\

47	&NGC~7172		&OH		&21		&$207 \pm 50$			&-			&-		&		\\

\hline \hline

\end{tabular}  

{$^{\rm A}$ Sources which have been observed both by CO and OH molecules.}\\

{$^{\rm B}$ References:  1= \citet{2014A&A...562A..21C}, 2= \citet{2011ApJ...733L..16S}, 3= \citet{2011ApJ...735...88A}, 4= \citet{2002ApJ...580L..21W},  5= \citet{2012A&A...546A..68A}, 6=\citet{2013A&A...549A..51F}, 7= \citet{2006ApJ...644..862S}, 8=\citet{2012ApJ...752...38T}, 9=\citet{1996A&A...305..421M}, 10=\citet{1985ApJ...298L..31S}, 11=\citet{1997ApJ...485..552M}, 12=\citet{1995A&A...297..643W}, 13= \citet{2006AJ....132.2383T}, 14=\citet{2015A&A...580A..35G}, 15=\citet{2015ApJ...811...15L}, 16=\citet{2014A&A...565A..46D}, 17=\citet{2014A&A...564A.128L}, 18=\citet{2013A&A...558A.124C}, 19=\citet{2015Natur.519..436T}, 20=\citet{2013ApJ...776...27V}, 21=\citet{2016ApJ...826..111S}   }\\

{$^{\rm C}$ Different authors have used +ve and -ve notations to denote outflow velocities (blue shifted) with respect to the systemic velocity. To avoid confusion and maintain uniformity, we have considered the modulus of the velocities}

}
\end{table*}


\begin{table*}

{\footnotesize
\centering
  \caption{Details of X-ray observations. \label{Table:xray}}
  \begin{tabular}{llllllllllll} \hline\hline 

Index   &  Source 		&Telescope	&observation	& Observation 		&Exposure	&Net Exposure	&Total		 \\ 
	&			&		&ID		&Date			&(ks)		&(ks)		&counts	\\  \hline \\ 

1	&IRAS~F08572+3915	&\xmm{}		&0200630101	&13-04-2004		&29		&14		&2.28e+02	\\			

2	&IRAS~F10565+2448	&\xmm{}		&0150320201	&17-06-2003		&32		&25		&1.307e+03\\

3	&IRAS~23365+3604	&\chandra{}	&4115		&03-02-2003		&10		&10		&74\\

4	&Mrk~273		&\xmm{}		&0722610201	&04-11-2013		&23		&6		&9.81e+02\\

5	&Mrk~876		&\xmm{}		&0102040601	&14-11-2002		&13		&0.1		&3.06e+02	\\

6	&I~Zw~1			&\xmm{}		&0743050301	&19-01-2015		&141		&20		&1.81e+05\\

7	&MrK~231		&\xmm{}		&0770580501	&28-05-2015		&26		&21		&3.63e+03\\		

8	& NGC~1266		&\xmm{}		&0693520101	&23-07-2012		&139		&96		&9.80e+03  	\\

9	&M~82			&\xmm{}		&0206080101	&21-04-2004		&104		&62		&3.08e+05		     \\

10	&NGC~1377		&\chandra{}	&16086		&10-12-2013		&48		&44		&2.06e+02		    	\\

11	&NGC~6240		&\xmm{}		&0147420201	&14-03-2003		&42		&4		&2.61e+03		     	\\

12	&NGC~3256		&\xmm{}		&0300430101	&06-12-2005		&134		&97		&5.54e+04			     	\\

13	&NGC~3628		&\xmm{}		&0110980101	&27-11-2000		&65		&38		&6.29e+03		    	\\

14	&NGC~253		&\xmm{}		&0152020101	&19-06-2003		&141		&-	&{--}	    	\\

15	&NGC~6764		&\chandra{}	&9269		&20-01-2008		&20		&20		&5.89e+02		    	\\

16	&NGC~1068		&\xmm		&0740060201	&10-07-2014		&64		&44		&5.26e+05		     	\\

17	&IC~5063		&\chandra{}	&7878		&15-06-2007		&35		&34		&5.37e+03		     	\\

18	&NGC~2146		&\xmm{}		&0110930101	&26-08-2001		&27		&12		&6.34e+03		    	\\	

19	&IRAS~17208-0014 	&\xmm{}		&0081340601	&19-02-2002		&19		&12		&6.89e+02			\\		

20	&NGC~1614	   	&\chandra{}	&15050		&21-11-2012		&16		&16		&8.33e+02    	\\	

21	&IRAS~05083+7936	&\xmm{}		&009400101	&11-09-2001		&33	     	&26		&9.59e+02	\\	

22	&IRAS~13451+1232	&\chandra	&836		&24-02-2000		&28	     	&25		&1.42e+03\\

23	&3C~293			&\chandra	&12712		&16-11-2010		&69		&68		&2.12e+03	     		\\

24	&NGC~1433		&\chandra	&16345	        &04-03-2015		&49		&49		&3.30e+02	\\ 

25	&IRAS~13120-5453	&\xmm{} 	&0693520201	&20-02-2013		&129		&85		&6.15e+03			\\

26	&IRAS~14378-3651	&\chandra	&7889		&25-06-2007		&14		&14		&1.05e+02	\\

27	&IRAS~F11119+3257	&\chandra	&3137		&30-06-2002		&19		&18		&2.89e+03	     	\\


28	&IRAS~F01572+0009	&\xmm		&0101640201	&29-07-2000		&15		&5		&6.42e+03			 \\
29	&IRAS~F05024-1941	&\xmm		&0405950401	&07-02-2007		&42		&26		&6.89e+02   	\\

30	&IRAS~F05189-2524	&\xmm		&0722610101	&02-10-2013		&38		&30		&1.48e+04		\\
31	&IRAS 07251-0248	&\chandra	&7804		&01-12-2006		&16		&15		&7.80e+01		 \\

32	&IRAS~F07599+6508	&\xmm		&0094400301	&24-10-2001		&23		&16		&7.12e+02		\\

33	&IRAS 09022-3615	&\xmm		&0670300401	&23-04-2012		&33		&16		&1.16e+03	\\
34	&IRAS~F09320+6134	&\xmm		&0085640201	&12-11-2001		&35		&26		&1.49e+03	    	\\

35	&IRAS~F12072-0444	&\chandra	&4109		&01-02-2003		&10		&10		&7.3e+01		     \\

36	&IRAS~F12112+0305	&\xmm		&0081340801	&30-12-2001		&23		&18		&4.92e+02		     	\\


37	&IRAS~F14348-1447	&\xmm		&0081341401	&29-07-2002		&22		&15		&7.97e+02     \\

38	&IRAS~F14394+5332	&\xmm		&0651100301	&06-07-2015		&13		&8		&3.31e+02		     \\


39	&IRAS~F15327+2340	&\xmm		&0205510201	&14-01-2005		&35		&6.3		&5.09e+02	     \\

40	&IRAS~F15462-0450	&\chandra	&10348		&23-04-2009		&15		&15		&6.08e+02		\\

41	&IRAS~F19297-0406	&\chandra	&7890		&18-06-2007		&17		&-		&-	     \\
42	&IRAS 19542+1110	&\chandra	&7817		&10-09-2007		&15		&15		&4.27e+02	      \\

43	&IRAS~F20551-4250	&\xmm		&0081340401	&21-04-2001		&22		&11		&1.07e+03	     \\


44	&IRAS~F23233+2817  	&\xmm		&0553870101	&15-12-2008		&80		&52		&2.07e+03	     	\\

45	&NGC~5506		&\xmm		&0554170101	&02-01-2009		&90		&61		&7.94e+05		      \\

46	&NGC~7479		&\xmm		&0301651201	&24-06-2005		&16		&0.57		&9.8e+01	     	\\

47	&NGC~7172		&\xmm		&0414580101	&24-04-2007		&58 		&28		&9.81e+04		\\

\hline \hline

\end{tabular}  

}
\end{table*}


\begin{table}
{\fontsize{7}{7}\selectfont
\centering
  \caption{The X-ray properties of the sources with MO. \label{Table:flux}}
  \begin{tabular}{lccccccccccccccc} \hline\hline 

Index	&Source 		&$\rm F_{0.6-2\kev}$	&$\rm F_{2-10\kev}$		&$\rm L_{0.6-2\kev}$	&$\rm L_{2-10\kev}$	&$\rm KT_1$		&$\rm KT_2$ 	&$\Gamma$	&$\nh$					&$\chi^2/dof$\\ 
	&			&$\funit$		&$\funit$			&$\lunit$		&$\lunit$		&APEC			&APEC		&		&$\cmsqi$	\\ \hline

1	&IRAS~F08572+3915	&--			&--				&--			&--			&--			&--		&--			&--						&--		\\

2	&IRAS~F10565+2448	&$0$			&$-13.345_{-0.05}^{+0.05}$	&$0$			&$41.25\pm0.05$			&--			&--		&$2.17_{-0.23}^{+0.23}$	&0					&$72/54 \sim 1.54$	\\	

	  3	&IRAS~23365+3604&--			&--				&--			&--			&--			&--		&--			&--			&--		\\	

4	&Mrk~273		&$-13.44\pm0.05$	&$-12.35_{-0.05}^{+0.05}$	&$41.43\pm 0.04$	&$42.25\pm 0.05$		&--			&-		&$1.98\pm0.32$		&$<10^{20}$				&$70/37 \sim 1.90$	\\		

5	&Mrk~876		&$0$			&$-11.47_{-0.05}^{+0.05}$	&$0$			&$44.11\pm 0.04$	&--			&--		&$2.04_{0.18}^{+0.17}$	&0					&$14/11 \sim 1.27$		\\		

6	&I~Zw~1			&$-10.99\pm 0.01$	&$-11.21_{-0.005}^{+0.005}$	&$43.84\pm 0.01$	&$43.63\pm 0.02$	&--			&--		&$2.37_{-0.04}^{+0.08}$	&$4.5\times 10^{20}$			&$307/219 \sim 1.41$	\\		

7	&MrK~231		&$-10.64\pm 0.02$	&$-12.14_{-0.05}^{+0.05}$	&$43.93\pm 0.02$	&$42.42\pm 0.04$	&$0.25\pm 0.10$		&$0.80\pm0.10$	&$1.8^*$		&$0.91_{-0.17}^{+0.17}\times 10^{22}$	&$181/117 \sim 1.55$		\\

8	& NGC~1266		&$-10.95\pm 0.01$	&$-13.15_{-0.04}^{+0.04}$	&$41.98\pm 0.01$	&$39.78\pm 0.05$	&$0.08\pm 0.02$		&$0.16\pm0.05$	&$1.79_{-0.3}^{+0.3}$	&$0.78_{-0.02}^{+0.02}\times 10^{22}$	&$231/124 \sim 1.87$		\\		

9	&M~82			&$-$			&$-10.90_{-0.002}^{+0.002}$	&--			&$40.32\pm 0.06$	&$0.82\pm 0.12$		&--		&$1.83_{-0.03}^{+0.03}$	&$0.29_{-0.1}^{+0.1}\times 10^{22}$	&$818/239 \sim 3.42$		\\		

	  10&NGC~1377$^1$	&--			&--				&--			&--				&--			&--		&--			&--					&--		\\		

11	&NGC~6240		&$-12.78\pm 0.05$	&$-11.80_{-0.03}^{+0.03}$	&$41.87\pm 0.02$	&$42.28\pm 0.05$	&$0.03\pm 0.01$		&--		&$1.90_{-0.27}^{+0.27}$	&$0$					&$149/77 \sim 1.94$		\\		

12	&NGC~3256		&$-13.51\pm 0.02$	&$-12.35_{-0.01}^{+0.01}$	&$41.06\pm 0.02$	&$40.97\pm 0.03$	&$0.62\pm 0.22$		&$1.10\pm0.02$	&$2.40_{-0.06}^{+0.06}$	&$0.24_{-0.02}^{+0.02}\times 10^{22}$	&$443/193 \sim 2.32$		\\		

13	&NGC~3628		&$-20.30\pm 0.02$	&$-12.26_{-0.02}^{+0.02}$	&$0$			&$40.22\pm 0.04$	&$0.10\pm 0.02$		&--		&$1.50_{-0.08}^{+0.08}$	&$0.30_{-0.04}^{+0.04}\times 10^{22}$	&$172/154 \sim 1.12$		\\		

	  14&NGC~253		&--			&--				&--			&--			&--			&--		&--			&--				&--		\\

15	&NGC~6764		&$-13.23\pm 0.04$	&$-13.19_{-0.04}^{+0.06}$	&$40.16\pm 0.04$	&$39.85\pm0.05$		&$0.98\pm 0.05$		&--		&$>2.28$	&$0.25_{-0.09}^{+0.21}\times 10^{22}$	&$67/72 \sim 0.93$		\\		

16	&NGC~1068		&--			&$-11.29_{-0.005}^{+0.005}$	&--			&$40.95\pm0.11$		&--			&--		&$1.92_{-0.2}^{+0.2}$	&$0$					&$505/175 \sim 2.88$		\\		

17	&IC~5063		&$-13.35\pm0.03$	&$-10.79_{-0.02}^{+0.02}$	&$40.25\pm 0.03$	&$42.57\pm 0.03$	&$0.81\pm 0.14$		&$<2.4$		&$1.8^*$		&$21.47_{-0.92}^{+0.92}\times 10^{22}$	&$612/490 \sim 1.25$		\\		

18	&NGC~2146		&$-12.71\pm 0.03$	&$-12.00_{-0.02}^{+0.02}$	&$39.50\pm 0.03$	&$40.20\pm 0.03$	&$0.77\pm 0.15$		&--		&$1.59_{-0.10}^{+0.12}$	&$0.20_{-0.11}^{+0.11}\times 10^{22}$	&$329/119 \sim 2.77$		\\		

19	&IRAS~17208-0014	&$-13.88\pm 0.15$	&$-12.98_{-0.2}^{+0.2}$		&$40.68\pm 0.17$	&$41.59\pm 0.15$	&$0.76\pm 0.23$		&--		&$1.8^*$		&$6.23_{-5}^{+18}\times 10^{22}$	&$14/27$	\\ 	

20	&NGC~1614		&$-13.35\pm 0.20$	&$-12.86_{-0.03}^{+0.03}$	&$40.33\pm 0.20$	&$40.82\pm 0.18$	&$0.79\pm 0.13$		&--		&$1.8^*$		&$0.24_{-0.22}^{+0.24}\times 10^{22}$	&$108/110 \sim 0.98$		\\	   	

21	&IRAS~05083+7936	&$-13.98\pm 0.13$	&$-13.28_{-0.02}^{+0.12}$	&$40.79\pm 0.13$	&$41.49\pm 0.11$	&$0.75\pm 0.11$		&--		&$1.8^*$		&$4.89_{-3.52}^{+4.92}\times 10^{22}$	&$79/38 \sim 2.08$		\\	

22	&IRAS~13451+1232	&$-14.16\pm 0.08$	&$-11.89_{-0.02}^{+0.02}$	&$41.63\pm 0.08$	&$43.64\pm 0.09$	&$0.61\pm 0.17$		&--		&$1.69_{-0.20}^{+0.30}$	&$3.23_{-0.52}^{+0.52}\times 10^{22}$	&$182/230 \sim 0.79$		\\	

23	&3C~293			&$-14.06\pm 0.03$	&$-11.78_{-0.09}^{+0.01}$	&$41.07\pm 0.03$	&$42.85\pm 0.04$	&$1.06 \pm 0.29$	&$<2.11$	&$<1.4$		&$7.42_{-0.82}^{+0.82}\times 10^{22}$	&$262/320 \sim 0.820$		\\		

24	&NGC~1433		&$-13.43\pm 0.07$	&$-13.32_{-0.07}^{+0.09}$	&$38.91\pm 0.07$	&$39.02\pm 0.08$&$0.52\pm 0.10$		&--		&$1.8$		&$0$					&$44/51 \sim 0.86$		\\		

25	&IRAS~13120-5453	&$-13.98\pm 0.02$	&$-12.76_{-0.03}^{+0.03}$	&$40.32\pm 0.02$	&$41.54\pm 0.05$	&$1.24\pm 0.1$		&--		&$1.50_{-0.15}^{+0.08}$	&$0.22_{-0.07}^{+0.03} \times 10^{22}$	&$223/146 \sim 1.53$		\\	

26&IRAS~14378-3651$^1$		&--			&--				&--			&--			&--		&--			&--					&--			\\

27	&IRAS~F11119+3257	&$-14.19\pm 0.05$	&$-11.79_{-0.01}^{+0.01}$	&$41.76\pm 0.05$	&$44.16\pm 0.11$	&$0.40\pm 0.15$		&--		&$1.83_{-0.13}^{+0.13}$	&$0.83_{-0.09}^{+0.09}\times 10^{22}$	&$282/280 \sim 1.00$		\\

28	&IRAS~F01572+0009	&$-14.73\pm 0.20$	&$-12.07_{-0.02}^{+0.02}$	&$41.07\pm 0.21$	&$43.73\pm 0.12$	&$0.07\pm 0.02$		&--		&$2.26_{-0.10}^{+0.10}$	&$<0.009 \times 10^{22}$		&$103/85 \sim 1.21$		\\
	
29	&IRAS~F05024-1941	&$-14.72\pm 0.20$	&$-13.46_{-0.2}^{+0.3}$		&$41.24\pm 0.21$	&$42.50\pm 0.18$	&$<2.16$		&--		&$1.8$		&$0$					&$25/25 \sim 1.00$		\\	

30	&IRAS~F05189-2524	&$-13.14\pm 0.02$	&$-11.25_{-0.005}^{+0.005}$	& $41.42\pm 0.02$	&$43.31\pm 0.11$	&$0.09\pm 0.02$		&--		&$1.86_{-0.10}^{+0.11}$	&$6.61_{-0.70}^{+0.70}\times 10^{22}$	&$286/209 \sim 1.37$		\\
	
31	&IRAS 07251-0248	&--			&--				&--			&--			&--			&--		&--			&--					&--		\\	

32	&IRAS~F07599+6508	&$-13.90\pm 0.08$	&$<-11.34$			&$41.81\pm 0.08$	&$43.37\pm 0.12$	&$0.11\pm 0.02$		&$0.75\pm0.02$	&$1.8^*$		&$0$					&$17/27 \sim 0.64$		\\

33	&IRAS 09022-3615	&$-14.15\pm 0.45$	&$-12.70_{-0.09}^{+0.09}$	&$40.73\pm 0.45$	&$42.19\pm 0.42$	&$0.64\pm 0.32$		&--		&$1.8^*$		&$0$					&$63/48 \sim 1.32$		\\	
34	&IRAS~F09320+6134	&$-13.53\pm 0.03$	&$-11.83_{-0.07}^{+0.07}$	&$40.97\pm 0.04$	&$42.67\pm 0.05$	&$0.85\pm 0.21$		&--		&$1.8^*$		&$0$					&$82/62 \sim 1.34$		\\	
	
35	&IRAS~F12072-0444	&$-14.25\pm 0.03$	&$-13.05_{-0.4}^{+0.4}$		&$41.33\pm 0.20$	&$42.53\pm 0.18$	&$<0.99$		&--		&$1.8^*$		&$0$					&$1.4/2 \sim 0.71$	\\	

36	&IRAS~F12112+0305	&$-14.69\pm 0.05$	&$-13.47_{-0.4}^{+0.4}$		&$40.39\pm 0.20$	&$41.60$		&$<0.68$		&--		&$1.8^*$		&$0$					&$20/17 \sim 1.19$		\\	


37	&IRAS~F14348-1447	&$0$			&$-13.22_{-0.11}^{+0.13}$	&$0$			&$41.96\pm 0.12$	&--			&--		&$1.8^*$		&$0$					&$32/32 \sim 1.0$		\\

38	&IRAS~F14394+5332	&$0$			&$-13.41_{-0.1}^{+0.1}$		&$0$			&$41.97\pm 0.11$	&--			&--		&$1.8^*$		&$0$					&$12/10 \sim 1.2$		\\


39	&IRAS~F15327+2340	&$-13.40\pm 0.20$	&$-12.66_{-0.20}^{+0.11}$	&$40.67\pm 0.21$	&$41.17\pm 0.15$	&$0.73\pm 0.15$		&--		&$1.8^*$		&$0$					&$38/18 \sim 2.15$		\\	

40	&IRAS~F15462-0450	&$-14.23\pm 0.22$	&$-12.38_{-0.05}^{+0.05}$	&$41.11\pm 0.26$	&$42.97\pm 0.23$	&$0.25\pm 0.14$		&--		&$1.8^*$		&$0$					&$72/95 \sim 0.77$		\\	

41	&IRAS 19297-0406	&--			&--				&--			&--			&--		&--			&--				\\	

42	&IRAS 19542+1110	&$-14.03\pm 0.09$	&$-12.32_{-0.05}^{+0.05}$	&$40.90\pm 0.18$	&$42.61\pm 0.14$	&$>1.92$		&--		&$1.8^*$		&$0$					&$68/61 \sim 1.12$		\\	

43	&IRAS~F20551-4250	&$-13.62\pm 0.18$	&$-12.99_{-0.15}^{+0.15}$	&$40.94\pm 0.18$	&$41.57\pm 0.15$	&$0.60\pm 0.25$		&--		&$1.8^*$		&$0$					&$52/39 \sim 1.34$		\\	


44	&IRAS~F23233+2817  	&$-18.61\pm 0.05$	&$-13.95_{-0.05}^{+0.05}$	&$0$			&$41.50\pm 0.15$	&$<0.04$		&--		&$1.8^*$		&$0$					&$71/70 \sim 1.0$\\

45	&NGC~5506		&$-12.12\pm 0.01$	&$-9.866_{-0.001}^{+0.001}$	&$40.86\pm 0.01$	&$43.11\pm 0.05$	&--			&--		&$1.78_{-0.01}^{+0.01}$	&$3.09_{-0.03}^{+0.03}\times 10^{22}$	&$453/261 \sim 1.74$		\\		

46	&NGC~7479		&--			&--   				&--			&--			&--			&--		&--			&--					&--		\\		

47	&NGC~7172		&$-13.24\pm 0.03$	&$-10.16_{-0.01}^{+0.01}$	&--			&$42.91\pm 0.06$	&$0.75\pm 0.14$		&$<2.5$		&$1.56_{-0.04}^{+0.08}$	&$7.79_{-0.20}^{+0.30}\times 10^{22}$	&$267/269 \sim 1.07$		\\

\hline \hline

\end{tabular}  

{$^*$ Sources for which the powerlaw slope $\Gamma$ could not be constrained and hence fixed to $\Gamma=1.8$.}\\

}
\end{table}


\begin{table*}

{\footnotesize
\centering
	\caption{The $\lhard$ luminosity of the MOX sources obtained using different methods and the final list of values used in the correlations. \label{Table:Finallhard}}
  \begin{tabular}{llllllllllll} \hline\hline 

	  Index   &  Source 		&$\log \lhard$		&$\log \lhard$		&$\log \lhard$	& 		 \\ 
		&			&(XMM/Chandra)$^1$	&(\nustar{})$^2$	&Final selection$^5$\\
		&			&$\lunit$		&$\lunit$		&$\lunit$	\\  
	(1)	& (2)			&(3)			&(4)			&(5)		\\ \hline \\

	  1	&IRAS~F08572+3915(CT)	&$^\dagger 41.38^{\rm a}$&$-$			&43.38$^\#$			\\	
                                                                                        
	  2	&IRAS~F10565+2448	&$41.25$		&$-$			&41.25			\\
                                                                                        
	  3	&IRAS~23365+3604	&$^\dagger 41.51^{\rm b}$&$-$			&41.51			\\
                                                                                        
	  4	&Mrk~273		&$42.25$		&$42.93^{\rm A}$	&42.93			\\	
                                                                                        
	  5	&Mrk~876		&$44.11$		&$-$			&44.11			\\	
                                                                                        
	  6	&I~Zw~1			&43.62			&$-$			&43.62			\\	
                                                                                        
	  7	&MrK~231		&42.42			&$42.47^{\rm B}$	&42.47			\\
                                                                                        
	  8	& NGC~1266		&39.78			&$-$			&39.78			\\
                                                                                        
	  9	&M~82			&40.32			&$-$			&40.32			\\
                                                                                        
	  10	&NGC~1377(CT)		&$^\dagger39.00^{\rm c}$&$-$			&41.00$^\#$			\\
                                                                                        
	  11	&NGC~6240(CT)		&42.28			&$43.84^{\rm C}$	&43.84			\\
                                                                                        
	  12	&NGC~3256		&40.96			&$40.04^{\rm D}$	&40.96			\\
                                                                                        
	  13	&NGC~3628		&40.22			&$-$			&40.22			\\
                                                                                        
	  14	&NGC~253		&$^\dagger39.00^{\rm d}$&$39.47^{\rm E}$	&39.47			\\
			                                                                
	  15	&NGC~6764		&39.85			&$-$			&39.85			\\	
                                                                                        
	  16	&NGC~1068(CT)		&40.95			&$43.84^{\rm F}$	&43.84		\\		
                                                                                        
	  17	&IC~5063		&42.57			&$-$			&42.57			\\
                                                                                        
	  18	&NGC~2146		&40.20			&$-$			&40.20			\\
                                                                                        
	  19	&IRAS~17208-0014 	&41.58			&$-$			&41.58			\\
                                                                                        
	  20	&NGC~1614	   	&40.82			&$-$			&40.82			\\
                                                                                        
	  21	&IRAS~05083+7936	&41.49			&$-$			&41.49			\\
                                                                                        
	  22	&IRAS~13451+1232	&43.64			&$-$			&43.64			\\
                                                                                        
	  23	&3C~293			&42.85			&$-$			&42.85			\\
                                                                                        
	  24	&NGC~1433		&39.01			&$-$			&39.01				\\
                                                                                        
	  25	&IRAS~13120-5453(CT)	&41.54			&$43.09^{\rm G}$	&43.09			\\
                                                                                        
	  26	&IRAS~14378-3651	&$^\dagger41.53^{\rm e}$&$-$			&41.96			\\
		                                                                        
	  27	&IRAS~F11119+3257	&44.16			&$-$			&44.16			\\
                                                                                        
	28	&IRAS~F01572+0009	&43.73			&$-$			&43.73			\\	
	29	&IRAS~F05024-1941	&42.50			&$-$			&42.50			\\
                                                                                        
	30	&IRAS~F05189-2524	&43.31			&$43.56^{\rm H}$	&43.56			\\
	31	&IRAS 07251-0248	&$^\dagger43.20^{\rm f}	$&$-$			&43.20			\\
	                                                                                
	32	&IRAS~F07599+6508	&43.37			&$42.70^{\rm I}$	&43.37			\\

	33	&IRAS 09022-3615	&42.18			&$43.14^{\rm J}$	&43.14			\\
	34	&IRAS~F09320+6134	&42.67			&$-$			&42.67			\\
                                                                                        
	35	&IRAS~F12072-0444	&42.53			&$-$			&42.53		       \\
                                                                                        
	36	&IRAS~F12112+0305	&41.60			&$-$			&41.60			\\

	37	&IRAS~F14348-1447(CT)	&41.96			&$-$			&43.96$^\#$		\\

	38	&IRAS~F14394+5332	&41.97			&$-$			&41.97			\\

	39	&IRAS~F15327+2340	&41.16			&$-$			&41.16			\\
                                                                                        
	40	&IRAS~F15462-0450	&45.96			&$-$			&42.96			\\
                                                                                        
	41	&IRAS~F19297-0406	&$^\dagger41.25^{\rm g}$&$-$			&41.25			\\	
	42	&IRAS 19542+1110	&42.62			&$-$			&42.61			\\

	43	&IRAS~F20551-4250(CT)	&41.57			&$-$			&43.57$^\#$			\\

	44	&IRAS~F23233+2817  	&41.50			&$-$			&41.50			\\

	45	&NGC~5506		&43.11			&$-$			&43.11			\\
                                                                                        
	46	&NGC~7479		&$^\dagger42.00^{\rm *}$&$-$			&42.00			\\
                                                                                        
	47	&NGC~7172		&42.90			&$-$			&42.90			\\

\hline \hline

\end{tabular}

\begin{flushleft}
	{$^{\#}$The C-thick sources for which we multiplied the $\lhard$ obtained in Column 3 by a factor of 100.}\\
	{$^{\dagger}$}	The $\lhard$ of the MOX sources estimated using the hardness ratio method.\\
	{Columns 1 \& 2:} The source indices and names.\\
	{Column 3:} The $\lhard$ values of the MOX sources obtained using X-ray spectral fits and HR method using \xmm{} and \chandra{} observations.\\
	References for HR method: a=\citet{2010ApJ...725.1848T}, b=\citet{2011A&A...529A.106I}, c=\citet{2016A&A...590A..73A}, d=\citet{2016A&A...592L...3K}, e=\citet{2011A&A...529A.106I},f=\citet{2011MNRAS.415..619N},g=\citet{2011A&A...529A.106I}\\
	{Column 4:} The intrinsic $\lhard$ values obtained using \nustar{} observations. \\
	References for \nustar{} observations: A=\citet{2015ApJ...814...56T}, B=\citet{2014ApJ...785...19T,2017ApJ...836..155R}, C=\citet{2016A&A...585A.157P}, D=\citet{2015ApJ...806..126L}, E=\citet{2013ApJ...771..134L},F=\citet{2016MNRAS.456L..94M}, G=\citet{2015ApJ...814...56T}, H=\citet{2015ApJ...814...56T}, I=\citet{2014ApJ...794...70L}, J=\citet{2017ApJ...835..179O} \\
	{Column 5:} The final set of $\lhard$ values of the MOX sources used in the correlations and analysis throughout this work.\\
	{$^*$ This was obtained using broad band X-ray spectroscopy using \xmm{} observations by \citet{2011MNRAS.413.1206B}.}
\end{flushleft}

}
\end{table*}


\begin{table*}

{\footnotesize
\centering
  \caption{The $\lhard$ of the MOX sources calculated using two methods. \label{Table:l2to10}}
  \begin{tabular}{llllllllllll} \hline\hline \\

	  Index &  Source 		&$\lhard^a$				&$\lhard^{c}$		 		 \\
	  	&			&					&$12\, \mu$m flux			\\
		&			&($\lx$)				&($\lmu$)			\\
		&			&$\lunit$				&$\lunit$				\\  \hline \\

	  1	&IRAS~F08572+3915 	&43.38			&$44.27$\\	
                                              
	  2	&IRAS~F10565+2448	&41.25			&$42.71$\\	
                                              
	  3	&IRAS~23365+3604	&41.51			&$43.61$\\	
                                              
	  4	&Mrk~273		&42.93			&$43.48$	\\		
                                              
	  5	&Mrk~876		&44.11			&$44.52$\\		
                                              
	  6	&I~Zw~1			&43.62			&$44.76$\\		
                                              
	  7	&MrK~231		&42.47			&$44.76$	\\		
                                              
	  8	& NGC~1266		&39.78			&$41.90$		\\		
                                              
	  9	&M~82			&40.32			&$40.24$			\\		
                                              
	  10	&NGC~1377		&41.00			&$41.99$\\		
                                              
	  11	&NGC~6240		&43.84			&$43.81$	\\		
                                              
	  12	&NGC~3256		&40.96			&$40.94$	\\		
                                              
	  13	&NGC~3628		&40.22			&$40.19$\\		
                                              
	  14	&NGC~253		&39.47			&--	\\		
                                              
	  15	&NGC~6764		&39.85			&$40.99$	\\		
                                              
	  16	&NGC~1068		&43.84			&$42.95$			\\		
                                              
	  17	&IC~5063		&42.57			&$43.46$	\\		
                                              
	  18	&NGC~2146		&40.20			&$39.83$	\\		
                                              
	  19	&IRAS~17208-0014 	&41.58			&$42.70$	\\	
                                              
	  20	&NGC~1614		&40.82			&--	\\	   	
                                              
	  21	&IRAS~05083+7936	&41.49			&--	\\	
                                              
	  22	&IRAS~13451+1232	&43.64			&$44.63$\\	
                                              
	  23	&3C~293			&42.85			&--\\	
                                              
	  24	&NGC~1433		&39.01			&--\\		
                                              
	  25	&IRAS~13120-5453	&43.09			&$43.55$\\	
                                              
	  26	&IRAS~14378-3651	&41.96			&$43.41$\\	
                                              
	  27	&IRAS~F11119+3257	&44.16			&$45.08$\\

28	&IRAS~F01572+0009	       	&43.73			&$44.76$\\
29	&IRAS~F05024-1941		&42.50			&$44.31$	\\	
                                              
30	&IRAS~F05189-2524		&43.56			&$44.34$	\\
31	&IRAS 07251-0248		&43.20		&$43.63$\\	
                                              	
32	&IRAS~F07599+6508		&43.37		&$45.09$ \\

33	&IRAS 09022-3615		&43.14		&$43.99$	\\	
34	&IRAS~F09320+6134		&42.67		&$43.59$	\\	
                                              	
35	&IRAS~F12072-0444		&42.53		&$44.57$\\	
                                              	
36	&IRAS~F12112+0305		&41.60		&$43.46$\\

37	&IRAS~F14348-1447		&43.96      &$43.54$\\

38	&IRAS~F14394+5332		&41.97      &$44.10$	\\

39	&IRAS~F15327+2340		&41.16		&$42.43$\\	
                                              	
40	&IRAS~F15462-0450		&42.96		&$44.18$\\	
                                              	
41	&IRAS~F19297-0406		&41.25		&$43.64$\\	
42	&IRAS 19542+1110		&42.61		&$43.34$\\

43	&IRAS~F20551-4250		&43.57		&$43.84$\\

44	&IRAS~F23233+2817  		&41.50		&$44.27$\\

45	&NGC~5506			&43.11		&$43.22$\\	
                                              	
46	&NGC~7479			  &42.		&$43.11$\\		
	                                      	
47	&NGC~7172			&42.90		&$43.00$	\\		
                                        	
\hline \hline\\

\end{tabular}

$^a$ The $2-10\kev$ luminosity obtained in Table \ref{Table:Finallhard} column 7. \\
$^c$ The $2-10\kev$ luminosity obtained from the 12 $\mu$m flux as described in section \ref{subsec:IRflux}.\\

	}
\end{table*}


\begin{table*}

{\footnotesize
\centering
  \caption{Total bolometric luminosity, AGN luminosity and the X-ray bolometric correction fraction of the sources. \label{Table:lbol}}
  \begin{tabular}{llllllllllll} \hline\hline 

	  Index   &  Source 		&$\log \lbol$	&$\alpha_{\rm AGN}$	&$\log L_{\rm AGN}$	&References		 	&$\log(\lx/L_{\rm AGN})^a$	&$\log(\lmu/L_{\rm AGN})^a$								 \\ 
		&			&$\lunit$	&(in $\%$)		&			&for $\alpha_{\rm AGN}$		&										\\  \hline \\

	  1	&IRAS~F08572+3915	&45.78			&70.4		&45.62			&1				&-2.24		&-1.36									\\	
                                                                                                                                                                        
	  2	&IRAS~F10565+2448	&45.68			&47.1		&45.35			&1				&-4.10		&-2.64										\\
		                                                                                                                                                        
	  3	&IRAS~23365+3604	&45.80			&44.6		&45.45			&1				&-3.93		&-1.83										\\
		                                                                                                                                                        
	  4	&Mrk~273		&45.79			&34.2		&45.32			&1				&-2.39		&-1.84										\\	
                                                                                                                                                                        
	  5	&Mrk~876		&45.87			&92.6 		&45.83			&1				&-1.72		&-1.32										\\	
                                                                                                                                                                        
	  6	&I~Zw~1			&45.59			&90.1		&45.54			&1				&-1.92		&-0.97										\\	
                                                                                                                                                                        
	  7	&MrK~231		&46.18			&80.5		&46.08			&1				&-3.62		&-1.32										\\
                                                                                                                                                                        
	  8	& NGC~1266		&43.91			&25		&43.30			&2				&-3.52		&-1.40											\\
                                                                                                                                                                        
	  9	&M~82			&44.58			&0.09		&41.53			&2				&-1.21		&-1.29											\\
                                                                                                                                                                        
	  10	&NGC~1377		&43.63			&20		&42.93			&2				&-1.93		&-0.94										\\
                                                                                                                                                                        
	  11	&NGC~6240		&45.48			&78		&45.37			&1				&-1.53		&-1.56										\\
                                                                                                                                                                        
	  12	&NGC~3256		&45.12			&0.07		&41.96			&2				&-1.00		&-1.02									\\
                                                                                                                                                                        
	  13	&NGC~3628		&43.83			&0.09		&40.78			&2				&-0.56		&-0.59										\\
                                                                                                                                                                        
	  14	&NGC~253		&44.06			&0.04		&40.66			&2				&$-$		&-										\\
			                                                                                                                                                
	  15	&NGC~6764		&43.99			&1.7		&42.22			&2				&-2.37		&-1.23										\\	
                                                                                                                                                                        
	  16	&NGC~1068		&44.95			&9.7		&43.94			&2				&-0.09		&-0.98										\\		
                                                                                                                                                             	           
	  17	&IC~5063		&44.34			&90		&44.29			&2				&-1.72		&-0.83									\\
                                                                                                                                                                        
	  18	&NGC~2146		&44.61			&0.03		&41.08			&2				&-0.88		&-1.25									\\
                                                                                                                                                                        
	  19	&IRAS~17208-0014 	&46.08			&5		&44.77			&1				&-3.19		&-2.07									\\
                                                                                                                                                                        
	  20	&NGC~1614	   	&45.34			&0		&0			&4				&--		&--										\\
                                                                                                                                                                        
	  21	&IRAS~05083+7936	&45.63			&0		&0			&1				&--		&--										\\
                                                                                                                                                                        
	  22	&IRAS~13451+1232	&45.96			&80.6		&45.87			&1				&-2.23		&-1.26									\\
                                                                                                                                                                        
	  23	&3C~293			&-			&-		&-			&-				&-- 		&--										\\
                                                                                                                                                                        
	  24	&NGC~1433		&-			&-		&-			&-				&--		&--									\\
                                                                                                                                                                        
	  25	&IRAS~13120-5453	&45.88			&33.4		&45.40			&1				&-2.31		&-1.85									\\
                                                                                                                                                                        
	  26	&IRAS~14378-3651	&45.75			&21.1		&45.07			&1				&-3.11		&-1.66									\\
		                                                                                                                                                        
	  27	&IRAS~F11119+3257	&46.29			&80		&46.19			&1				&-2.03		&-1.11									\\

                                                                                                                                                                        
28	&IRAS~F01572+0009	&46.26				&64.6		&46.07			&1				&-2.34		&-1.31											\\	
29	&IRAS~F05024-1941	&46.01				&7.3		&44.87			&1				&-2.37		&-0.56										\\
                                                                                                                                                                        
30	&IRAS~F05189-2524	&45.80				&71.7		&45.65			&1				&-2.09		&-1.31										\\
31	&IRAS 07251-0248	&46.03				&30.0		&45.51			&1				&-2.30		&-1.87										\\
	                                                                                                                                                                
32	&IRAS~F07599+6508	&46.17				&87.6		&46.11			&1				&-2.74		&-1.02									\\

33	&IRAS 09022-3615	&45.93				&54.9		&45.66			&1				&-2.53		&-1.67									\\
34	&IRAS~F09320+6134	&45.63				&56.4		&45.38			&1				&-2.71		&-1.79									\\
                                                                                                                                                                        
35	&IRAS~F12072-0444	&46.04				&74.8		&45.91			&1		       		&-3.38		&-1.34									\\
                                                                                                                                                                        
36	&IRAS~F12112+0305	&45.96				&17.8		&45.21			&1				&-3.61		&-1.75									\\
                                                                                                                                                                        
                                                                                                                                                                        
37	&IRAS~F14348-1447	&45.98				&17.4		&45.22			&1				&-1.26		&-2.68									\\

38	&IRAS~F14394+5332	&45.75				&62.5		&45.54			&1				&-3.57		&-1.44										\\

                                                                                                                                                                        
39	&IRAS~F15327+2340	&45.80				&5.8		&44.56			&1				&-3.40		&-2.13										\\
                                                                                                                                                                        
40	&IRAS~F15462-0450	&45.85				&60.6		&45.63			&1				&-2.67		&-1.45									\\
                                                                                                                                                                        
41	&IRAS~F19297-0406	&46.02				&23.4		&45.38			&1				&-4.13		&-1.74									\\	
42	&IRAS 19542+1110	&45.70				&25.5		&45.11			&1				&-2.49		&-1.76									\\

43	&IRAS~F20551-4250	&45.69				&56.9		&45.44			&1				&-1.87		&-1.60									\\
                                                                                                                                                                        

44	&IRAS~F23233+2817  	&45.69				&44.6		&45.33			&1				&-3.84		&-1.06									\\

45	&NGC~5506		&44.21				&93.3		&44.18			&3				&-1.06		&-0.95									\\
                                                                                                                                                                        
46	&NGC~7479		&43.49				&83.7		&43.41			&3				&-1.41		&-0.30									\\
                                                                                                                                                                        
47	&NGC~7172		&44.37				&92.4		&44.33			&3				&-1.44		&-1.33									\\

\hline \hline

\end{tabular}
  
{References: 1=\citet{2013ApJ...776...27V}  2=\citet{2014A&A...562A..21C}, 3= \citet{2016ApJ...826..111S}, 4=\citet{2009PASP..121..559A} }\\
{$^a$ See Section \ref{subsec:bestlhard} for the definition of $\lx$ and $\lmu$.}

}
\end{table*}


\begin{table*}

{\footnotesize
\centering
  \caption{Correlation results between parameters $x$ and $y$ ($y=ax+b$). \label{Table:corr}}
  \begin{tabular}{llllllllllll} \hline\hline

Correlation 			& $a$	\hspace{1.5cm} &Dev($a$)\hspace{1.5cm}&$b$ \hspace{1.5cm}&Dev($b$) \hspace{1.5cm}&$R_{\rm S}$ \hspace{1.5cm}&$P_{\rm null}$ \hspace{1.5cm}&Data points \\ \hline

$L_{2-10\kev}$ vs MO vel	&$150$		&$18$		&$-5901$	&$766$		&$0.56$		&$\rm 4 \times 10^{-5}$	 &47	\\ 
$L_{2-10\kev}$ vs MO $\mout$	&$0.55$		&$0.07$		&$-21$		&$2.89$		&$0.76$		&$\rm 1.1 \times 10^{-5} $	&25	\\ 

	  $\lmu$ vs MO vel	&$180$		&$24$		&$-7382$	&$1063$		&$0.70$		&$\rm 2.6 \times 10^{-7}$	 &43	\\ 
	  $\lmu$ vs MO $\mout$	&$0.51$		&$0.06$		&$-19$		&$2.40$		&$0.81$		&$\rm 3.8 \times 10^{-6} $	&22	\\

$L_{0.6-2\kev}$ vs MO vel	&$173$		&$37$		&$-6686$	&$1520$		&$0.59$		&$\rm 4 \times 10^{-4}$	&31	\\ 
$L_{0.6-2\kev}$ vs MO $\mout$	&$0.52$		&$0.08$		&$-19.58$	&$3.31$		&$0.83$		&$\rm 2.5\times 10^{-5}$	&17\\

$L_{\rm AGN}$      vs MO vel		&$155$		&$18$		&$-6511$	&$813$		&$0.70$		&$\rm 1.48\times 10^{-7}$	&43	\\ 
$L_{\rm AGN}$    vs MO $\mout$	&$0.45$		&$0.04$		&$-17.93$	&$1.93$		&$0.86$		&$\rm 3.8\times 10^{-7}$	&25	\\

$\lsb$      vs MO vel		&$197$		&$61$		&$-8437$	&$2775$		&$0.40$		&$\rm 0.008$	&43	\\ 
$\lsb$     vs MO $\mout$	&$1.02$		&$0.16$		&$-43.99$	&$7.22$		&$0.57$		&$\rm 0.005$	&22	\\

\hline \hline

\end{tabular}  

}
\end{table*}

\clearpage

\appendix


\section{A. The Best fit spectra and models}

In this section we show the best fit data for the sources in the MOX sample, along with the best fit model and the residuals after the data have been fitted with the model. For sources with counts $\le 200$ we have shown the spectra for viewing purpose only as we have used HR method to calculate the luminosity.

\newpage

\begin{figure}
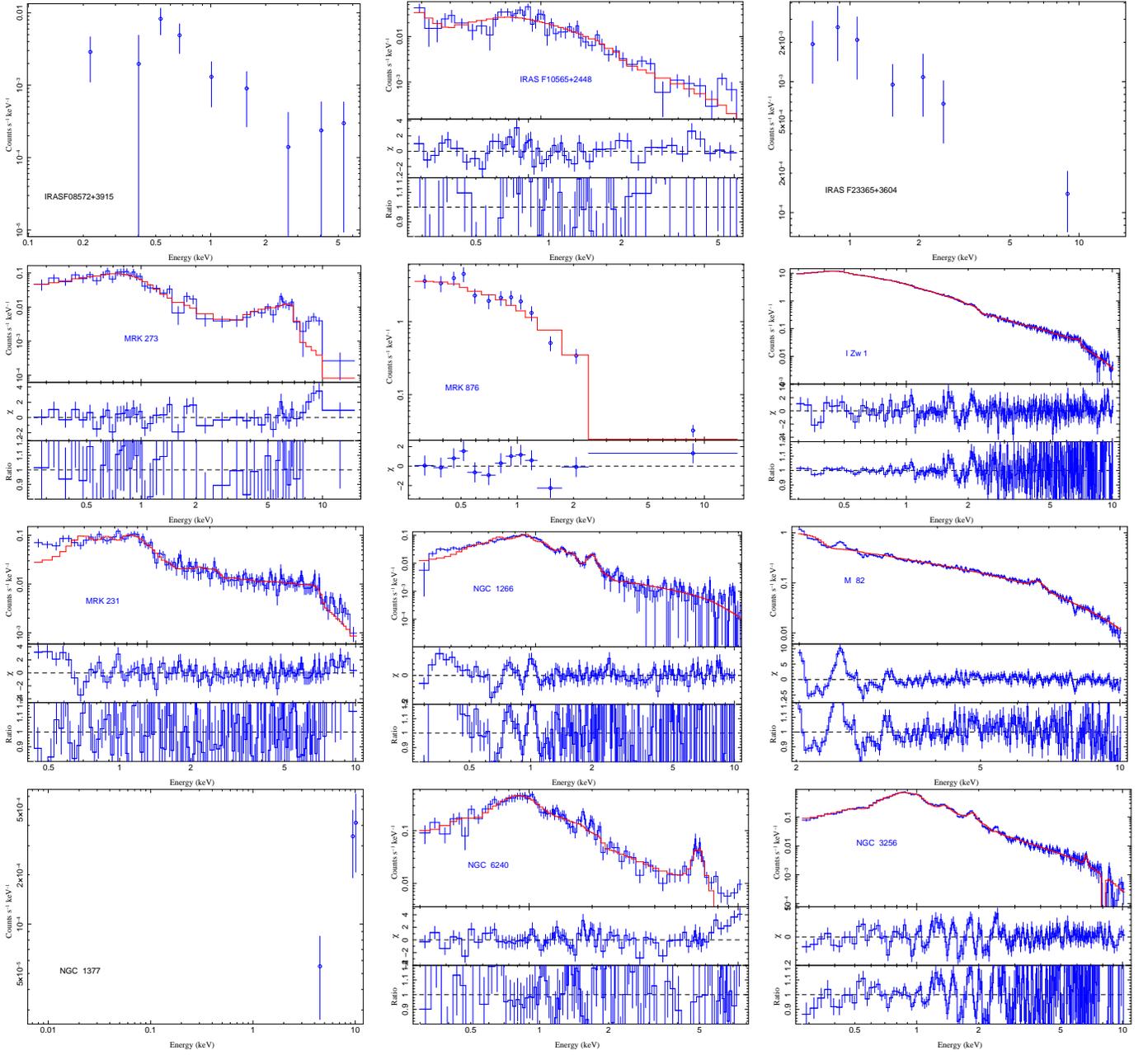

  \centering

\vbox{

\hbox{
\includegraphics[width=4.2cm,angle=-90]{IRASF08572.ps} 
\includegraphics[width=4.2cm,angle=-90]{IRASF10565_multiplot.ps} 
\includegraphics[width=4.2cm,angle=-90]{IRASF23365.ps}
}

\hbox{
\includegraphics[width=4.2cm,angle=-90]{MRK273_multiplot.ps}
\includegraphics[width=4.2cm,angle=-90]{MRK876_bestfit.ps}
\includegraphics[width=4.2cm,angle=-90]{Izw1_multiplot.ps}
}

\hbox{
\includegraphics[width=4.2cm,angle=-90]{MRK231_multiplot.ps}
\includegraphics[width=4.2cm,angle=-90]{NGC1266_multiplot.ps}
\includegraphics[width=4.2cm,angle=-90]{M82_multiplot.ps}
}

\hbox{
\includegraphics[width=4.2cm,angle=-90]{NGC1377.ps}
\includegraphics[width=4.2cm,angle=-90]{NGC6240_multiplot.ps}
\includegraphics[width=4.2cm,angle=-90]{NGC3256_multiplot.ps}
}

 }

\caption{The X-ray data, the best fit model and the residuals of the sources. For the sources where the low counts do not allow us to carry out standard fitting, we have plotted just the X-ray data. See Section \ref{sec:data-analysis} for details.}   \label{fig:7}
\end{figure}


\clearpage
\begin{figure}
  \centering

\vbox{

\hbox{
\includegraphics[width=4.2cm,angle=-90]{NGC3628_multiplot.ps} 
\includegraphics[width=4.2cm,angle=-90]{NGC6764_multiplot.ps} 
\includegraphics[width=4.2cm,angle=-90]{NGC1068_multiplot.ps}
}

\hbox{
\includegraphics[width=4.2cm,angle=-90]{IC5063_multiplot.ps}
\includegraphics[width=4.2cm,angle=-90]{NGC2146_multiplot.ps}
\includegraphics[width=4.2cm,angle=-90]{IRAS17208_multiplot.ps}
}

\hbox{
\includegraphics[width=4.2cm,angle=-90]{NGC1614_multiplot.ps}
\includegraphics[width=4.2cm,angle=-90]{IRAS05083_multiplot.ps}
\includegraphics[width=4.2cm,angle=-90]{IRAS13451_multiplot.ps}
}

\hbox{
\includegraphics[width=4.2cm,angle=-90]{3C293_multiplot.ps}
\includegraphics[width=4.2cm,angle=-90]{NGC1433_multiplot.ps}
\includegraphics[width=4.2cm,angle=-90]{IRAS13120_multiplot.ps}
}

 }

\caption{Continued from Fig. \ref{fig:7}. } \label{fig:8}
\end{figure}


\clearpage

\begin{figure}
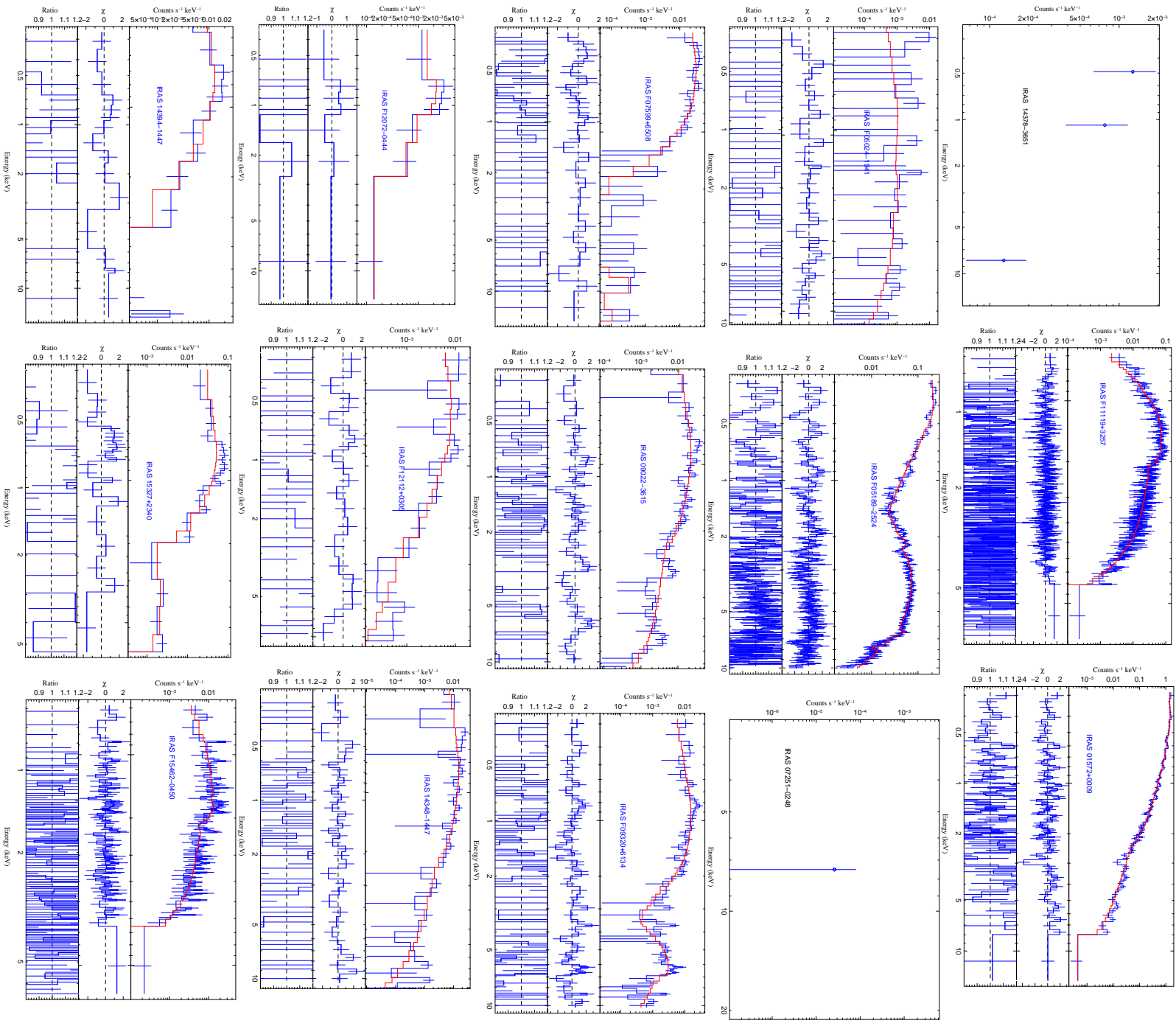

  \centering 

\vbox{

\hbox{
\includegraphics[width=4.2cm,angle=-90]{IRAS14378.ps} 
\includegraphics[width=4.2cm,angle=-90]{IRASF11119_multiplot.ps} 
\includegraphics[width=4.2cm,angle=-90]{IRAS01572_multiplot.ps} 
 }

\hbox{
 \includegraphics[width=4.2cm,angle=-90]{IRASF05024_multiplot.ps} 
\includegraphics[width=4.2cm,angle=-90]{IRASF05189_multiplot.ps} 
\includegraphics[width=4.2cm,angle=-90]{IRAS07251.ps} 

}

\hbox{
\includegraphics[width=4.2cm,angle=-90]{IRASF07599_multiplot.ps} 
\includegraphics[width=4.2cm,angle=-90]{IRAS09022_multiplot.ps} 
\includegraphics[width=4.2cm,angle=-90]{IRASF09320_multiplot.ps}
}

\hbox{
 
\includegraphics[width=4.2cm,angle=-90]{IRASF12072_multiplot.ps} 
\includegraphics[width=4.2cm,angle=-90]{IRASF12112_multiplot.ps} 
\includegraphics[width=4.2cm,angle=-90]{IRASF14348_multiplot.ps}
}

\hbox{
 
\includegraphics[width=4.2cm,angle=-90]{IRAS14394_multiplot.ps} 
\includegraphics[width=4.2cm,angle=-90]{IRAS15327_multiplot.ps}
\includegraphics[width=4.2cm,angle=-90]{IRASF15462_multiplot.ps} 
}

}
\caption{Same as Fig. \ref{fig:7} except that for these sources we only have outflow velocity estimates and not the molecular gas mass outflow rate. } \label{fig:10}
\end{figure}


\clearpage
\begin{figure}
  \centering 

\vbox{

\hbox{

\includegraphics[width=4.2cm,angle=-90]{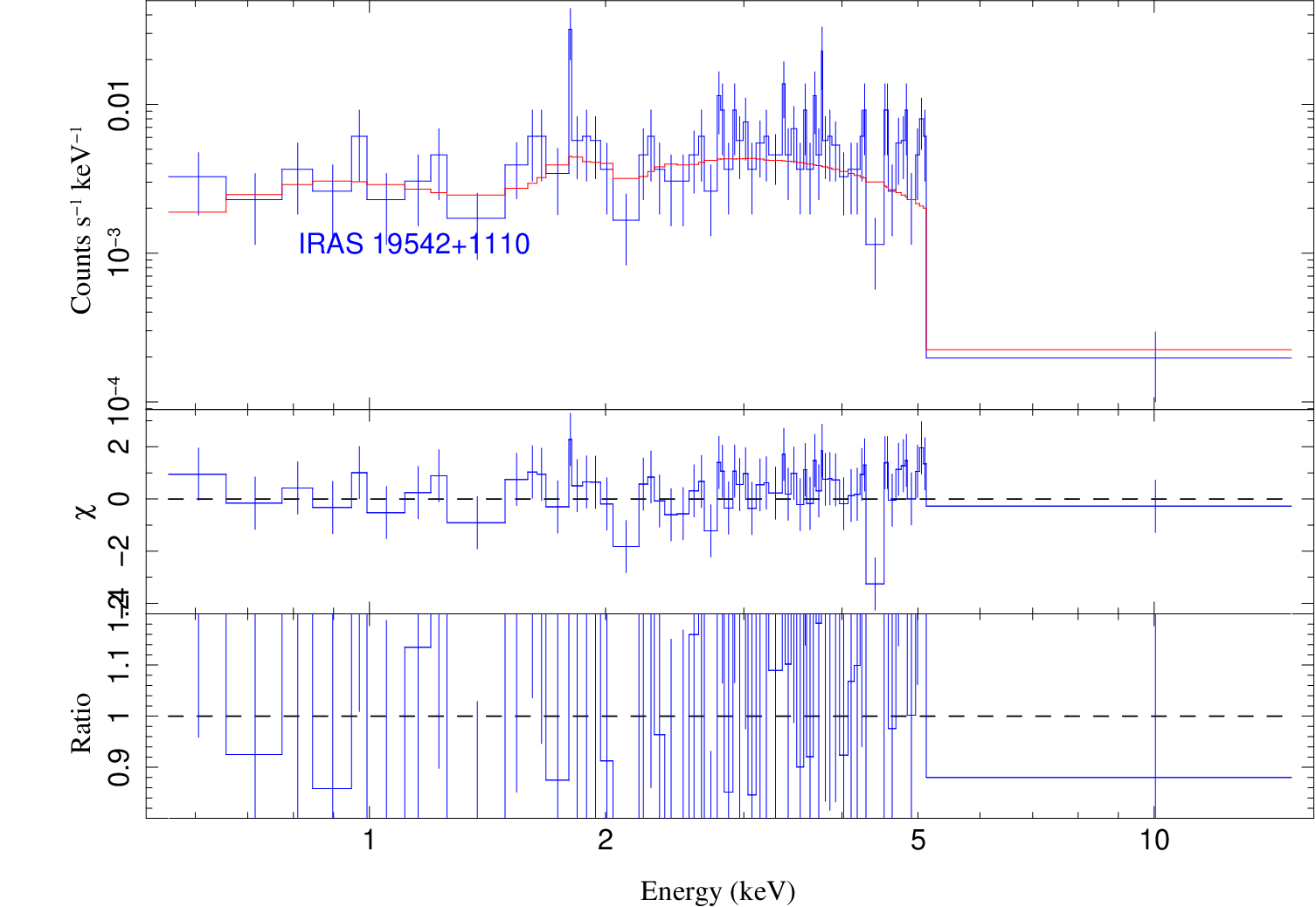} 
\includegraphics[width=4.2cm,angle=-90]{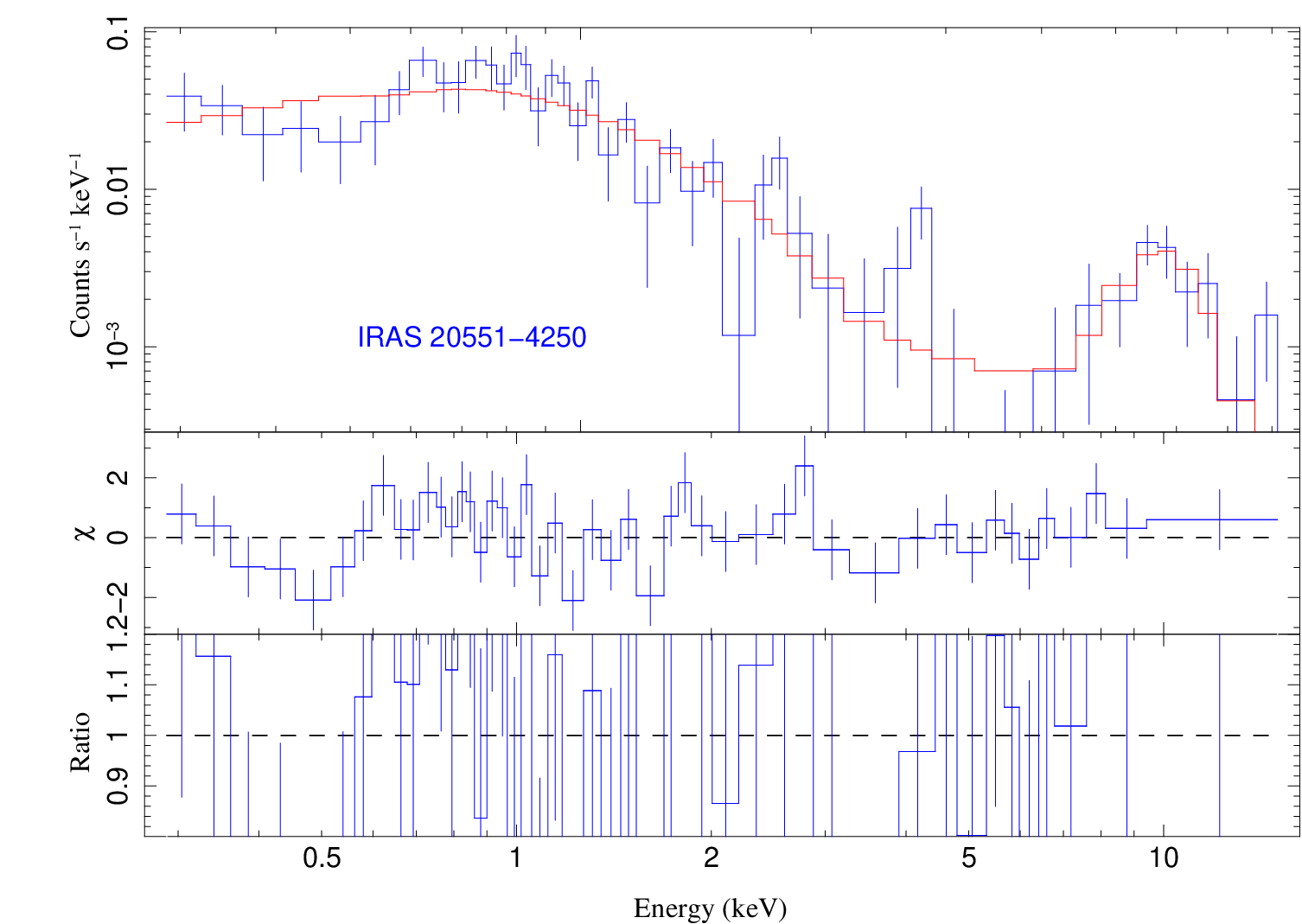} 
\includegraphics[width=4.2cm,angle=-90]{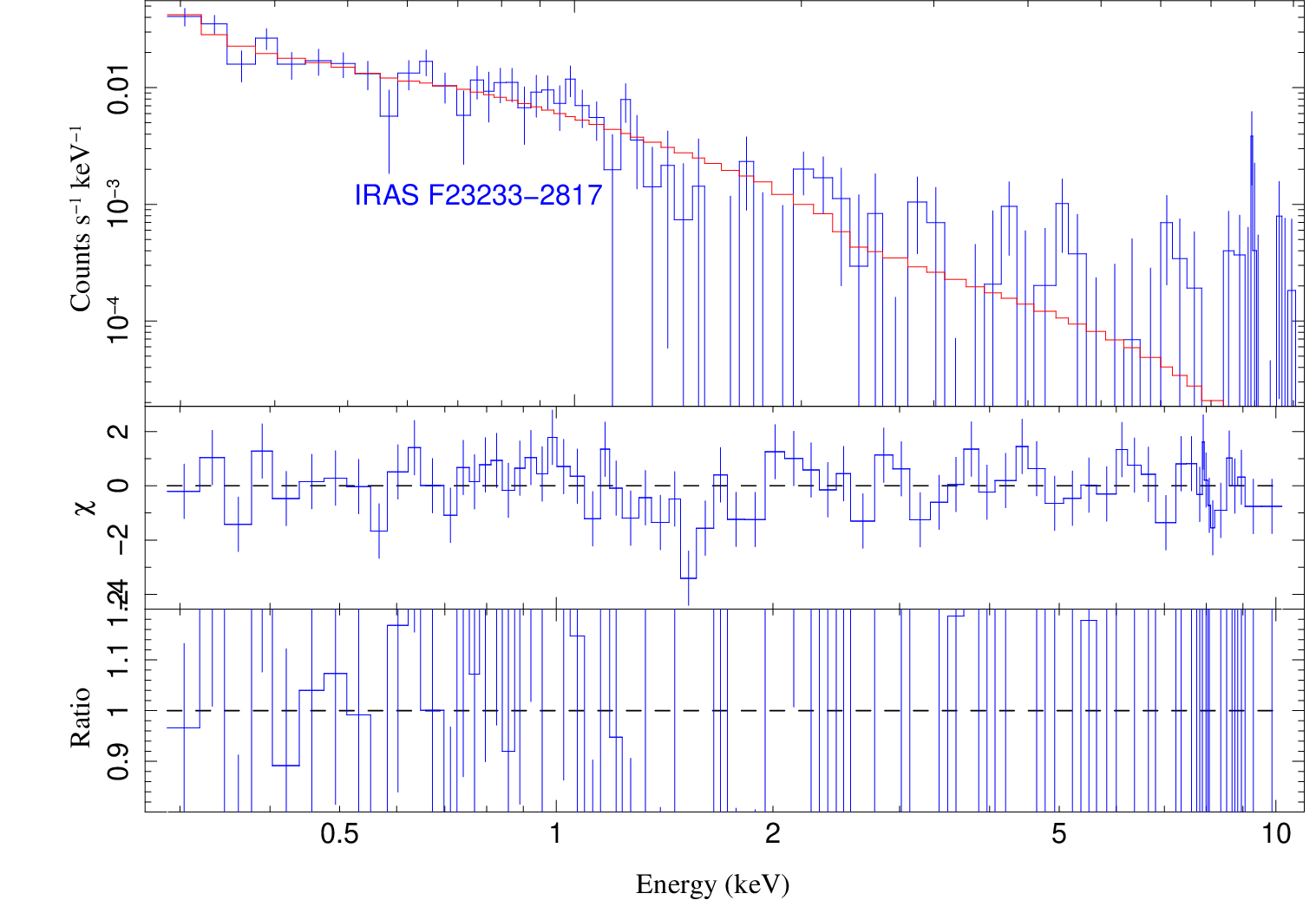}}

\hbox{
 
\includegraphics[width=4.2cm,angle=-90]{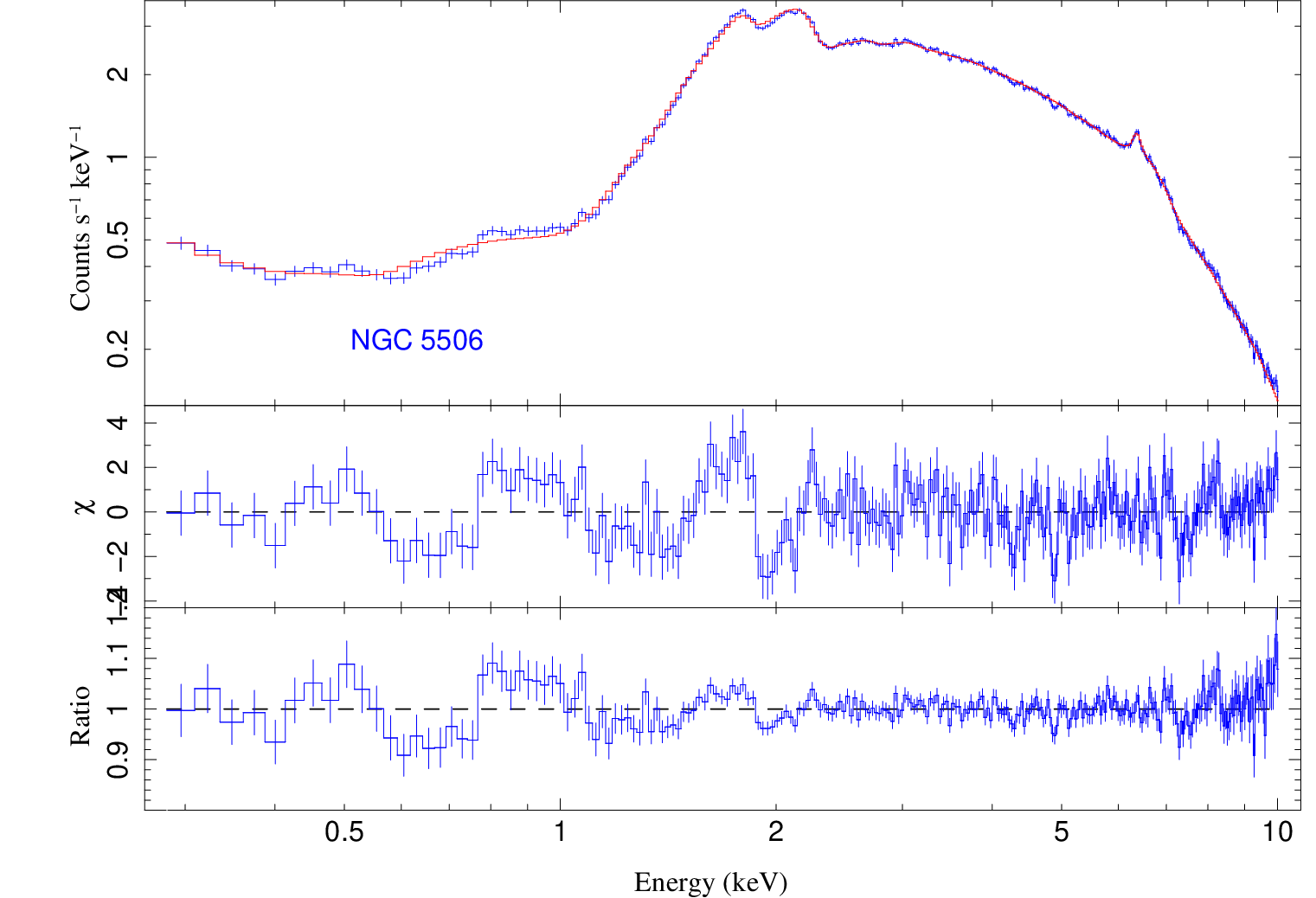} 
\includegraphics[width=4.2cm,angle=-90]{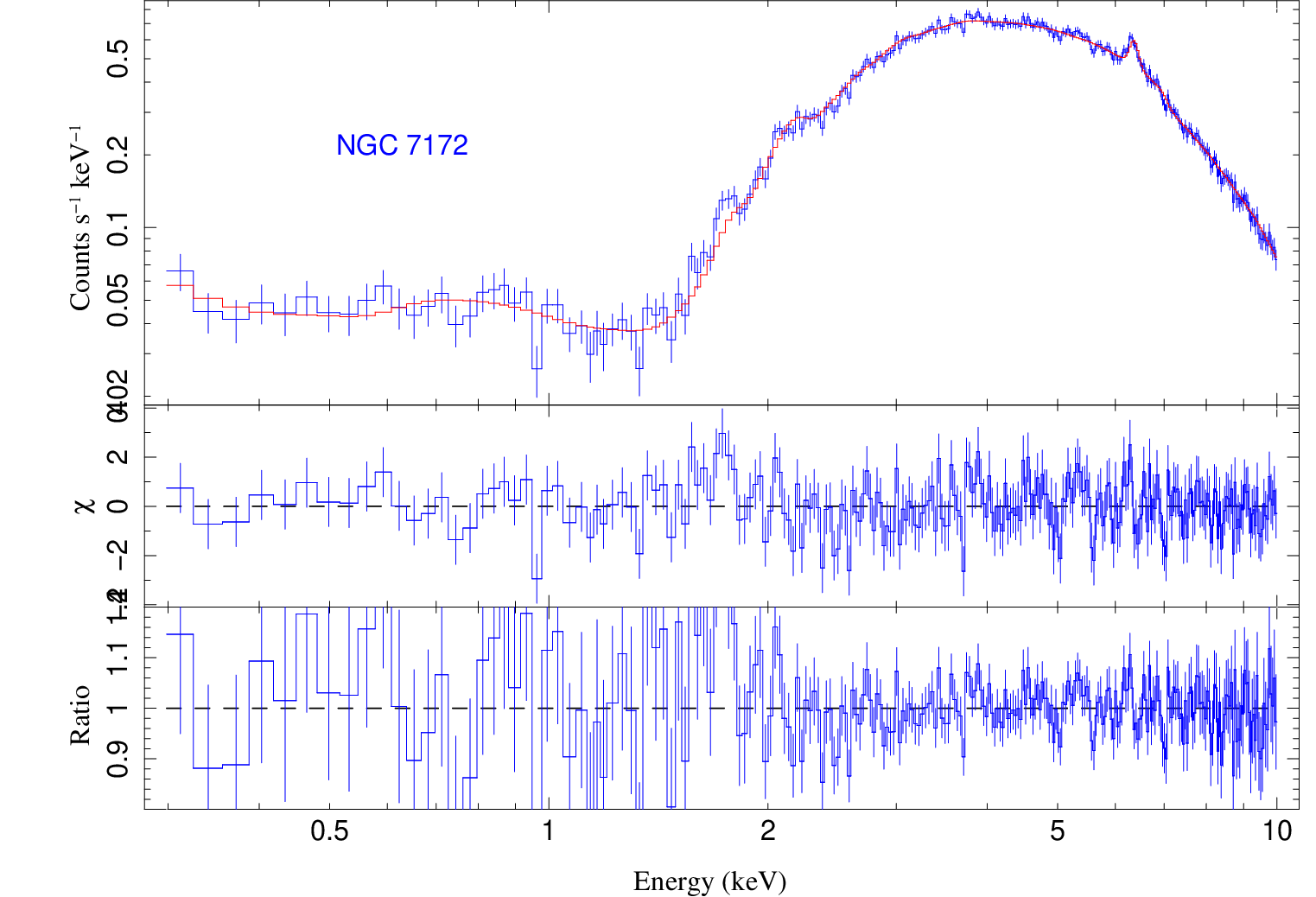} 

}

}
\caption{Same as Fig. \ref{fig:10} } \label{fig:11}
\end{figure}


\clearpage
\section{B. Description of the individual sources in the MOX sample.}

1. IRAS~F08572+3915: This is a double nucleus ULIRG. The source has been  identified as Compton thick \citep{2010ApJ...725.1848T}, and the previous studies have estimated an absorption column density of $\sim 10^{25} \cmsqi$. The powerlaw photon index $\Gamma=-0.43$ is not constrained due to low counts. The authors classified this as a weak ULIRG. This source has also been studied by \citet{2011A&A...529A.106I} using \chandra{} observations. The source luminosity estimated by the authors are $L_{\rm SX}= 8.0\times 10^{40} \lunit$ and $L_{\rm HX}= 2.0\times 10^{41} \lunit$ in the soft and the hard band respectively. Nustar hard X-ray studies were carried out by \citet{2015ApJ...814...56T} and the authors conclude that the source is X-ray weak and could not be detected in any of the \nustar{} energy bands. The molecular outflows in this source has been detected using IRAM PDBI telescope with the CO(1-0) emission line \citep{2014A&A...562A..21C}.

In our analysis, we found that this source has very low counts and hence HR method was used to calculate the $2-10\kev$ luminosity.\\

\noindent 2. IRAS~F10565+2448: This is a pair of interacting spiral galaxies, and is a heavily obscured source. \citet{2010ApJ...725.1848T} studied this source using \chandra{} observation and have estimated an absorption column density of $0.05_{-0.04}^{+0.07}\times 10^{22} \cmsqi$, implying a Compton thin obscurer. The estimated powerlaw $\Gamma=1.62_{-0.13}^{+0.14}$. \citet{2011A&A...529A.106I} studied the \chandra{} observation of this source and have estimated a luminosity of $L_{\rm SX}= 1.21\times 10^{41} \lunit$ and $L_{\rm HX}= 1.6\times 10^{41} \lunit$. The authors mention that the hard X-ray emission is point like but the soft X-ray emission is much more extended up to 7". Nustar hard X-ray studies carried out by \citet{2015ApJ...814...56T} could not detect the source in any energy band of \nustar{}.

We could not constrain the intrinsic neutral absorption for this source. The best fit photon index is $\Gamma= 2.17_{-0.23}^{+0.23}$. In addition we required an absorption edge at $0.34 \kev$. \\

\noindent 3. IRAS~23365+3604:\citet{2011A&A...529A.106I} studied the \chandra{} data of this source and found that it is a heavily obscured source. A faint X-ray source is present in the nucleus, which however could not be studied properly because of the short exposure of the observation ($\sim 10\ks$). The hard X-ray color HR=-0.22 points to the fact that this object is an AGN, which is Compton thick. This source was also studied by \citet{2005ApJ...633..664T}, who found $\Gamma=1.10_{-0.25}^{+0.35}$ and an absorption column density of $50_{-27}^{+39}\times 10^{20} \cmsqi$.

Due to lack of counts, HR method was used to calculate the $2-10\kev$ luminosity .\\

\noindent 4. MRK~273:Classified as Seyfert 2 (NED). \cite{2015A&A...579A..90H} classified the candidate as a changing look candidate with both Compton thick and thin signatures available from different observations. Nustar hard X-ray studies were carried out by \citet{2015ApJ...814...56T} and an intrinsic luminosity of $\lhard=8.55\times 10^{42}\lunit$ was derived.

In our study, we detected a broad Fe K line which was modeled by diskline profile. Soft X-ray emission lines were modeled using a Gaussian profile. We could not detect any neutral absorption intrinsic to the source. The powerlaw slope $\Gamma <1.58$ is very flat.\\

\noindent 5. MRK~876: This is a Seyfert 1 galaxy and has a strong AGN at its centre. Early studies by \citet{1995A&A...296...90E} confirm the source to be variable in X-rays and UV. The Swift/XRT data studied by \citet{2015ApJ...798L..14B} found a broad Fe K$\alpha$ emission line. The Fe K line was also studied by \citet{2010A&A...524A..50D} using \xmm{} data. \citet{2005A&A...432...15P} had studied this source as a part of a sample of PG quasars, and reported $L_{2-10}=1.78\times 10^{44} \lunit$.  

In our work, the data quality being poor, we could not constrain any intrinsic neutral absorption. The spectra just required an absorbed powerlaw, and the slope could be constrained. We derived similar X-ray luminosity as those of \citet{2005A&A...432...15P}. The \nustar{} observation of this source is not yet made public.\\

\noindent 6. I~Zw~1: This is a narrow line Sy 1 galaxy, and is highly variable. An extensive X-ray study of this source has been done by \citet{2007MNRAS.377..391G} and \citet{2005A&A...432...15P} studied the source as a part of a sample of PG quasars.

For this source we detected two components of warm absorbers. There was also the presence of a broad Fe K emission line and a neutral intrinsic absorption column.\\

\noindent 7. MRK~231: Obscured source with a strong AGN, studied by \citet{2010ApJ...725.1848T}. A connection between MO and UFO is found in this source by \citet{2015A&A...583A..99F}. The authors confirm an energy conserving mechanism responsible for creating the molecular outflows from the UFOs. Chandra imaging and spectroscopy has been carried out by \citet{2014ApJ...790..116V}. A Nustar hard X-ray view of this source has been carried out by \citet{2015ApJ...814...56T}. A separate study using Nustar data focussing only on this source has been carried out by \citet{2014ApJ...785...19T}, and the authors measured an X-ray luminosity of $\lhard=3.94\times 10^{42}\lunit$. The authors concluded that this source is a Compton thin AGN.

We found that this source has a complex spectrum which required one component of warm absorber, one component of thermal emission ({\it APEC}), a neutral intrinsic absorption as well as a broad Fe K emission line. The powerlaw slope is flat and its lower value is pegged at $\Gamma=1.5$.\\

\noindent 8. NGC~1266: A nearby lenticular galaxy, harbouring an AGN which powers a massive MO detected in this source which harbours an AGN \citep{2015ApJ...798...31A}. Apart from extensive analysis of Chandra and XMM data, a multiwaveband study was carried out by the authors, weher they detected a soft emission from starburst, a powerlaw and Fe K line from the AGN. The intrinsic absorbtion column density estimated for this source from IR studies of \citep{2015ApJ...798...31A} is $\nh=3\times 10^{24} \cmsqi$, almost 3 orders of magnitude higher than that found using X-ray studies. Supression of Star formation in this S.B galaxy is studied by \citet{2015ApJ...798...31A}. There is \nustar{} observation of this source but there is no published study.

In our study we found that this source has a complex spectrum which required a soft thermal component ({\it APEC}) along with a warm absorber, a neutral intrinsic absorber, and soft X-ray emission lines at $1.48\kev$ and $1.85\kev$ in the observer's frame. The Fe K line was not detected due to poor SNR.\\

\noindent 9. M~82: This is a starburst dominated galaxy. \citet{2014MNRAS.437L..76L} studied the nuclear region of the source with 500 ks Chandra data. Fe K $\alpha$ line is detected, and most of the hard X-ray emission $2-8\kev$ has a thermal origin. A weakly broadened Fe K line was detected by \citet{2011MNRAS.418.1973C}.

The spectrum is complex with several discrete emission features in the soft X-ray band. We could not obtain a good statistical fit to the data with the baseline models used in this work. We detected narrow Fe K emission. A neutral intrinsic absorber has also been detected.\\

\noindent 10. NGC~1377: Chandra and Swift data not published, hence no previous studies available for this source. 

The source photon counts being weak, the HR method was used in our work.\\

\noindent 11. NGC~6240: Mostly Compton thick galaxy merger. \citet{2016A&A...585A.157P} studied the source with Nustar data, and concluded that this source could be an early merger stage galaxy with two nuclei separated, and an intrinsic source luminosity of $\lhard=7\times 10^{43}\lunit$. Both active and obscured Compton thick material present. \citet{2014ApJ...781...55W} detected fast shock heated gas within 5 kpc of the central region. \citet{2013ApJ...765..141N}  and \citet{2013A&A...549A..51F} have studied the Chandra data and detected a soft X-ray halo, and also CO emission lines. \citet{2010ApJ...725.1848T} and \citet{2011A&A...529A.106I} have studied the source in a sample. \citet{2005ApJ...629..739N} have studied \xmm{} observation of this source and found that starburst emission dominates the soft X-ray $0.5-3\kev$ energy range.

In our study the X-ray spectra required a broad Fe K emission line, along with a soft X-ray emission line at $0.89\kev$. We could not constrain the intrinsic neutral absorption.\\

\noindent 12. NGC~3256: Powerful starburst galaxy studied by \citet{2015ApJ...806..126L} with Chandra and Nustar data. Nature of X-ray emission is unclear as no obvious AGN signature was found. This galaxy was studied by \citet{2004MNRAS.352.1335J} and was referred to as starburst merger galaxy, and a hard X-ray bolometric correction was estimated to be $\sim 10^{-5}$. The $\lhard\sim 10^{40}\lunit$ has been estimated mostly from the ULXs and crowded X-ray sources, and not an AGN.

In our study we found that the X-ray spectrum is complex. It required two thermal components in the soft X-rays ({\it APEC}), one neutral intrinsic absorber, and three Gaussian emission lines for three Fe K emission lines at different ionisation states.\\

\noindent 13. NGC~3628: \citet{2012ApJ...752...38T} studied this starburst galaxy and found connection between MO and emission line plasma in X-rays. A study of the source was carried out by \citet{2001ApJ...560..707S} using Chandra data, where they find a luminous X-ray source 20'' away from the nucleus. 

The spectrum required one thermal component ({\it APEC}), neutral intrinsic absorption and a high energy absorption in the Fe K band which was modeled using an inverted Gaussian. The powerlaw slope is pegged at $\Gamma= 1.5$.\\

\noindent 14. NGC~253: A highly variable starburst galaxy studied by Nustar \citet{2013ApJ...771..134L}. Nustar and Chandra data reveal that the nuclear region contains three bright X-ray point sources which are ULXs and not an AGN, and highly obscured with a column density of $\log\nh=23 \cmsqi$.  The Fe K line complex was studied by \citet{2011ApJ...742L..31M} and found several highly ionised Fe K emission lines.

Due to low photon counts, HR method was employed in our work.\\

\noindent 15. NGC~6764: This is an AGN + Starburst galaxy and the chandra data is studied by \citet{2008ApJ...688..190C}.  

The hard X-ray band $>2 \kev$ has very few counts, hence the powerlaw slope upperlimit could not be constrained, $\Gamma>2.28$. The soft X-ray emission was modeled using {\it Apec}.\\

\noindent 16. NGC~1068: Compton thick Sy 2 galaxy was studied by Nustar data by \citet{2015ApJ...812..116B}. Multi component X-ray reflectors were needed to fit the data. \citet{2014ApJ...780..121K} studied the source with Chandra data, and found the amount of mass of gas necessary for the emission in X-rays is $\sim 3.7\times 10^5 \msol$. \citet{2011ApJ...738..147S} studied the Fe K line emission of the source. \citet{2016MNRAS.456L..94M} studied the source using Nustar data unveiling the obscured source.

This source could not be modeled with the baseline model. The spectra shows a very uniqe broad Fe K emission line, typical of Compton thick objects.\\

\noindent 17. IC~5063: Classified as narrow line Sy 2 radio galaxies and the Suzaku data are studied by \citet{2011ApJ...738...70T}. There is a Nustar data but not published. \citet{2012ApJ...748..130M} also studied the source in a sample of Sy 2 sources which have exhibited broad Fe K line in reflected spectra. The source is classified as Compton thin by \citet{2012ApJ...748..130M}.

Absorbed source with a concave spectra in the hard X-rays and diffuse soft X-ray emission.  An absorbed powerlaw and a blackbody for the diffuse soft X-ray emission could fit the data. There could be X-ray contribution from radio jets.\\

\noindent 18. NGC~2146: The Chandra observation of this starburst galaxy was carried out by \citet{2005PASJ...57..135I}. There were 6 ultra-luminous point sources detected in the field of view.

In our work we found that the spectra can be modeled by an absorbed powerlaw only.\\

\noindent 19. IRAS~17208-0014: \citet{2010ApJ...725.1848T} and \citet{2011A&A...529A.106I} have studied this source. This is a luminous ULIRG.

In our work the powerlaw slope could not be constrained due to low photon counts.\\

\noindent 20. NGC~1614: ULIRG, and a star forming galaxy studied in the multiwaveband by \citet{2014ApJ...786..156H}. This is also detected as a merger remnant by \cite{2017ApJ...835..174S} using ALMA data. The nature of the dominant emitting mechanism at the centre is still under debate, however AGN presence may not be needed to describe the spectral properties. Possibly a compact starburst ($r\le 90\pc$) is present. The total IR luminosity is $L= 4\times 10^{11}\lsol$ \citet{2009PASP..121..559A,2014ApJ...786..156H}. An upper limit to AGN luminosity is given by the authors, $L_{\rm AGN} \le 4.5\times 10^{11} \lsol$.

In our study, the powerlaw slope could not be constrained due to low photon counts.\\

\noindent 21. IRAS~05083+7936: This is an absorbed quasar. \citet{2014MNRAS.444.2580B} had studied this absorbed quasar. Nustar hard X-ray studies carried out by \citet{2015ApJ...814...56T}.

In our work, the powerlaw slope could not be constrained due to low photon counts.\\

\noindent 22. IRAS~13451+1232: A Seyfert 2 galaxy. \citet{2013ApJ...777...27J}, \citet{2010ApJ...725.1848T} and \citet{2014ApJ...787...61L} studied the source in X-rays. 

We found that the hard X-ray photon count is poor, however, the powerlaw slope and the absorption column could be constrained.\\

\noindent 23. 3C~293:\citet{2015ApJ...801...17L} studied the Jet-ISM interaction of this radio loud source and states how the molecular gas is heated by the jets. Only Chandra data available for this source.\\

\noindent 24. NGC 1433:Only Chandra data available which is not published.

              In our study we found the source has very poor data counts. The powerlaw slope as well as the intrinsic neutral absorption column density were frozen to a value of $1.5$ and $0.07 \times 10^{22} \cmsqi$, as they could not be constrained.\\

\noindent 25. IRAS~13120-5453: 129 ks \xmm{} data is not published yet. Nustar hard X-ray studies carried out by \citet{2015ApJ...814...56T} who confirms it as Compton thick AGN.

We found that this source has a narrow Fe K emission line.\\

\noindent 26. IRAS~14378-3651: Nustar hard X-ray studies carried out by \citet{2015ApJ...814...56T} and the authors do not detect any source X-ray flux beyond $10\kev$. The \chandra{} spectra have been studied by \citet{2011A&A...529A.106I}.\\

\noindent 27. IRAS~F11119+3257: Studied by \citet{2015Natur.519..436T} who found strong molecular outflows in IR using Herschel PACS as well as ultrafast outflows in the Xrays using Suzaku telescope. \citet{2017ApJ...850..151T} observed this source with \nustar{} and detected a similar observed flux as ours.

We found that this is a bright source and could be modeled with a simple absorbed powerlaw and a very weak bbody for the soft emission.\\

\noindent 28. IRAS~F01572+0009: This source is also known as MRK~1014 alias PG0157+001. It has been studied by \citet{2014A&A...567A.142R} and a $2-10\kev$ luminosity of $10^{43.80} \lunit$ has been estimated using the same dataset used by us. The authors have detected reflection from distant matter and used the model {\it Mytorus}.  

We found that this is a low count source which could be modeled with a simple absorbed powerlaw and a bbody for the soft emission. The data did not require a neutral intrinsic absorber.\\

\noindent 29. IRAS~F05024-1941: This is a ULIRG and was studied by \citet{2010ApJ...725.1848T} and the authors have used HR method to calculate the flux.  

The source has very low photon counts. The powerlaw slope and the neutral absorber $\nh$ could not be constrained.\\

\noindent 30. IRAS~F05189-2524: A ULIRG studied by \citet{2010ApJ...725.1848T} and the authors found $L_{2-10\kev}=2.3 \times 10^{43} \lunit$.	 

The soft emission was modeled using two bbody components. The spectra required two Gaussian emission lines for two Fe K complex. The spectra is concave indicating neutral absorption.\\

\noindent 31. IRAS 07251-0248:  This source is a ULIRG and has been studied by \citet{2011A&A...529A.106I} as well as by \citet{2010ApJ...725.1848T} both of whom found the source to be Compton thin AGN. However, the source could not be detected in the X-rays  (Chandra) by the authors (as well as by us) due to its weakness and obscuration. The X-ray flux of the source estimated from the infra-red analysis is $F_{2-10\kev}=7.4 \times 10^{-12} \funit$ \citep{2011MNRAS.415..619N}.  \\	

\noindent 32. IRAS~F07599+6508: This is a Compton thin absorbed ULIRG. \citet{2010ApJ...725.1848T} studied this source and calculated a luminosity of $L_{2-10\kev}=1.12 \times 10^{42} \lunit$ using HR method. \citet{2004AJ....127..758I} studied a sample of four ULIRGs with detectable broad near-infrared-emission lines produced by AGN, one of the sources being IRAS~F07599+6508. Using spectral analysis they could constrain the luminosity of the AGN $L_{2-10\kev}=8 \times 10^{41} \lunit$. 

Very low photon counts did not allow us to constrain the powerlaw slope or the $\nh$ of the intrinsic absorber.\\

\noindent 33. IRAS 09022-3615: This is a LIRG and \citet{2011A&A...529A.106I} calculated an X-ray luminosity of $L_{2-10\kev}=2.0 \times 10^{42} \lunit$. The authors found that the hard X-ray source is marginally resolved.		

Very low photon counts did not allow us to constrain the powerlaw slope or the $\nh$ of the intrinsic absorber.\\
	
\noindent 34. IRAS~F09320+6134: The alternative name of this source is UGC~05101 and was studied in a large sample of ULIRGs by \citet{2012ApJS..203....9U}. The authors have also estimated the infrared luminosity for all the sources. The X-ray flux estimated is $F_{2-10\kev}=8.64 \times 10^{-9} \funit$ which is higher than our estimate of $10^{-11.83} \funit$.	

Very low photon counts did not allow us to constrain the powerlaw slope or the $\nh$ of the intrinsic absorber.
\\

\noindent 35. IRAS~F12072-0444: This is a ULIRG and was studied by \citet{2010ApJ...725.1848T} who have calculated a luminosity of $L_{2-10\kev}=1.5 \times 10^{41} \lunit$.	
Very low photon counts did not allow us to constrain the powerlaw slope or the $\nh$ of the intrinsic absorber.\\

\noindent 36. IRAS~F12112+0305: This is a ULIRG and was studied by \citet{2010ApJ...725.1848T} and \citet{2011A&A...529A.106I} who have calculated a luminosity of $L_{2-10\kev}=1.5 \times 10^{41} \lunit$ and $4\times 10^{41} \lunit$ respectively.	

Very low photon counts did not allow us to constrain the powerlaw slope or the $\nh$ of the intrinsic absorber.\\	


\noindent 37. IRAS~F14348-1447: This is a ULIRG studied by \citet{2010ApJ...725.1848T} who estimated a luminosity of $L_{2-10\kev}=7.4 \times 10^{41} \lunit$  

Very low photon counts did not allow us to constrain the powerlaw slope or the $\nh$ of the intrinsic absorber.\\

\noindent 38. IRAS~F14394+5332: This source has not been studied before.

Very low photon counts did not allow us to constrain the powerlaw slope or the $\nh$ of the intrinsic absorber.\\


\noindent 39. IRAS~F15327+2340: This source is also known as ARP~220 and i a starburst galaxy. \citet{2012ApJ...758...82L} have worked on a sample of sources including this source where the authors have attempted to disentangle the AGN and starburst contribution of the sources in the $0.5-2 \kev$ soft X-ray band.
Also studied by \citet{2010ApJ...725.1848T} and they calculated a luminosity of $L_{2-10\kev}=1.2 \times 10^{41} \lunit$.	

Very low photon counts did not allow us to constrain the powerlaw slope or the $\nh$ of the intrinsic absorber.\\

\noindent 40. IRAS~F15462-0450: ULIRG studied by \citet{2010ApJ...725.1848T} and they calculated a luminosity of $L_{2-10\kev}=1.3 \times 10^{43} \lunit$ using HR method.	
Very low photon counts did not allow us to constrain the powerlaw slope or the $\nh$ of the intrinsic absorber.\\

\noindent 41. IRAS 19297-0406: This source is a very faint source in X-rays and studied as a part of C-goals by \citet{2011A&A...529A.106I}  who calculated an X-ray luminosity of $L_{2-10\kev}=1.8 \times 10^{41} \lunit$ using  the HR method. 

No significant counts. Hence we used previous studies to estimate the X-ray luminosity.	\\

\noindent 42. IRAS 19542+1110 : This source is a very faint source in X-rays and studied as a part of C-goals by \citet{2011A&A...529A.106I}  who calculated an X-ray luminosity of $L_{2-10\kev}=1.0 \times 10^{42} \lunit$ using  the HR method. 	

Very low photon counts did not allow us to constrain the powerlaw slope or the $\nh$ of the intrinsic absorber.\\	

\noindent 43. IRAS~F20551-4250: Also known as ESO 286-IG19 and studied by \citet{2011A&A...529A.106I}  who calculated an X-ray luminosity of $L_{2-10\kev}=2.1 \times 10^{41} \lunit$.	

Very low photon counts did not allow us to constrain the powerlaw slope or the $\nh$ of the intrinsic absorber.\\

\noindent 44. IRAS~F22491-1808: Both the papers \citet{2011A&A...529A.106I} and \citet{2010ApJ...725.1848T} discuss this source and the luminosity calculated by them are $0.6 \times 10^{41}$	and $2.1 \times 10^{41} \lunit$ respectively.

Very low photon counts did not allow us to constrain the powerlaw slope or the $\nh$ of the intrinsic absorber.\\


\noindent 45. NGC~5506: \citet{2010MNRAS.406.2013G} studied the source which is an X-ray obscured Seyfert galaxy with a broad Fe K$\alpha$ emission line. The maximum value of the luminosity quoted for this source is $L_{2-10\kev}=1.1 \times 10^{43} \lunit$.
The source is an obscured Seyfert 1 galaxy.\\	

\noindent 46. NGC~7479: \citet{2009A&A...500..999A} studied the source in a sample of Sy1, Sy2 and Compton thick galaxies, where they infer an X-ray luminosity of $L_{2-10\kev}=2.8 \times 10^{40} \lunit$. The source is a Seyfert 2 galaxy with significant obscuration.	

No significant counts. Hence we used previous studies to estimate the X-ray luminosity.\\		

\noindent 47. NGC~7172: \citet{1998MNRAS.298..824G} studied this Seyfert 2 source and calculated a luminosity of $L_{2-10\kev}=1.5 \times 10^{43} \lunit$.		
The source is an obscured Seyfert 1 galaxy.\\


\clearpage

\section{C. Sources in MOX sample without \nustar{} observations, as on December 2017.}

In this section we list the 23 sources in the MOX sample which have not been observed by \nustar{}. The sources are:IRAS~23365+3604, IZw1, NGC~1377, NGC~3628, NGC~2146, NGC~6764, IRAS~17208-0014, NGC~1614, IRAS05083+7936, IRAS13451+1232, 3C293, NGC1433, IRASF01572+0009, IRASF05024-1941, IRAS07251-0248, IRAS 09022-3615, IRASF12112+0305, IRASF14348-1447, IRASF14394+5332, IRASF15462-0450, IRASF19297-0406, IRAS19542+1110, and IRASF23233+2817.

\vspace{2cm}

$Acknowledgements:$  Author SL is grateful to Silvia Martocchia for supplying the X-ray bolometric corrections and AGN bolometric luminosities for the PG quasars and the high redshift bright quasar sample (WISSH). Author SL is also grateful to Daniel Stern, Sylvain Veilleux, Richard Mushotzky and Richard Rothschild for sharing interesting ideas involved in this work. This research has made use of the NASA/IPAC Extragalactic Database (NED) which is operated by the Jet Propulsion Laboratory, California Institute of Technology, under contract with the National Aeronautics and Space Administration.

\bibliographystyle{apj}
\bibliography{mybib}

\end{document}